\documentclass[aps,pre,epsf,superscriptaddress,amsmath,amssymb,amsfonts,twocolumn,showpacs,showkeys]{revtex4-1}
\usepackage{amsmath,amssymb,amsfonts,bm,comment}
\usepackage{graphics,float,tikz}
\usepackage[caption=false]{subfig}
\usepackage[colorlinks=true,
	linkcolor=blue,
	filecolor=magenta,      
	urlcolor=blue,
	citecolor=blue]{hyperref}
\usepackage{physics,xparse,mathtools,isotope}
\usepackage[table-parse-only]{siunitx}
\usepackage{multirow,array,relsize,lipsum,longtable,tabularx}
\newcommand{\im}{\mathrm{i}}

\begin{document}
\title{Pattern Formation and Evidence of Quantum Turbulence in Binary Bose-Einstein Condensates Interacting with a Pair of Laguerre-Gaussian Laser Beams }

\author{Madhura Ghosh Dastidar}
\affiliation{Department of Physics, Indian Institute of Technology Kharagpur, India}
\author{Subrata Das}
\email{subratappt@iitkgp.ac.in}
\affiliation{Department of Physics, Indian Institute of Technology Kharagpur, India}
\author{Koushik Mukherjee}
\affiliation{Department of Physics, Indian Institute of Technology Kharagpur, India}

\author{Sonjoy Majumder}
\email{sonjoym@phy.iitkgp.ac.in}
\affiliation{Department of Physics, Indian Institute of Technology Kharagpur, India}
\date{\today}

\begin{abstract}

	We theoretically investigate the out-of-equilibrium dynamics in a binary Bose-Einstein condensate confined within two-dimensional box potentials. One species of the condensate interacts with a pair of oppositely wound, but otherwise identical Laguerre-Gaussian laser pulses, while the other species is influenced only via the interspecies interaction. Starting from the Hamiltonian, we derive the equations of motion that accurately delineate the behavior of the condensates during and after the light-matter interaction. Depending on the number the helical windings (or the magnitude of topological charge), the species directly participating in the interaction with lasers is dynamically segmented into distinct parts which collide together as the pulses gradually diminish. This collision event generates nonlinear structures in the related species, coupled with the complementary structures produced in the other species, due to the interspecies interaction. The long-time dynamics of the optically perturbed species is found to develop the Kolmogorov-Saffman scaling law in the incompressible kinetic energy spectrum, a characteristic feature of the quantum turbulent state. However, the same scaling law is not definitively exhibited in the other species. This study warrants the usage of Laguerre-Gaussian beams for future experiments on quantum turbulence in Bose-Einstein condensates.

\end{abstract}
\maketitle

\section{Introduction}\label{Introduction}

Light-matter interaction utilizing laser beams is an essential tool to explore the physics of ultracold atoms~\cite{Chu1998, Cohen1998, Phillips1998, Ritsch2013}. Most research activities in that direction fall into two main categories. One category involves the quantum gases being coupled to the electromagnetic mode of a high-quality optical resonator~\cite{Slama2007, Ritsch2013}. This picture of light-matter interaction considers the quantum nature of both light and matter on equal footing~\cite{Walther_2006, Mekhov2015, Dutra2005} thereby facilitating the most promising and controllable implementation of various phenomena involving quantum many-particle physics~\cite{Cooper2019, Mivehvar2021}. In another category, spontaneous emission of the involved laser light is drastically suppressed, leading to the creation of an optical dipole potential resulting from the induced Stark shift~\cite{GRIMM200095, Garraway2000, Frese2000}. This latter regime, where the underlying electromagnetic field is treated classically, and its spatial profile plays the most significant role in determining the form of the dipole potential, has laid the foundation of the trapping and the manipulation of cold bosonic and fermionic gases~\cite{Cornell2002, Ketterle2002, Bloch2008, Giorgini2008} and mesoscopic particles~\cite{Gordon1980}.

A special type of electromagnetic wave is the Laguerre-Gaussian (LG) laser beam that possesses a transverse phase cross-section of  $e^{i l \phi}$, where $\phi$ is the angular variable, and $l$ is the phase winding number (often referred to as topological charge) of the beam~\cite{Allen1992, Alison2011, Miles2017, DENNIS2009}. Over the years, remarkable success has been achieved in the field of generation~\cite{Ren2010, Ruffato2014, Sueda2004, Beijersbergen1993, Beijersbergen1994, Heckenberg1992} and manipulation~\cite{Mair2001, ALLEN1999, Yao_2011, Franke-Arnold2008, Shen2019, Padgett2017} of LG beams, paving the way to further studies of light-matter interaction~\cite{Allen1996, Babiker2002, Lloyd2012, Power1995, Araoka2005, Bougouffa2020, Mashhadi_2017, Quinteiro2017, Mukherjee_2017}. In that direction, the interaction of LG beams with atomic Bose-Einstein condensate(BEC) is well-studied, both in and beyond the paraxial limit~\cite{Andersen2006, Mondal2014, Mondal2015, Bhowmik2016, Kanamoto2007, Wright2008, Bhowmik_2018_paraxial, Bhowmik_2018, Das_2020}. The optical trap created by the LG beam has been used to confine BEC and investigate the quantized vortex states~\cite{Tempere2001}. Moreover, such LG modes transfer orbital angular momentum from the optical field to the BEC, thus creating a single vortex~\cite{Marzlin1997, Simula2008, Nandi2004} or a superposition of vortices in the latter~\cite{Bhowmik2016}. On-demand particle transfer between two BECs has also been demonstrated, using a pair of LG and Gaussian laser beams~\cite{Mukherjee2021}.

The impacts of LG beam on the BEC have been discussed majorly in the context of coherent particle transfer between different states and vortex imprinting~\cite{Mukherjee2021}. One largely unexplored arena is exploiting the interference phenomenon using two LG beams~\cite{HUANG2016132, Tao2006, YANG20113597, Franke_2007} incident onto the BEC in 2D, and thereby, triggering out-of-equilibrium dynamics in the latter. The out-of-equilibrium dynamics in BEC are interesting in their own right.  Many intrinsic condensate parameters~\cite{chin2010, Thorsten2006} and the dimensionality of the trapping potential~\cite{ Vogels2001} of the condensate can be remarkably controlled in the experiments~\cite{ Bloch2008}. Consequently, a plethora of density-wave-pattern~\cite{Nicolin2011, Staliunas2002, Engels2007, Maity2020,Zhang2020} and non-linear structure forming phenomena~\cite{Anglin1999, Mukherjee_2020, KEVREKIDIS_2004, KEVREKIDIS2004, Fetter_2001} stemming from dynamical means such as periodic modulation~\cite{Maity2020} or quenching a system parameter~\cite{Law2010} has been vastly investigated. Such non-equilibrium dynamics often showcase the spectral scaling law indicating the emergence of quantum turbulence~\cite{ Allen_2014, TSATSOS20161, White2012, White4719, Madeira2020} that is phenomenologically very much similar to classical turbulence~\cite{Kraichnan_1980}. Infact, the convincing evidences of this similarity have been demonstrated in the experiments~\cite{Navon2016, Navon2019}.

Motivated by the studies mentioned above, we attempt to develop a light-matter interaction mechanism that allows us to dynamically imprint the interference profiles of a pair of LG beams onto the BEC and subsequently unravel the emergent density patterns and energy transport phenomenon~\cite{Horng2009}. Our starting point is a binary Bose-Einstein condensate, made of two different atomic elements (alias species A and species B), confined in 2D box potential of the same size. Species A interacts with the pair of LG pulses applied simultaneously, while species B does not directly participate in this interaction. Such selective interaction can be experimentally performed by employing the so-called "tune-in" and "tune-out" approaches~\cite{LeBlanc2007}. Moreover, the involved LG pulses possess charges of the same magnitude but an opposite sign, and as a result, their interference produces azimuthally varied potentials~\cite{Alexander2004}(see the discussion below). We consider overlapping BECs~\cite{Ao1998} whose interspecies interaction can be tuned through Feshbach resonance~\cite{chin2010}. This framework allows exploring more non-trivial structures, such as vortex~\cite{KASAMATSU2005} and vortex-bright~\cite{Law2010}, and the energy exchange between the species for varying interaction strengths. Note that our objective in this paper is to identify the density patterns that evolve dynamically due to the interference of two pulses onto species A. Non-uniform density profiles stemming from  harmonic confinement can make this identification extremely difficult. Hence, we choose a box potential rendering the initial density profiles of the BEC uniform. Most importantly, the 2D box potentials have already been experimentally implemented~\cite{Ville2018, Chomaz2015} and are accessible to the current state-of-the-art experimental setting.

We consider two separate cases involving different magnitudes of charges of the LG laser pulses. For single unit charge, species A is separated into two parts after the pulses are applied, and the segments merge as the pulses gradually diminish. This merging process creates soliton stripes which eventually break into vortex-antivortex pairs~\cite{Ma2010}. When a finite interspecies interaction is considered (within the miscible regime), the non-linear structures in species A are coupled by complementary structures in species B. For example, in species B, a bright soliton is created precisely at the same position where a vortex is located in species A~\cite{Ma2010}. Moreover, we assert that a transfer of incompressible kinetic energy~\cite{nore1997kolmogorov} from species A to species B is responsible for the vortex creation in the latter. The corresponding incompressible spectrum develops $k^{-5/3}$ and $k^{-4}$ power-law scalings at low and high momentum, respectively~\cite{Saffman_1971}. However, such scaling is not apparent in species B. The pattern formation and the spectral scaling law compose the main products of this work. We observe that species A is segmented into four parts for the double unit charge and gives rise to a similar to the above-described overall dynamics, but the relevant phenomenology is less prominent than that of the unit charged one.

The rest of the paper is arranged as follows. Sec.~\ref{THEORY} describes our setup, the Hamiltonian and the equations governing the dynamics during and after the incidence of light pulses.  Here, we also discuss  the  observables which we manipulate to monitor the dynamics of the system. In Sec~\ref{Dynamics of a binary mixture with tunable miscibility}, we illustrate the particle density profiles of the condensates during the dynamics, for both singly and doubly charged LG beams. We calculate the time evolution of the different parts of the kinetic energy and their densities in Sec.~\ref{KEC}. We characterize the scaling laws for the incompressible and compressible kinetic-energy spectra in Sec.~\ref{spectra}.
We conclude the paper in Sec.~\ref{Conclusion}, discussing the implications and scope of our results. We present the spatial and temporal profiles of the electric field vectors associated with the LG laser pulses in Appendix~\ref{field}, where we also highlight the strength of interaction between the BEC and the optical beams. Finally, Appendix~\ref{ap:EOM} provides the detailed derivation of the equations of motion using the Hamiltonian describing our system.

.   \begin{figure}[h]
	\centering
	\includegraphics[width=0.9\linewidth]{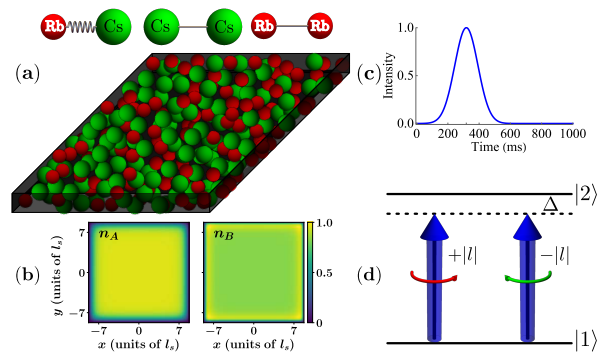}
	\caption{Schematic of the considered setup. (a) Binary BEC consisting of $^{87}$Rb and $^{133}$Cs atoms is confined within a quasi-2D box potential. (b) Initial densities of particles in the condensates for species A ($n_A$) and Species B ($n_B$). (c) A pair of LG pulses, having the same temporal width and peak position, (d) winding numbers with the same magnitude $\abs{l}$ but opposite orientations, couple $\ket{1}$ and $\ket{2}$, the two internal states of the $^{87}$Rb atom. The intraspecies interaction of each species is fixed, while that of interspecies is tunable via Feshbach resonance.}
	\label{schmetic}
\end{figure}

\section{Setup}\label{THEORY}
\subsection{Hamiltonian and Equations of motion}
We consider a binary mixture of BECs comprising an equal number of $^{87}$Rb and $^{133}$Cs atoms~\cite{McCarron2011} and both being confined in a 2D homogeneous box potential of the same size; see the schematic~\ref{schmetic}(a).  To trigger the light-matter interaction, only the $^{87}$Rb atoms are exposed to a pair of LG laser pulses possessing winding numbers $l_1$ and $l_2$, respectively, whereas the $^{133}$Cs atoms are assumed to be uninfluenced by the external light sources. The laser pulses are considered to be propagating parallel and co-linear to the $z$-axis, and therefore, both the laser pulses and BEC share the same transverse ($x$-$y$) plane. In particular, we consider that the applied laser pulses couple two internal states, namely, $\ket{1}$ and $\ket{2}$, as shown in  Fig.~\ref{schmetic}(d). The frequencies of the pulses are considered to be far-detuned from the resonant frequency between the two aforementioned internal states. Since the detuning is large herein, photon scattering can be neglected compared to the dipole force. Consequently, a negligible amount of population resides at $\ket{2}$, which allows us to eliminate the excited state adiabatically~\cite{GRIMM200095}.

Let $\hat{\Psi}^\dagger_{A,k}$[$\hat{\Psi}_{A,k}$] and $\hat{\Psi}^\dagger_B$[$\hat{\Psi}_B$] be the bosonic creation[annihilation] operators for the atoms of species A in $\ket{k}(k = 1, 2)$ state and those of species B, respectively.

The Hamiltonian of the system reads~\cite{Zeng1995}
\begin{widetext}
	\begin{equation}\label{ham_two_com}
		\begin{split}
			H &  =   \int \mathrm{d}\vb{r} \Bigg [ \hat{\Psi}^{\dagger}_{A,1}(\vb{r}, t)\hat{h}_{A} \hat{\Psi}_{A,1}(\vb{r}, t) +
				\hbar \Delta \hat{\Psi}_{A, 2}^\dagger(\vb{r},t) \hat{\Psi}_{A,2}(\vb{r},t)  + \hat{\Psi}^{\dagger}_{B}(\vb{r}, t)\hat{h}_{B} \hat{\Psi}_{B}(\vb{r}, t) +  \Big (\hbar\Omega(\vb{r},t)e^{\im l_1 \phi}\hat{\Psi}_{A, 2}^\dagger(\vb{r},t)\hat{\Psi}_{A,1}(\vb{r},t) \\ & + \hbar\Omega(\vb{r},t)e^{\im l_2 \phi}\hat{\Psi}_{A,2}^\dagger(\vb{r},t)\hat{\Psi}_{A, 1}(\vb{r},t) + \mathrm{h.c.} \Big ) +
				\frac{U_{AA}}{2}\hat{\Psi}_{A,1}^\dagger(\vb{r},t)\hat{\Psi}_{A,1}^\dagger(\vb{r},t)\hat{\Psi}_{A,1}(\vb{r},t)\hat{\Psi}_{A,1}(\vb{r},t) +
				\frac{U_{BB}}{2}\hat{\Psi}_B^\dagger(\vb{r},t)\hat{\Psi}_B^\dagger(\vb{r},t)\\&
				\times \hat{\Psi}_B(\vb{r},t)\hat{\Psi}_B(\vb{r},t) +
				\frac{U_{AB}}{2}\hat{\Psi}_{B}^\dagger(\vb{r},t)\hat{\Psi}_{A,1}^\dagger(\vb{r},t)\hat{\Psi}_B(\vb{r},t)\hat{\Psi}_{A,1}(\vb{r},t) \Bigg ].
		\end{split}
	\end{equation}
\end{widetext}

The first and second terms in the above Eq.~\eqref{ham_two_com} correspond to the kinetic energy of the atoms in the states $\ket{1}$ and $\ket{2}$, respectively, of species A, and the third term determines the kinetic energy of species B. The notation $\hat{h}_{p}$ denotes the kinetic energy operator of the $p$-th ($p = A, B$) species. The forth and fifth terms designate the energies stemming from the interaction of the species A with the LG laser modes having $+l_1$ and $+l_2$ units of charge, respectively. The interatomic interaction energies within the species A and species B are characterized by the sixth and seventh term, respectively, of Eq.~\eqref{ham_two_com}. The last term  denotes the interaction energy between an atom of species A and that of species B.  The coupling between the states $\ket{1}$ and $\ket{2}$ through the LG laser mode of the winding number $l_{\sigma}$ is characterized by the   absolute value of the Rabi frequency $\abs{\Omega_{l_\sigma}(\vb r, t)}$($\sigma = 1, 2$), which  can be expressed via the relation $|\Omega_{l_\sigma}(\vb{r},t)|^2 = \Omega^2_{0} e^{-((t-\tau)^2/T^2)}(x^2 + y^2)^{|l_{\sigma}|}e^{-2((x^2+y^2)/w^2)}$(see the appendix~\ref{field}). In the previous expression, both laser modes have the same profile, with the temporal peak position at $\tau$, pulse width $T$ and beam waist $w$.  Moreover, the coefficients of the intra- and interspecies interaction strengths are given by $U_{pp} = 4 \pi \hbar^2a_{pp}/m_{p}$ and $U_{pq} = 2 \pi \hbar^2 a_{pq}/m_{pq}$, respectively~\cite{Pethick2008}. Here, $a_{pp}$ and $a_{pq}$ are  the intra- and interspecies $s$-wave scattering lengths of atoms, $m_{p}$ is the mass of the atom of the $p$-th($p=A,B$) species, and $m_{pq} = m_{p}m_{q}/(m_{p} + m_{q})$ is the reduced mass. Utilizing the commutation relations obeyed by the bosonic field operators and Heisenberg equations of motion, one can derive the dynamical equations for $\hat{\Psi}^\dagger_{A, 1}$, $\hat{\Psi}^\dagger_{A, 2}$ and $\hat{\Psi}^\dagger_{B}$. Then, setting $\partial \hat{\Psi}_{A,2}^\dagger/\partial t = 0$, as negligible number of atoms of species A populate the state $\ket{2}$ due to large detuning $\Delta$, and replacing the field operator $\hat{\Psi}_p(\vb r, t)$ by $\Psi_p(\vb r, t)$(to this end we will use $\Psi_{A}$ to refer to the state $\Psi_{A,1}$ just for the notational convenience), we obtain (see the Appendix~\ref{ap:EOM})
\begin{align}
	\label{GP_eqn1}
	 & i\pdv{\Psi_A(\vb{r},t)}{t} = \big[ -\frac{1}{2}\laplacian_\perp + \mathcal{G}_{AA} \abs{\hat{\Psi}_A(\vb{r},t)}^2 + \mathcal{G}_{AB} \abs{\Psi_B(\vb{r},t)}^2  \nonumber \\
	 & -  \mathcal{V}(t)r^{(|l_1| + |l_2|)}e^{-2\frac{r^2}{w^2}} \cos^2{\frac{(l_1-l_2)\phi}{2}}\big]
	\Psi_A (\vb{r},t),
\end{align}
and
\begin{align}
	\label{GP_eqn2}
	i\pdv{\Psi_B(\vb{r},t)}{t} = & \big[ -\frac{1}{2}\laplacian_\perp + \mathcal{G}_{BB} \abs{\hat{\Psi}_B(\vb{r},t)}^2 + \mathcal{G}_{AB} \abs{\Psi_A(\vb{r},t)}^2 \big]\nonumber \\&
	.\times \Psi_B (\vb{r},t).
\end{align}

Here, $r^2 = x^2 + y^2$, $\mathcal{V}(t) = \mathcal{V}_{\rm max}\exp[ -(t - \tau)^2/T^2 ]$, with $\mathcal{V}_{\rm max} = \Omega^2_0 m_A l^2_s/(\hbar\Delta)$ being related to maximum Rabi-frequency. The intra- and interspecies coupling strengths are $\mathcal{G}_{pp} = 2N\sqrt{2\pi \lambda}a_{pp}/l_{s}$ and $\mathcal{G}_{AB} = N\sqrt{2 \pi \lambda}a_{AB}/l_{s}$, respectively~\cite{Soumik2017, Mukherjee_2020}. Here, $N$ is the particle number in each species. The length scale $l_s$  determines the size of the box potential, and $\lambda = l_{s}/l_z$ is an anisotropy parameter. The system is tightly confined in the $z$-direction by applying a strong trapping frequency $\omega_z/(2\pi)$ so that $l_z= \sqrt{\hbar/(m\omega_z)}$ is much smaller than $l_s$. Note that the Eqs.~\eqref{GP_eqn1} and ~\eqref{GP_eqn2} are the dimensionless forms of Eq.~\ref{GP_eq_A1} and Eq.~\ref{GP_eq_B}(Appendix~\ref{ap:EOM}) . To cast the equations into dimensionless forms, we have scaled the spatial coordinates by $l_{s}$, time by $m_A l^2_{s}/\hbar$, and the condensate wavefunctions by $\sqrt{N/l^3_{s}}$.

\subsection{Details of the Numerical Simulations }
To investigate the LG modes assisted dynamics of the binary BECs, we numerically solve Eqs.~\eqref{GP_eqn1} and \eqref{GP_eqn2} by utilizing the well known split-time Crank-Nicolson method~\cite{muruganandam2009fortran}. Note that the very first step of the computation is to obtain an initial wavefunction (in the absence of the laser sources) for each species. To that purpose, we consider $\mathcal{V}_{\rm max} = 0$ in Eq.~\eqref{GP_eqn1} and  propagate the Eqs.~\eqref{GP_eqn1} and \eqref{GP_eqn2} in imaginary time, until the solution converges to an equilibrium state. Moreover, the normalization of the species' wavefunctions is preserved by applying the following transformation $\Psi_p \rightarrow \frac{\Psi_{p}}{\norm{\Psi_{p}}}$ to the $p$-species wavefunction at each instant of the imaginary time propagation. Having equipped with the initial state, we evolve the system in the presence of laser pulses. For this, we solve Eqs.~\eqref{GP_eqn1} and \eqref{GP_eqn2} in real time. The numerical computations are carried out in a square grid of $400 \times 400$ grid points with grid spacing $\Delta x = \Delta y  = 0.05$, and the time step of integration $\delta t = 10^{-4}$. The intracomponent scattering lengths of the two components are $a_{AA} = 100.04a_{0}$~\cite{McCarron2011} and $a_{BB} = 280a_{0}$, $a_0$ being the Bohr radius, and the total number of atoms in each species is $N = 10^4$. We choose $l_{s} = 2.8 \mu$m and $\lambda = 25$.

\subsection{Observables}
In the following, we introduce the observables  used in the remainder of our work to monitor various distinctive features of the system in the non-equilibrium state.\\
To visualize the spatially resolved nature of the system during and after the light-matter interaction, we resort to the particle densities $n_{p}$ of the $p$-th species, defined as $n_p(x, y; t)= \abs{\Psi_p}^2(x, y;t)$. The phase $\phi_p(x,y;t)$ of the $p$-th species' wavefunction provides the concrete signature of vortex formation during the dynamics. In particular, $\phi_p(x, y; t)$ changes by an amount $\kappa \times 2 \pi$ around each vortex possessing a $\kappa$ unit charge.  Moreover, the spatially resolved measure of the vorticity can be elucidated by vorticity of the $p$-th species~\cite{Mukherjee_2020},
\begin{align}
	\boldsymbol{\Omega}_p = \boldsymbol{\nabla} \times \boldsymbol{J}_{p},
	\label{vorticity_eq}
\end{align}
where $\boldsymbol{J}_{p} = \frac{i}{2}(\Psi^{*}_p \boldsymbol{\nabla} \Psi_p - \Psi_p \boldsymbol{\nabla} \Psi^{*}_{p})$ is the probability current. Defining the superfluid velocity~\cite{Pethick2008} $\boldsymbol{u}_{p} = \boldsymbol{\nabla}\phi_p$, and on applying the Madelung transformation~\cite{madelung1927quantentheorie} $\Psi_p = n_{p}e^{i\phi_p}$, the Eq.~\eqref{vorticity_eq} can be cast into a form,
\begin{align}
	\boldsymbol{\Omega}_p = \nabla\sqrt{n_p} \times \boldsymbol{u_p} + \sum_{j}^{}2 \pi k_j\sqrt{n_p}\delta(\bf{r} - \bf{r}_j)\hat{\bf{z}},
	\label{vorticity_eq2}
\end{align}
where the delta function determines the location of the $j$-th vortex with charge $k_j$. Interestingly, according to the Eq.~\eqref{vorticity_eq2}, the vortical content of the $p$ species arises from the two different origins. The first term in Eq.~\eqref{vorticity_eq2} is related to the generation of vorticity due to a density gradient, whereas the second term is the usual contribution in a superfluid made by the quantized vortices (and therefore has no classical analog).\\
In order to characterize the dynamics from the perspectives of the generated vortices and sound waves (acoustic waves) in the system, we calculate the total kinetic energy term of the $p$-th species, given by~\cite{Pethick2008},
\begin{align}
	\label{GP_kin}
	\frac{1}{2}\abs{\Psi_{p}}^2 = \frac{1}{2}\Big(n_{p}\boldsymbol{u}^2_{p} + \abs{\boldsymbol{\nabla}\sqrt{n_p}}^2 \Big).
\end{align}
The first and second terms in Eq.~\eqref{GP_kin} represent the kinetic ($E_{kin,p}$) and quantum pressure ($E_{q,p}$) energy, respectively, of the $p$-th species. The velocity vector $\sqrt{n}_{p}\boldsymbol{u}_{p}$ is decomposed into a solenoidal part (incompressible) $\boldsymbol{u}^{ic}_{p}$ and an irrotational (compressible) part $\boldsymbol{u}^{c}_{p}$, i.e., $\boldsymbol{u}_{p} = \boldsymbol{u}^{ic}_{p} + \boldsymbol{u}^{c}_{p}$ such that $\boldsymbol{\nabla}.\boldsymbol{u}^{ic}_{p} = 0$ and $\boldsymbol{\nabla} \times \boldsymbol{u}^{c}_{p} = 0$~\cite{Nore1997, Mithun2021}. This allows us to define the scalar potential $\Phi_p$ and the vector potential $\boldsymbol{A}_p$ obeying the relations $\sqrt{n}_{p}\boldsymbol{u}_p^{c} = \nabla \Phi_p$ and $\sqrt{n}_{p}\boldsymbol{u}_p^{ic} = \nabla \times \boldsymbol{A}_p$, respectively. Taking the divergence on both sides of the last expression, we arrive at the Poisson equation for the scalar potential, $\nabla^2\Phi_p = \nabla.(\sqrt{n_{p}} \boldsymbol{u}_p)$, which is solvable to determine $\Phi_p$ and subsequently, $\boldsymbol{A}_p$. Once $\Phi_p$ and $\boldsymbol{A}_p$ are known the components $\boldsymbol{u}_p^{ic}$ and $\boldsymbol{u}_p^c$ can be easily determined. Therefore, the incompressible and compressible kinetic energies are defined as
\begin{align}
	\label{kins_GP1}
	E^{\rm ic}_{{\rm kin},p} = \frac{1}{2}\int n_{p}\abs{(u^{ic}_{p})}^2dxdy,
\end{align}
and
\begin{align}
	\label{kins_GP2}
	E^{\rm c}_{{\rm kin},p} = \frac{1}{2}\int n_{p}\abs{(u^{c}_{p})}^2dxdy,
\end{align}
respectively. In the $k$-space (wave number), the angle-averaged incompressible and compressible kinetic energy spectrum of the $p$-th species is represented by~\cite{Mithun2021}
\begin{align}
	\label{kin_spec}
	E^{\rm ic[c]}_{{\rm kin},p}(k)=\frac{k}{2}\sum_{j = x, y}^{} \int_{0}^{2\pi}|F_j(\sqrt{n}\boldsymbol{u}_{j,p}^{ic[c]})|^2d\phi.
\end{align}
where $F_j(\sqrt{n}\boldsymbol{u}_p^{ic[c]})$ denotes the Fourier transform of $\sqrt{n}\boldsymbol{u}_{j,p}^{ic[c]}$ corresponding to the $j$-th component of $\boldsymbol{u}_p = (u_{x, p}, u_{y,p})$. Moreover, to perfrom the integral in Eq.~\eqref{kin_spec}, we numerically add the grid points with $(k^2_x+k^2_y)^{1/2}=k$, where $k_x$ and $k_y$ are the Cartesian components of the wave vector $k$.

\section{Characteristics of the Light-matter interaction }\label{Dynamics of a binary mixture with tunable miscibility}

In this section, we describe the dynamics ensued in the binary BEC when one species is triggered by a pair of LG laser pulses with topological charges $l_1 = +l$ and $l_2 = -l$, and having equal intensities characterized by the $\mathcal{V}_{max}$, keeping the other species unaffected. In particular, the system's evolution is investigated for two specific values of charge, namely, $\abs{l} = 1$ and $\abs{l} = 2$, and from zero to finite repulsive interspecies scattering length $a_{AB}$. The dynamics is first  analyzed, in order to gain an overview of the dynamical evolution, by employing the corresponding density $n_p(x, y, t)$, phase $\phi_p(x, y,t)$ and vorticity $\Omega_{p}(x, y, t)$ evolution of the participating components. These observables serve as the foundations upon which we carry out a  more detailed analysis resorting to the different parts of their kinetic energy to further our understanding of the system.

\subsection{Singly charged LG laser pulses}\label{OAM 1}
We begin discussing  our numerical results by analyzing the non-equilibrium dynamics of the 2D binary bosonic system when both laser pulses possess one unit charge but with opposite signs, namely, $l_1 = 1$ and $l_2 = -1$. In particular, two LG pulses are incident simultaneously along the $z$-direction, on the species A lying in the $x$-$y$ plane. Moreover, the interaction parameter $\mathcal{V}_{max} = 500$ and the beam waist $w = 30l_{s}$ have the same values for both pulses. First, setting $a_{AB} = 0$,  we probe the dynamics of the non-interacting species. Afterwards, we move to the case of finite interspecies interaction and draw a comparison with the former to identify the possible alterations.

\begin{figure}[h]
	\centering
	\includegraphics[width=\linewidth]{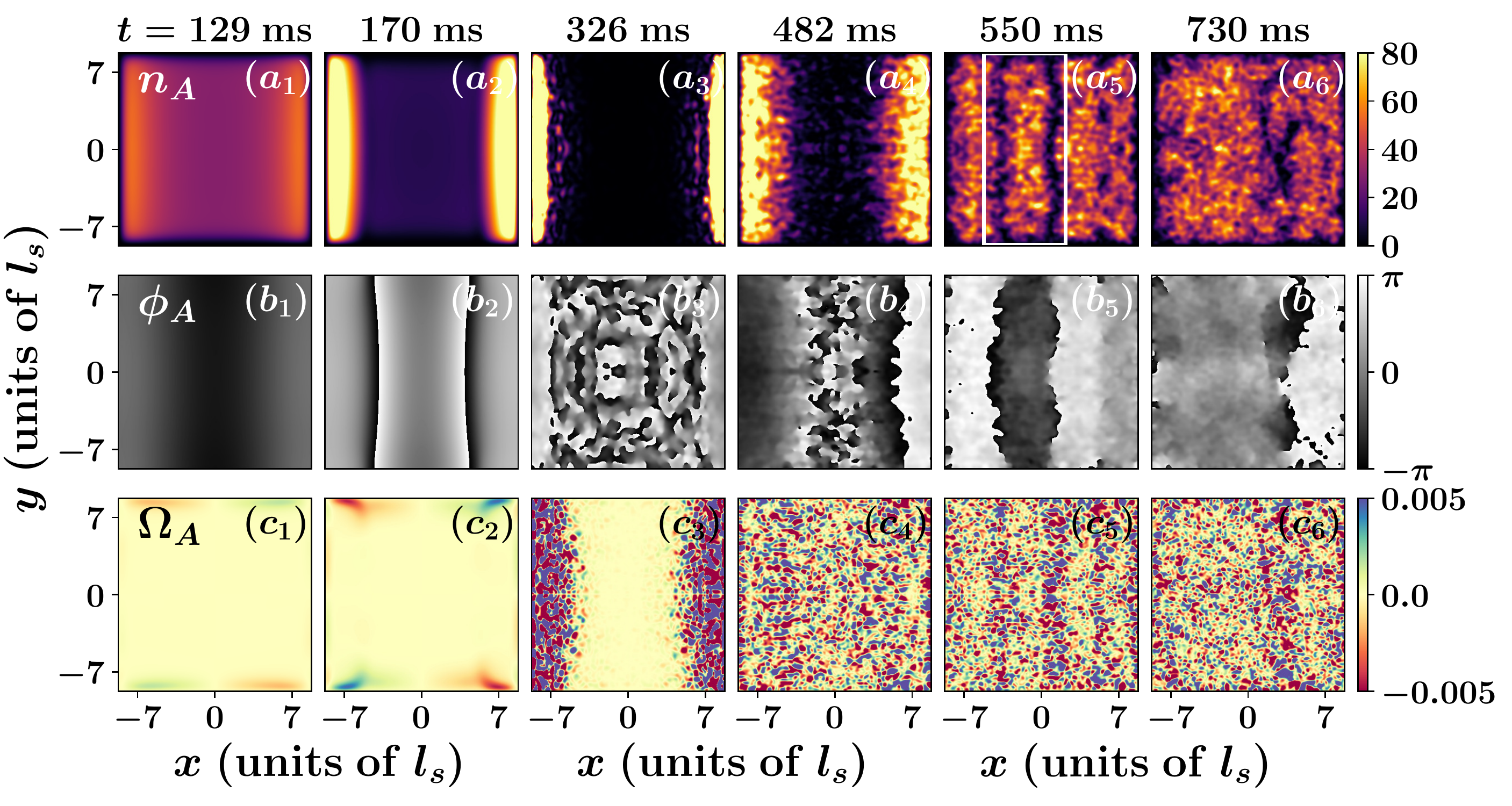}
	\caption{(Color online) ($a_1$)-($a_6$) Density, ($b_1$)-($b_6$) phase, and ($c_1$)-($c_6$) vorticity profiles of  species A at various time instants (see legends). The binary BEC consists of $N_A = N_B = 10^4$ atoms with zero interspecies interaction. Each species is confined within a 2D box potential characterized by the length scale $l_s$. The intra- and interspecies scattering lengths are $a_{AA} = 100.04a_0$, $a_{BB} = 280a_{0}$, and $a_{AB} = 0$, respectively. The dynamics is initiated by exposing species A to a pair of LG laser pulses with wingdings $l_1 = 1$ and $l_2 = -1$, respectively.}
	\label{Oam1_G0.eps}
\end{figure}

We illustrate evolution of the density, phase, and vorticity profiles of species A at different stages of the dynamics in Fig.~\ref{Oam1_G0.eps}. The initial state of the system, at $t = 0$ ms, showcases a uniform density distribution for both species, like the ones shown in Fig.~\ref{schmetic}(b). Since we considered the case of non-interacting species, the uniform density distribution of species B remained unaltered throughout our observation( not shown for the brevity).

Focusing on species A, we can notice the changes in the density profiles  [Figs.~\ref{Oam1_G0.eps}($a_1$)-($a_6$)] as the optical potential contributed by the two laser pulses gradually strengthens, modifying the initial uniform box geometry of the species. For instance, at $t = 129$ms when the pulses are just incident, the density of species A shows a tendency of separation into two vertical segments; see Fig.~\ref{Oam1_G0.eps}($a_1$). Particularly, owing to the Gaussian temporal profiles of the pulses, the segregation of the density of species A occurs gradually. Therefore, at $t = 170$ms, the species A is noticed to be further segregated into two distinguishable segments  [Fig.~\ref{Oam1_G0.eps}($a_2$)]. Note that the segregation is still not completed because LG pulses do not have the maximum amplitudes yet. Moving further in time, at $t = 326$ms, one can notice the broad dark region at the center, and species A is largely accumulated on both sides of the box~ [Fig.~\ref{Oam1_G0.eps}($a_3$)]. This phenomenon corresponds to the time instant when the pulses have finally reached the maximum amplitude. Most importantly, this dark region dividing species A is contributed  by the optical potential, which achieves the maximum values at $\phi = \pi/2$(or $3\pi/2$) when $l_1 = 1$ and $l_2 = -1$. The density deformation, as per the first term of Eq.~\eqref{vorticity_eq2}, gives rise to vortices within the two segments; see the corresponding vorticity profile in Fig.~\ref{Oam1_G0.eps}($c_3$). Note that these vortices are not associated with the typical phase singularities of the wavefunction of species A, which is evident from the continuous phase structure in the region (on the sides of the box) where the non-negligible density of species A exists  [Fig.~\ref{Oam1_G0.eps}($a_3$) and \ref{Oam1_G0.eps}($b_3$)]. After that, as the pulses progressively reach towards the end, the segments of species A gradually approach each other and finally collide, for example, see Fig.~\ref{Oam1_G0.eps}($a_4$). This collision takes place because the pulses' intensity gradually diminishes, and the corresponding optical potential is too weak to set up a barrier within species A. Furthermore, as can be seen at $t = 550$ ms   [Fig.~\ref{Oam1_G0.eps}($a_5$)], species A restores the original box geometry and forms what resembles a solitonic stripe~\cite{Mukherjee_2020} as a result of the collision event. Subsequently, these stripes break into quantized vortices [see Fig.~\ref{Oam1_G0.eps}($b_6$)] that exhibit very erratic behavior which persists even in the long time dynamics; see also the Figs.~\ref{Oam1_G0.eps}($c_5$) and \ref{Oam1_G0.eps}($c_6$). Finally, let us comment that since the interspecies interaction is zero, species B is not affected by the light-induced dynamics occurring in species A and retains its initial state throughout the time evolution (densities not shown for brevity).
\begin{figure}[h]
	\centering
	\includegraphics[width=\linewidth]{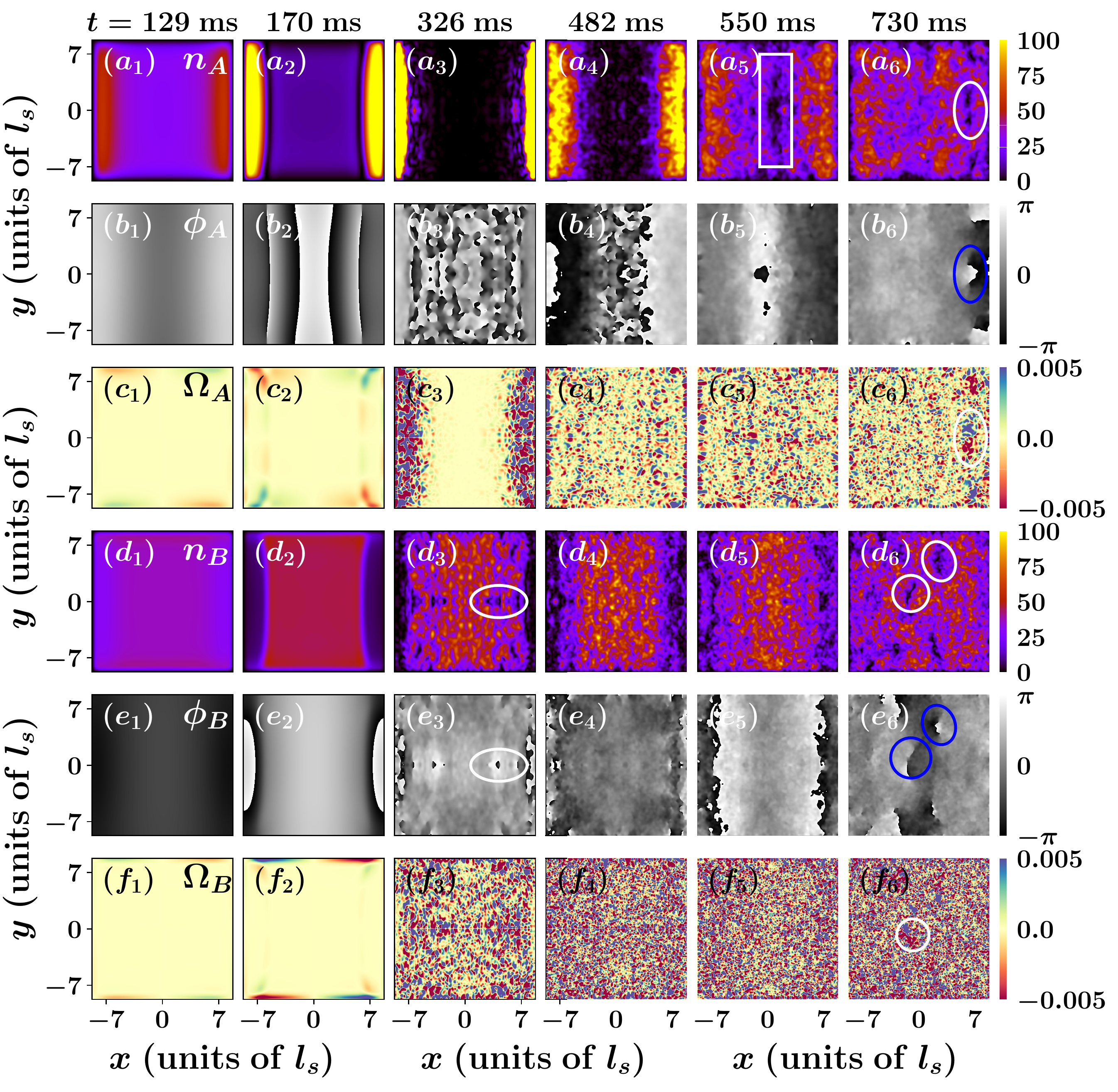}
	\caption{(Color online) Density profiles of ($a_1$)-($a_6$) species A and ($d_1$)-($d_6$) species B at various time instants (see legends). Shown also corresponding phase [vorticity] profiles ($b_1$)-($b_6$) [($c_1$)-($c_6$)] of species A and ($e_1$)-($e_6$) [($f_1$)-($f_6$)] of species B. The binary BEC consists of $N_A = N_B = 10^4$ atoms. Each species is confined within a 2D box potential characterized by the length scale $l_s$. The intra- and interspecies scattering lengths are $a_{AA} = 100.04a_0$, $a_{BB} = 280a_{0}$, and $a_{AB} = 100a_0$, respectively. The dynamics is initiated by exposing species A to a pair of LG laser pulses with wingdings $l_1 = 1$ and $l_2 = -1$, respectively.}
	\label{Oam1_G12.eps}
\end{figure}

Having explicated the behavior of the system in the absence of interspecies interaction above, it is now imperative to assess the effect of a finite interspecies interaction. Note that for the value,  $a_{AB} = 100a_0$, the scattering length that we consider in the subsequent discussion ensures that there is an overlap between both species in the initial state. The advantage of such a miscible state is two folds. First,  it captures  within species B the possible fingerprints of the light-matter interaction occurring in species A. Second, the role played by species B in modifying the optical potential induced dynamics of the other (species A) can also be elucidated in this setting. Figure~\ref{Oam1_G12.eps} presents the snapshots of the density, phase and vorticity profiles of both species at different stages of the dynamics. In particular, we have considered the same time instants as those in the case of no interaction between the components to compare two scenarios. In fact, at $t =129$ ms the density of species A displays the same tendency of separation as that for the $a_{AB} = 0$ case. Moreover, those segmented parts located on the sides of the box act as the local potential humps (related to the term $\mathcal{G}_{AB}\abs{\Psi_A}^2$) for the species B. Therefore, species B mostly occupies the lower density region of species A, a result which becomes increasingly evident as the pulses progress to reach the maxima~[compare Figs.~\ref{Oam1_G12.eps}($a_1$)-($a_3$) to Figs.~\ref{Oam1_G12.eps}($d_1$)-($d_3$)]. Furthermore, at $t=326$ms, as can be deduced from the Fig.~\ref{Oam1_G12.eps}($c_3$), vortex-antivortex pairs emanating from the density distortions~ [Fig.~\ref{Oam1_G12.eps}($a_3$)] develop within the segments of species A. Similarly, within species B, since it is pushed away from both sides towards the central region of the geometry, severe density distortion occurs, leading to the generation of vortex-antivortex structures [Figs.~\ref{Oam1_G12.eps}($d_3$) and \ref{Oam1_G12.eps}($f_3$)]. One can notice the emergence of the quantized vortices within species B; see the circles in Figs.~\ref{Oam1_G12.eps}($d_3$) and \ref{Oam1_G12.eps}($e_3$). Finally, when the optical potential is completely removed, for instance, at $t = 550$ ms, both the species are brought back to the initial box potential, and the solitonic stripes form within the components. Consequently, these stripes bend and break into quantized vortex-antivortex structures; see the circles in Figs.~\ref{Oam1_G12.eps}($a_6$), \ref{Oam1_G12.eps}($b_6$), \ref{Oam1_G12.eps}($d_6$) and \ref{Oam1_G12.eps}($e_6$). Furthermore, a closer inspection of the vorticity profiles  [Figs.~\ref{Oam1_G12.eps}($d_6$), \ref{Oam1_G12.eps}($f_6$)] reveals that vortices arising from the density distortions have lesser magnitudes than those of the quantized vortices[the circles in Figs.~\ref{Oam1_G12.eps}($d_6$), \ref{Oam1_G12.eps}($f_6$)]; see also the phase profiles. Indeed, in the long-time dynamics, both species are filled with vortex-antivortex structures, which we shall expound on in the later sections while characterizing the energetics of the system.

\subsection{Doubly charged LG laser pulses}
In the previous section, we have gained intuition regarding the fundamental features of the light-induced dynamics in species A and its impact on species B through the interspecies interaction. Next, we discuss the ensuing dynamics when the applied LG laser pulses carry a charge of two units, more precisely, $l_1 = 2$ and $l_2 = -2$. We consider the interaction parameter $\mathcal{V}_{max} =1500 $, the beam waist $w =30 l_{\rm s}$ to have the same values for both the pulses. As in the previous discussion, we first examine the behavior of the non-interacting species and then compare our findings to that of the interacting ones.

\begin{figure}[h]
	\centering
	\includegraphics[width=\linewidth]{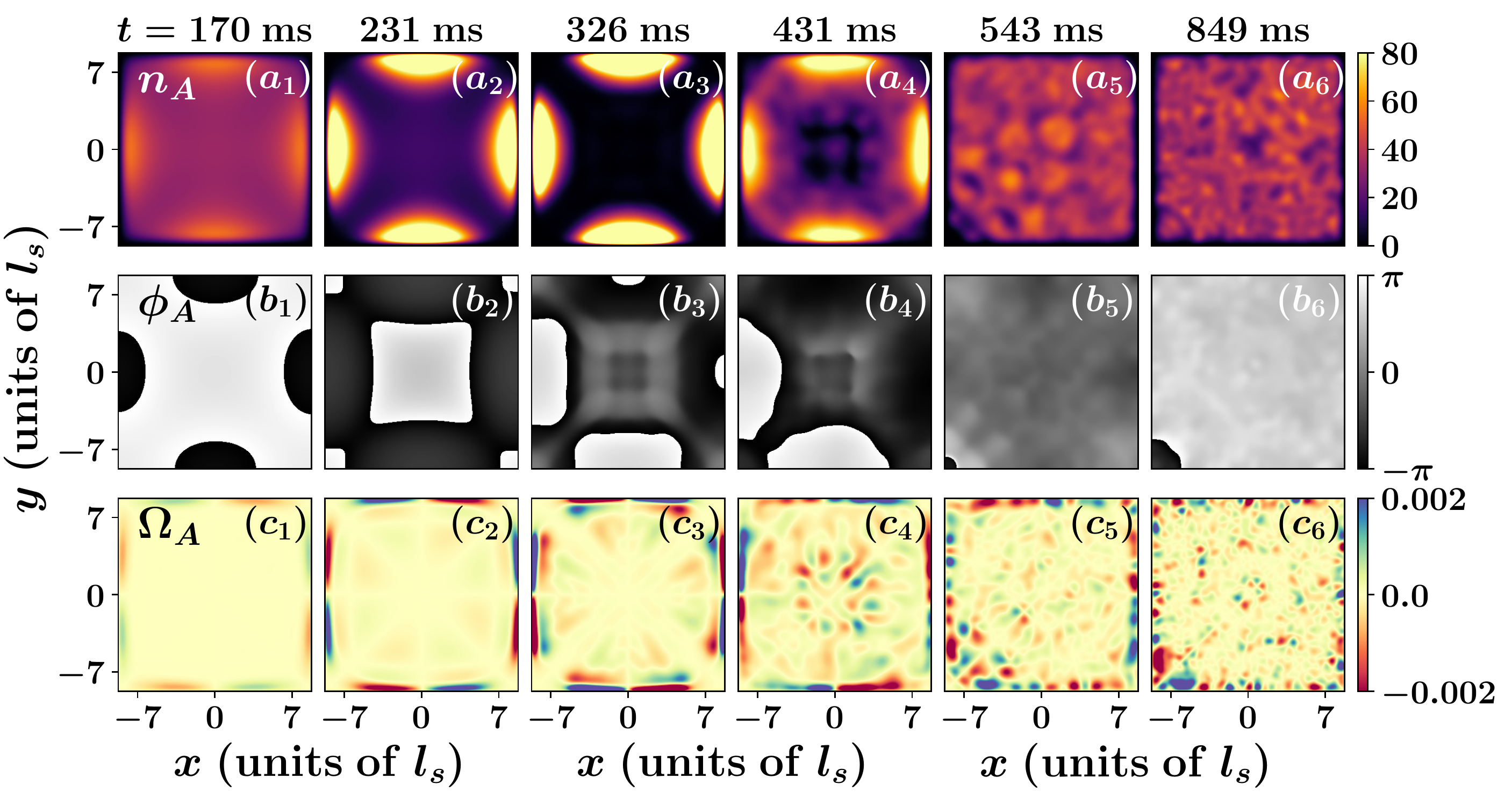}
	\caption{(Color online)($a_1$)-($a_6$) Density, ($b_1$)-($b_6$) phase and ($c_1$)-($c_6$) vorticity profiles of  species A at various time instants (see legends). The binary BEC consists of $N_A = N_B = 10^4$ atoms with zero interspecies interaction. Each species is confined within a 2D box potential characterized by the length scale $l_s$. The intra- and interspecies scattering lengths are $a_{AA} = 100.04a_0$, $a_{BB} = 280a_{0}$, and $a_{AB} = 0$, respectively. The dynamics is initiated by exposing species A to a pair of LG laser pulses with wingdings $l_1 = 2$ and $l_2 = -2$, respectively.}
	\label{Oam2_G0.eps}
\end{figure}

To visualize the spatially resolved dynamics of species A, for $a_{AB} = 0$, we evoke the density, phase, and vorticity profiles illustrated in Fig.~\ref{Oam2_G0.eps}. We find that the system phenomenologically exhibits the same dynamical behavior compared to the case of singly charged LG beams. Note that the last term in Eq.~\ref{GP_eqn1}, characterizing the optical potential, has maxima at $\phi = (2n  +1)\pi/4 $ when $l_1 = 2$ and $l_2 = -2$. Therefore, along the lines, $x = y$ and $x = -y$, two potential barriers contributed by the laser pulses are set up dynamically, and species A is segregated gradually into four distinct lobes. From Figs.~\ref{Oam2_G0.eps}($a_1$)-($a_6$), this is indeed obvious that the density of species A is sliced into four segments. As the pulses are gradually removed, those segments move to regain their initial composition. This ultimately leads to a collision among the segments that generates vortex-antivortex pairs  [Figs.~\ref{Oam2_G0.eps}($b_5$)-($b_6$) and Figs.~\ref{Oam2_G0.eps}($c_5$)-($c_6$)]. These pairs, being related to the density gradient, not the phase singularities, are not quantized in nature [see Figs.~\ref{Oam2_G0.eps}($b_5$)-($b_6$)].
\begin{figure}[h]
	\centering
	\includegraphics[width=\linewidth]{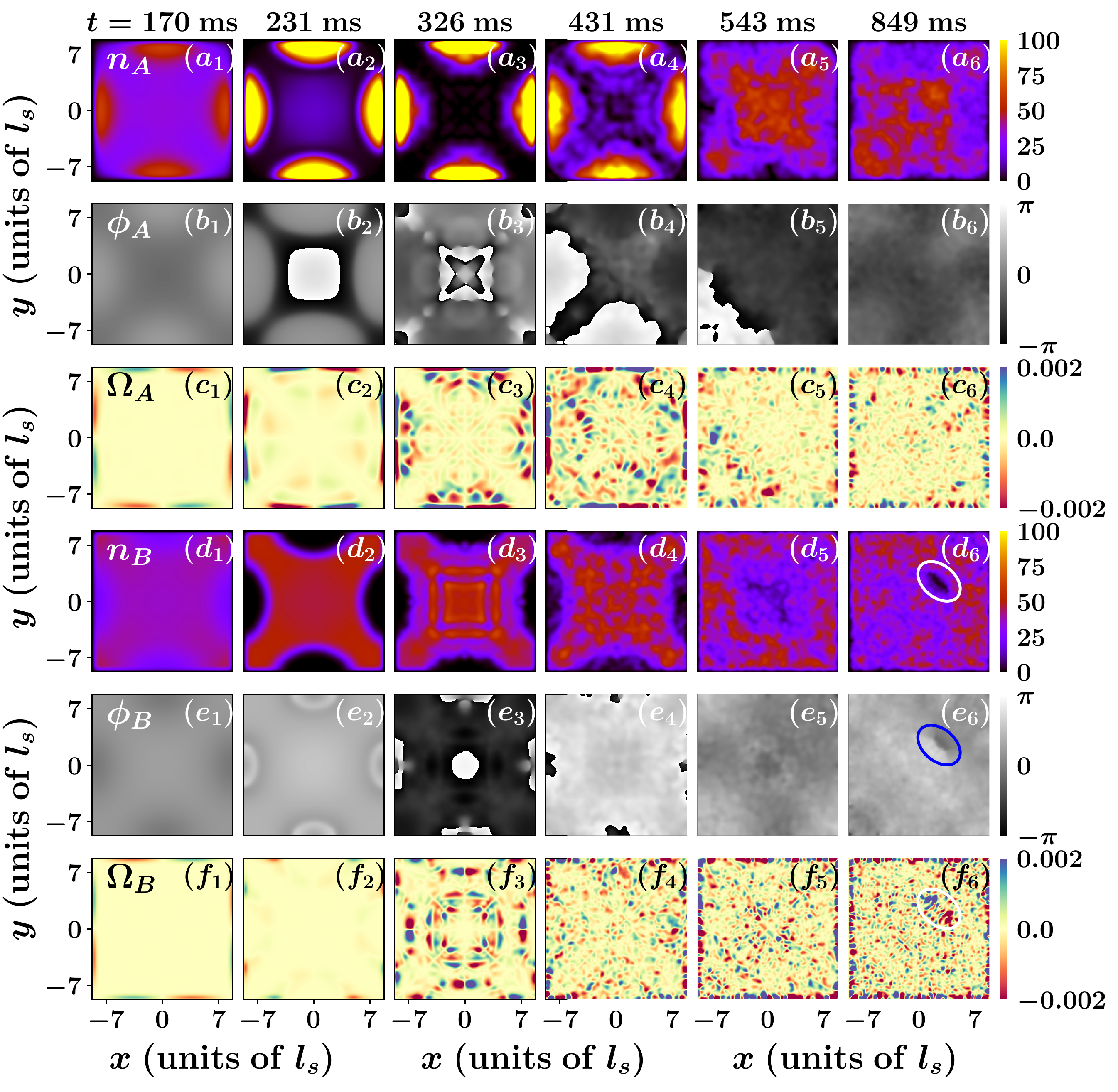}
	\caption{(Color online) Density profiles of ($a_1$)-($a_6$) species A and ($d_1$)-($d_6$) species B at various time instants (see legends). The corresponding phase[vorticity] profiles ($b_1$)-($b_6$) [($c_1$)-($c_6$)] of species A and ($e_1$)-($e_6$) [($f_1$)-($f_6$)] of species B. The binary BEC consists of $N_A = N_B = 10^4$ atoms. Each species is confined within a 2D box potential characterized by the length scale $l_s$. The intra-and interspecies scattering lengths are $a_{AA} = 100.04a_0$, $a_{BB} = 280a_{0}$, and $a_{AB} = 100a_0$, respectively. The dynamics is initiated by exposing species A to a pair of LG laser pulses with wingdings $l_1 = 2$ and $l_2 = -2$, respectively.}
	\label{Oam2_G12.eps}
\end{figure}
Finally, we analyze the light-induced dynamics for the finite interspecies interaction with the scattering length $a_{AB} = 100a_0$, particularly focusing on the same time instants surveyed for the case of non-interacting ones. As anticipated and observed in the previous case, the species A is sliced into four distinct parts by the pair of LG pulses. However, one noticeable difference is that segmented lobes of the species A possess higher densities when compared to those of the non-interacting case; see Figs.~\ref{Oam2_G0.eps}($a_1$)-($a_6$) to Figs.~\ref{Oam2_G12.eps}($a_1$)-($a_6$). The species B, expectedly, is affected so that a star-shaped density distribution complementing that of species A form within it  [Figs.~\ref{Oam2_G12.eps}($a_1$) and \ref{Oam2_G12.eps}($d_1$)]. Moreover, such density distribution of species B creates an effective potential(related to the term $\mathcal{G}_{AB}\abs{\psi_B}^2$ in Eq.~\eqref{GP_eqn1}) for species A. This potential, along with the potential originating from the LG pulses, pushes species A towards the edges, further enhancing its density. Note that due to the enhancement of the density, the collision among the segments is more severe for the  $a_{AB} = 100a_0$ case upon removing the light pulses. Consequently, more vortex-antivortex pairs, including the quantized ones [the circles in Figs.~\ref{Oam2_G12.eps}($d_6$), \ref{Oam2_G12.eps}($e_6$) and~\ref{Oam2_G12.eps}($f_6$) indicating the phase jump] are generated, enriching the vortical content of the system.

Before closing this section, let us briefly summarize our discussion. A pair of LG pulses possessing charges of equal magnitude $\abs{l}$ but the opposite sign can dynamically separate a homogeneous condensate into 2$\abs{l}$ segments and subsequently lead them to collide, generating vortex-antivortex pairs. If a second species is present with finite interspecies interaction, the densities of the segmented lobes of species A are modified, which can significantly alter the outcome of the collision event.

\section{Kinetic energy contributions }\label{KEC}
In the last section, we have highlighted the features of the light-assisted modulation dynamics manifesting within the density and vorticity profiles of the condensates. The theme of this section is the time evolution of the different parts of the kinetic energy, to shed further light on the previous discussion. Recall that the incompressible kinetic energy reveals the presence of the vortices associated with the phase singularities, whereas the compressible kinetic energy conveys the presence of the acoustic waves~\cite{Nore1997, Horng2009, Mithun2021, Mukherjee_2017}.
\begin{figure}[h]
	\centering
	\includegraphics[width=\linewidth]{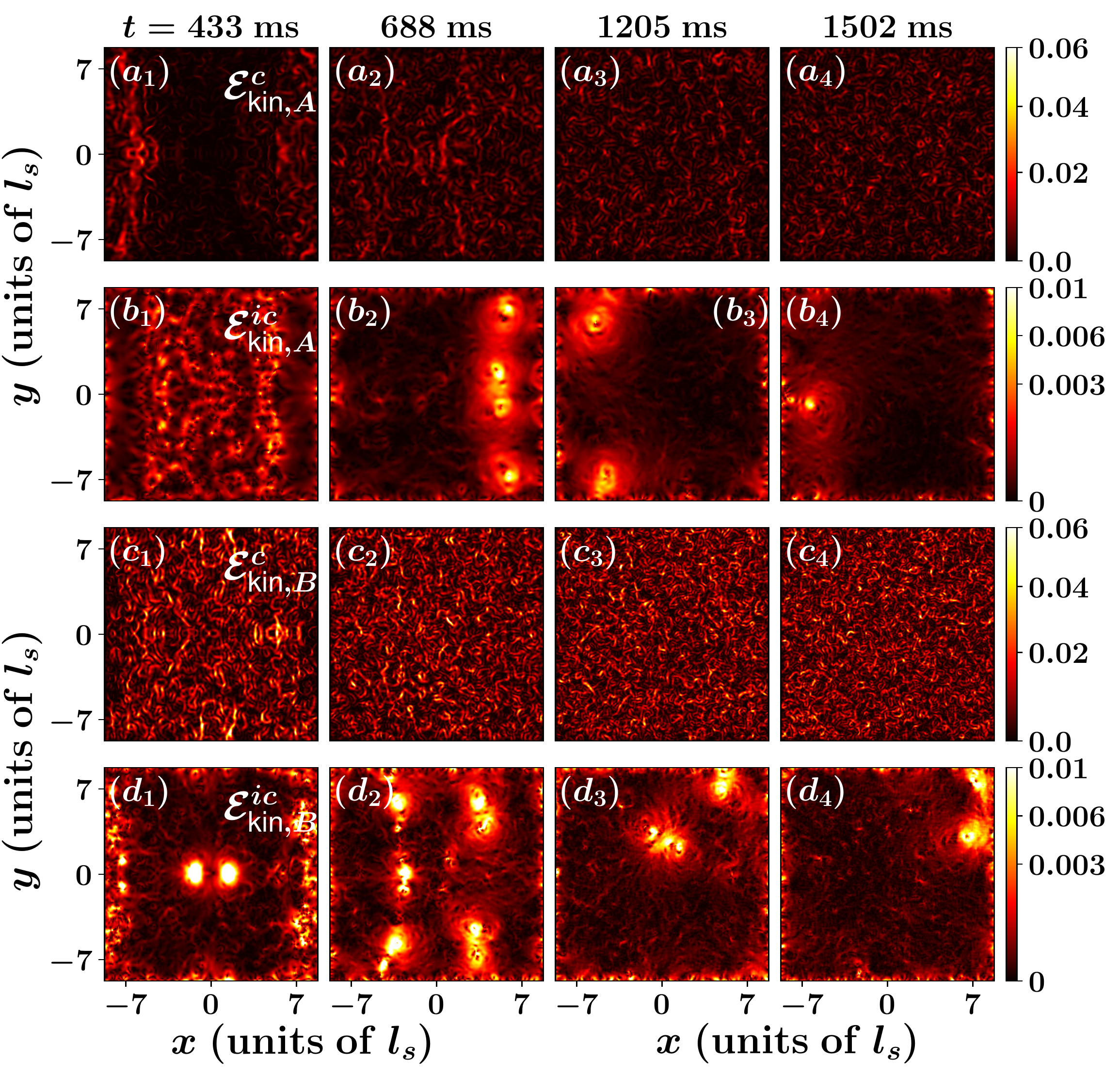}
	\caption{(Color online) Snapshots of the density distribution of  the compressible, ($a_1$)-($a_4$) $\mathcal{E}^{c}_{{\rm kin}, A}$ and ($b_1$)-($b_4$) $\mathcal{E}^{c}_{{\rm kin}, B}$, and incompressible, ($c_1$)-($c_4$) $\mathcal{E}^{ic}_{{\rm kin}, A}$ and ($d_1$)-($d_4$) $\mathcal{E}^{ic}_{{\rm kin}, B}$, parts of the kinetic energy of species A [($a_1$)-($a_4$), ($b_1$)-($b_4$)] and species B [($c_1$)-($c_4$), ($d_1$)-($d_4$)] at few specific time instants. The dynamics is triggered by the interaction between species A and a pair of LG laser pulses with windings $l_1 = 1$ and $l_2 = -1$, respectively. The intra-and interspecies scattering lengths are given by $a_{AA} = 100.04a_{0}$, $a_{BB} = 280a_{0}$, and $a_{AB} = 100a_0$, respectively.}
	\label{energy_1.eps}
\end{figure}
\begin{figure}[h]
	\centering
	\includegraphics[width=\linewidth]{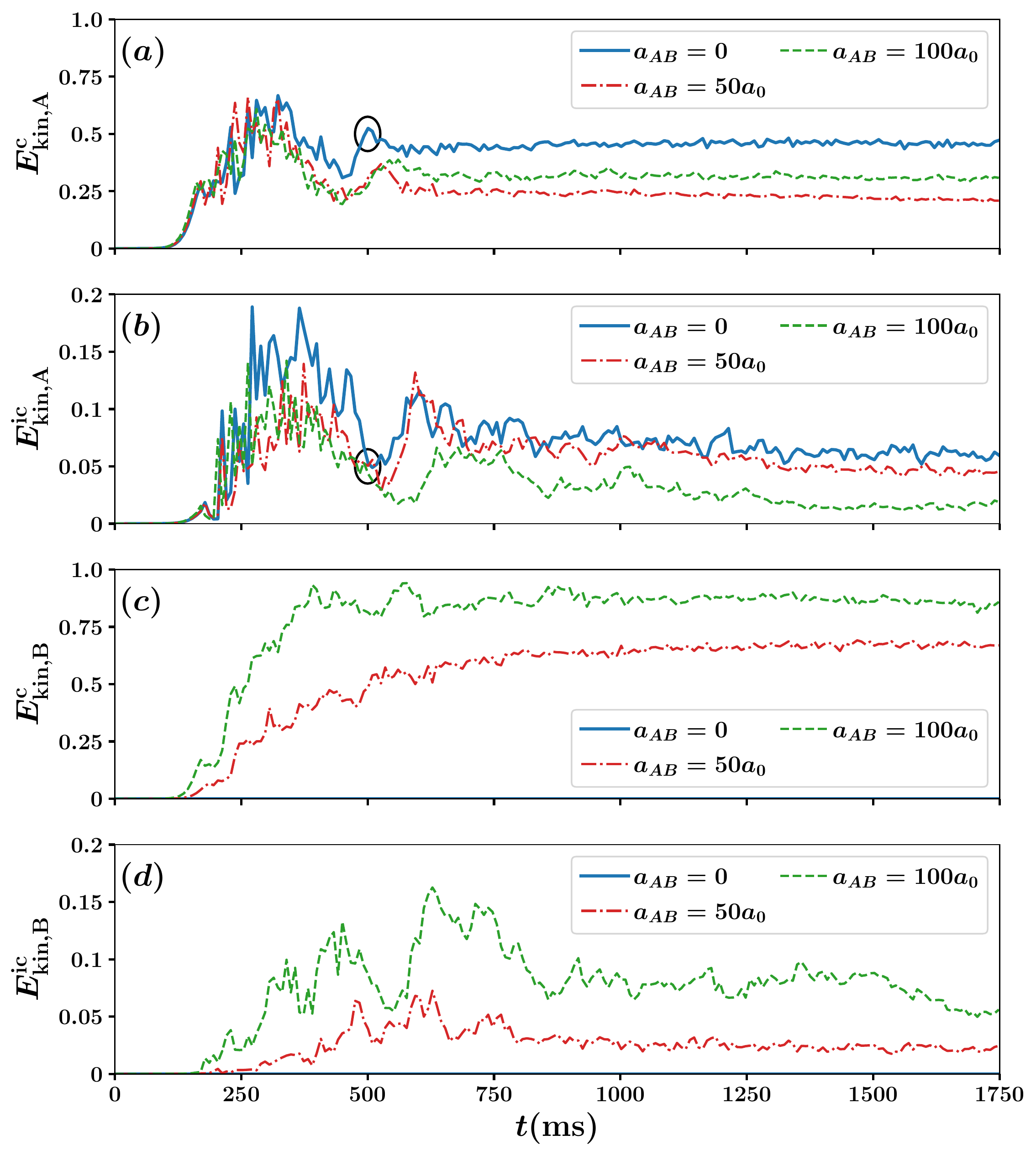}
	\caption{(Color online) Time evolution of the compressible, ($a$) $E^{c}_{{\rm kin}, A}$ and ($c$) $E^{c}_{{\rm kin}, B}$, and incompressible, ($b$) $E^{\rm ic}_{{\rm kin},A}$ and ($d$) $E^{\rm ic}_{{\rm kin},B}$, parts of the kinetic energy of the species A [($a$), ($b$)] and species B [($c$), ($d$)] for different interspecies scattering lengths $a_{AB}$ (see legend) during the evolution. Each species is confined within a uniform 2D box potential with fixed intraspecies scattering lengths $a_{AA} = 100.04a_0$, and $a_{BB} = 280a_{0}$, respectively. A pair of LG laser pulses windings $l_1 = 1$ and $l_2 = -1$ is made to interact with species A.}
	\label{energy_2.eps}
\end{figure}
\begin{figure}[h]
	\centering
	\includegraphics[width=\linewidth]{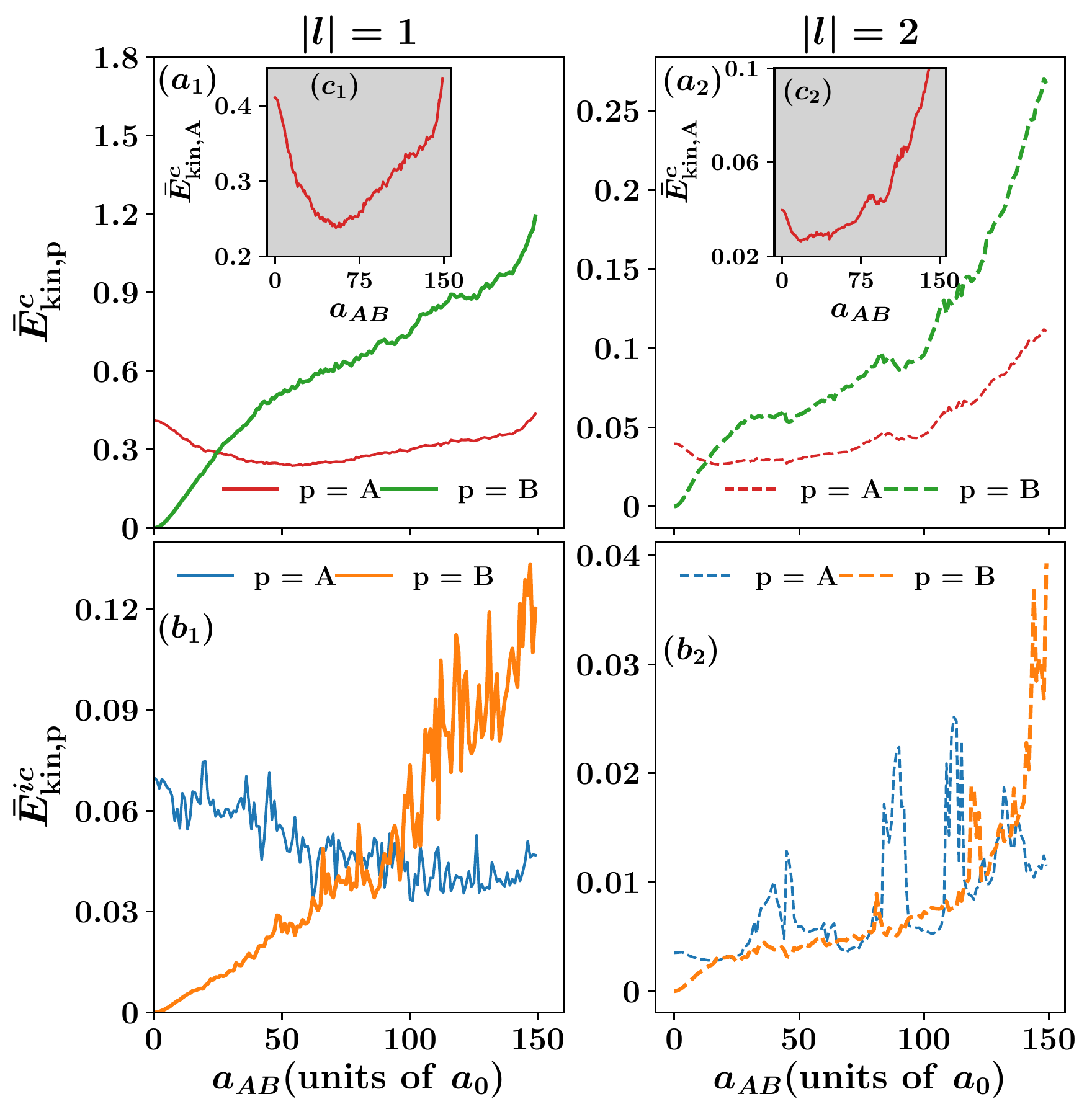}
	\caption{(Color online) The time average of the ($a_1$)-($a_2$) compressible, $\bar{E}^{\rm c}_{\rm kin, p}$ and ($b_1$)-($b_2$) incompressible, $\bar{E}^{\rm ic}_{\rm kin, p}$, parts of the kinetic energy of the $p-$th species[see legend] as a function of the interspecies scattering length $a_{AB}$ for the two different cases involving LG laser pulses with magnitude of the winding being  $\abs{l} = 1$ [($a$), ($c$)] and $\abs{l} = 2$ [($b$), ($d$)], respectively. The dynamics corresponding to each case is triggered by employing a pair of LG laser pulses, with winding of equal magnitude $\abs{l}$ but opposite orientation, to interact with species A. The insets show $\bar{E}^{\rm ic}_{\rm kin, p}$ as a function of $a_{AB}$ for ($c_1$) $\abs{l}=1$  and ($c_2$) $\abs{l}=2$.}
	\label{avgEn.eps}
\end{figure}

In Fig.~\ref{energy_1.eps} we display the compressible and incompressible kinetic energy densities, defined as $\mathcal{E}^{ic[c]}_{kin,p} = n_{p}|u^{ic[c]}_{p}|^2$ $(p = A, B)$, of the two interacting condensates for the interspecies scattering length $a_{AB}=100a_{0}$. The non-equilibrium dynamics
stems from the interaction of species A with a pair of laser pulses having winding $l_1 = 1$ and $l_{2}=-1$. At $t=433$ms  [Fig.~\ref{energy_1.eps}($a_1$)], we notice that the compressible kinetic energy density of the species A,  $\mathcal{E}^{c}_{{\rm kin}, A}$, is more pronounced only at the sides of the condensate. Notably, these acoustic waves spring from the density deformation, which is indeed evident from the Figs.~\ref{Oam1_G12.eps}($a_3$) and \ref{Oam1_G12.eps}$(a_4)$. However, shortly after the parts of species A collide, when the pulses vanish, the  $\mathcal{E}^{\rm c}_{{\rm kin}, A}$ is distributed all over the geometry indicating the further generation of the acoustic waves resulting from the collision [Figs~\ref{energy_1.eps}($a_3$)-($a_4$)]. Turning to the incompressible kinetic energy density of the species A, $\mathcal{E}^{\rm ic}_{{\rm kin}, A}$, we notice that, at $t = 433$ms  [Fig.~\ref{energy_1.eps}($b_1$)], it is majorly present in the location where the particle density of species A is substantially low [Figs.~\ref{Oam1_G12.eps}($a_3$) and \ref{Oam1_G12.eps}$(a_4)$]. Note that smaller particle density leads to larger healing length(that determines the size of the vortices) and, therefore, favoring the initial development of the vortices at the middle of the geometry. Subsequently, as most of the vortices annihilate each other, the $\mathcal{E}^{\rm ic}_{{\rm kin}, A}$ indicates the presence of only a few vortices in the species A [Fig.~\ref{energy_1.eps}($b_2$)]. These vortices, furthermore, in the long-time dynamics annihilate themselves producing the acoustic waves; see Figs.~\ref{energy_1.eps}($a_4$) and \ref{energy_1.eps}($b_4$).

Surprisingly, as evident from the Figs.~\ref{energy_1.eps}($c_1$)-($c_4$) and \ref{energy_1.eps}($d_1$)-($d_4$), the fabrics of the compressible and the incompressible energy densities are more intricate within species B when compared to those of species A, although the former is not directly acted upon by the light pulses. As already discussed in the Sec.~\ref{OAM 1}, the species B is compressed towards the $x=0$ plane, and subsequently, released producing accoustic waves with larger energy densities; for instance, compare between  Figs.~\ref{energy_1.eps}($c_1$)-($c_4$) and Figs.~\ref{energy_1.eps}($a_1$)-($a_4$). Furthermore, the incompressible energy density, $\mathcal{E}^{\rm ic}_{{\rm kin}, B}$, is predominantly distributed at the center, as evident from the two bright spots in Fig.~\ref{energy_1.eps}($d_1$). Afterwards, these bright spots break down into several parts, indicating the generation of several vortices. The final stage of the dynamics having significantly reduced $\mathcal{E}^{\rm ic}_{{\rm kin}, B}$, contributed by only few surviving vortices, as shown in Fig.~\ref{energy_1.eps}($c_4$).

It must be pointed out that the magnitude of the incompressible kinetic energy is quite insignificant compared to that of compressible kinetic energy [Figs~\ref{energy_2.eps}($a$), \ref{energy_2.eps}($b$), \ref{energy_2.eps}($c$), \ref{energy_2.eps}($d$) ]. Moreover, both kinetic energy contributions are prevailing in species B. Therefore, one could be curious about how the interspecies interaction is playing a role in the overall phenomenology.

In order to address the above question, we study the time evolution of the different parts of the kinetic energy (by integrating over the respective kinetic energy densities) in Fig.~\ref{energy_2.eps} for different interspecies scattering lengths $a_{AB}$. The time evolution of $E^{\rm ic[c]}_{{\rm kin}, A}(t)$ indicates that, for $t < 400$ ms and all values of $a_{AB}$, different kinetic components increase over time. This increase corresponds to the generation of acoustic waves and short-lived vortices stemming from the separation of the density of species A(see also the Fig.~\ref{Oam1_G12.eps}. For the later evolution time, $400$ ms $< t < 600$ ms, both the $E^{\rm ic}_{{\rm kin}, A}(t)$ and $E^{\rm c}_{{\rm kin}, A}(t)$ show decreasing tendency as the pulses continuously vanish  [Figs.~\ref{energy_2.eps}($a$) and \ref{energy_2.eps}($b$)]. Afterwards, for $t  > 700$ms, $E^{\rm c}_{{\rm kin}, A}(t)$ and  $E^{\rm ic}_{{\rm kin}, A}(t)$ oscillate around a mean value. Note that, at $t\approx 500$ms, when the pulses are present, $E^{\rm ic}_{{\rm kin}, A}(t)$ acquires a maximum value, when $E^{\rm c}_{{\rm kin}, A}(t)$
is minimized; see Figs.~\ref{energy_2.eps}($a$) and \ref{energy_2.eps}($b$). Therefore, the generation of vortices in the early stage of the dynamics can be attributed to converting acoustic energy into swirling energy. Similarly, when inspecting the longtime behavior of the system, one can notice that acoustic waves are continuously generated due to the annihilation of vortices by mutual collision or collision with the boundaries of the potential box.

Turning to the species B, for $a_{AB} = 0$, both $E^{\rm ic}_{{\rm kin}, B}(t)$ and $E^{\rm c}_{{\rm kin}, B}(t)$ remain zero throughout the dynamics. However, for nonzero $a_{AB}$, the density deformation occurring in species A also affects the density of the species B  [Figs.~\ref{energy_2.eps}($c$) and \ref{energy_2.eps}($d$)]. Consequently, $E^{\rm c}_{{\rm kin}, B}(t)$ initially increases gradually, and then saturates in the long time dynamics. On the other hand, $E^{ic}_{{\rm kin}, B}(t)$, at the early stage of the dynamics, displays a highly fluctuating and erratic behavior with an overall increasing tendency followed by a gradual decrease after $t \approx 750$ ms  [Figs.~\ref{energy_2.eps}($c$) and \ref{energy_2.eps}($d$)]. Moreover, with increasing $a_{AB}$ both $E^{\rm ic}_{{\rm kin}, B}(t)$ $E^{\rm c}_{{\rm kin}, B}(t)$ have increasing values at any instant of time, see Figs.~\ref{energy_2.eps}($c$) and Figs.~\ref{energy_2.eps}($d$). Furthermore, one can notice that $E^{\rm ic}_{{\rm kin}, A}(t)$ decreases for increasing $a_{AB}$ [Figs.~\ref{energy_2.eps}($b$)]. This suggests that swirling energy $E^{\rm ic}_{{\rm kin}, B}(t)$ develops at the expense of  $E^{\rm ic}_{{\rm kin}, A}(t)$. Infact, for $a_{AB} = 100a_{0}$, $E^{\rm ic}_{{\rm kin}, B}(t)$ becomes greater than $E^{\rm ic}_{{\rm kin}, A}(t)$ in the long time dynamics  [Fig.~\ref{energy_2.eps}($b$) and Fig.~\ref{energy_2.eps}($d$)]. Therefore, vortices located in species A not only shares their energy to the acoustic waves, but also contribute to creation of vortical structures in species B. On the other, when inspecting the time evolution of the $E^{\rm c}_{{\rm kin}, A}(t)$ as a function of $a_{AB}$, we notice that the corresponding behavior is not monotonic [Fig.~\ref{energy_2.eps}($a$)]. In particular, during the presence of the pulses, the increasing values of $E^{\rm ic}_{{\rm kin}, A}(t)$ are mostly independent of $a_{AB}$; see during $t < 350$ ms. However, after $t > 700$ms, for $a_{AB}=100a_{0}$, the values of $E^{\rm ic}_{{\rm kin}, A}(t)$ are lower than those corresponding to $a_{AB}=0$, but higher than those corresponding to $a_{AB} =50a_0$.

To unravel this peculiar behavior, we calculate the time average of different parts of the kinetic energy of the $p-$th species, defined as $\bar{E}^{\rm ic[c]}_{\rm kin, p} =(1/T_f)\int_{0}^{T_f}E^{\rm ic[c]}_{\rm kin, p}dt$, where $T_f$ is the total evolution time. In particular, such time-average serves as a measure of the overall content of a particular kinetic energy contribution developed during the dynamics. Inspecting the  Fig.~\ref{avgEn.eps}($a_1$) we find that $\bar{E}^{\rm ic[c]}_{{\rm kin}, A}$ decreases when $a_{AB}$ is increased from zero. It reaches the minimum value, $\bar{E}^{\rm ic}_{{\rm kin}, A} = 0.238 $ at $a_{AB} = 54a_{0}$, and then gradually increases with further increase of $a_{AB}$; see the inset Fig.~\ref{avgEn.eps}($c_1$) for a better visualization. On the contrary, $\bar{E}^{\rm c}_{{\rm kin}, B}$ exhibits a monotonic behavior, namely, an increasing tendency with $a_{AB}$  [Fig.~\ref{avgEn.eps}($a_1$)]. Infact, the value of $\bar{E}^{\rm c}_{{\rm kin}, B}$ surpasses that of $\bar{E}^{\rm c}_{{\rm kin}, A}$ when $a_{AB} = 26a_{0}$  [Fig.~\ref{avgEn.eps}($a_1$)]. The time average of the incompressible kinetic energies,  $\bar{E}^{\rm ic}_{{\rm kin}, A}$ and  $\bar{E}^{\rm ic}_{{\rm kin}, B}$, showcase an extremely fluctuating behavior with respect to $a_{AB}$, since they are the energies associated with the random chaotic motion of the vortices  [Fig.~\ref{avgEn.eps}($b_1$)]. Moreover, $\bar{E}^{\rm ic}_{{\rm kin}, A}$ possesses an increasing, while $\bar{E}^{\rm ic}_{{\rm kin}, B}$ possesses a decreasing tendency suggesting vortical energy transfer from species A to species B. Remarkably, at $a_{AB} \approx 93a_{0}$, the overall vortical content of species B exceeds that of species A  [Fig.~\ref{avgEn.eps}($b_1$)].

For the sake of completeness, we also present in Figs.~\ref{avgEn.eps}($a_2$) and~\ref{avgEn.eps}($b_2$) the time average of the kinetic energy contribution when the pair of laser beams have winding $l_1 =2$ and $l_2 = -2$. Remarkably enough, the overall phenomenology  for $\abs{l} = 2$ is same as that for $\abs{l}=1$. The major difference between the two cases is that $\bar{E}^{\rm ic[c]}_{\rm kin, p}$ is significantly suppressed for the former when compared to the latter. Moreover, $\bar{E}^{\rm c }_{{\rm kin}, B}$ exceeds $\bar{E}^{\rm c }_{{\rm kin}, A}$ at a smaller interspecies scattering length $a_{AB} = 13a_{0}$  [Fig.~\ref{avgEn.eps}($a_2$)]. On the other hand,  incompressible kinetic energy transfer between the two species is much more dramatic and depends on $a_{AB}$ in a non-trivial manner, since $\bar{E}^{\rm ic}_{{\rm kin}, A}$ and $\bar{E}^{\rm  ic}_{{\rm kin}, B}$ cross each other multiple times  [Fig.~\ref{avgEn.eps}($b_2$)]. However, a more detailed discussion of this behavior lie the beyond the scope of this manuscript.

\section{Incompressible and compressible kinetic energy spectra}\label{spectra}
\begin{figure}[h]
	\centering
	\includegraphics[width=0.97\linewidth]{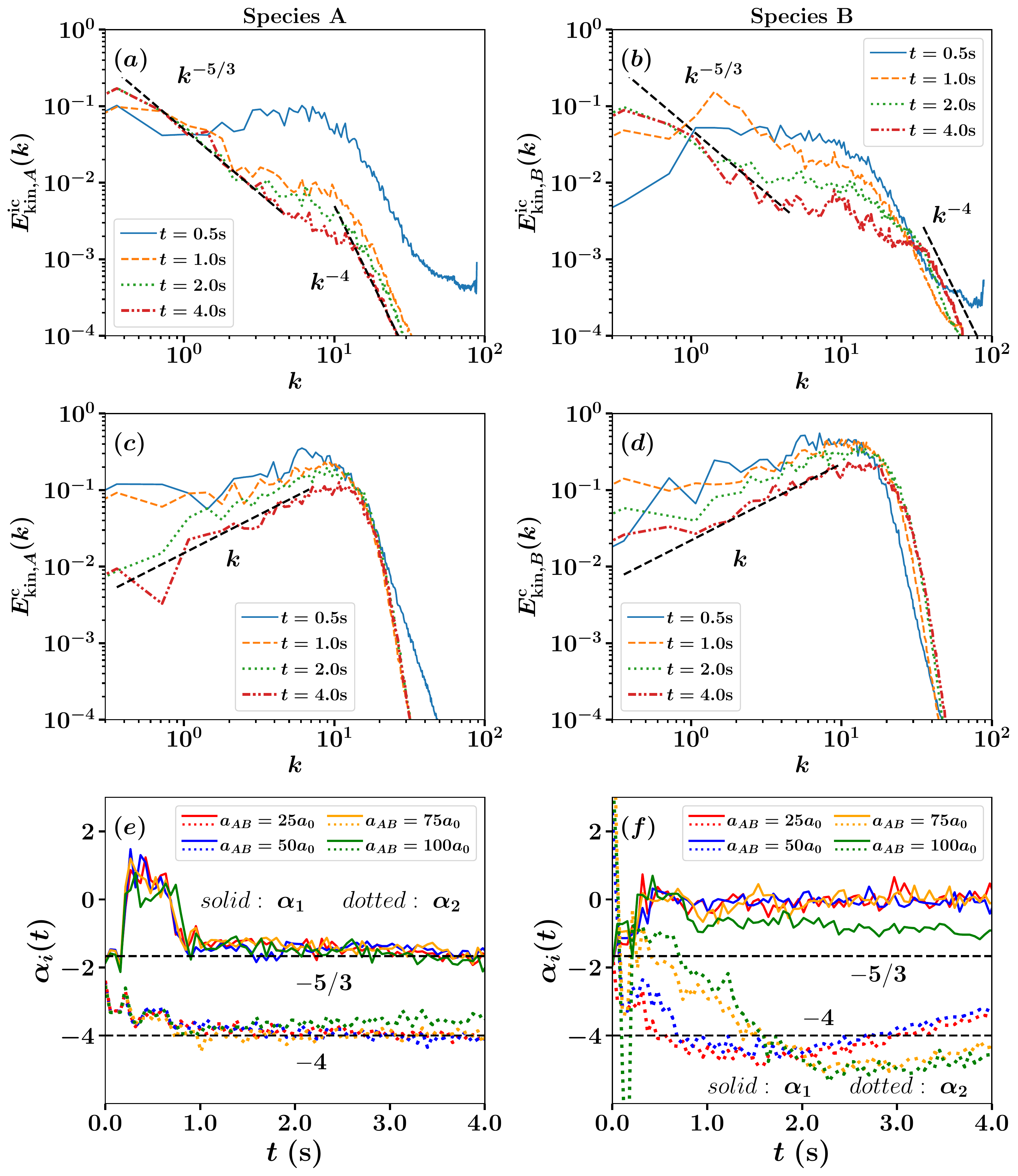}
	\caption{(Color online) Shows ($a$)-($b$) the incompressible $E^{\rm ic}_{{\rm kin}, p}(k)$ and ($c$)-($d$) compressible $E^{c}_{{\rm kin}, p}(k)$ kinetic energy spectra(with $p=A,B$) for the species A[($a$), ($c$)] and species B [($b$), ($d$)] at different instants of time (see the legends) for the interspecies scattering length $a_{AB}=100a_{0}$. We have also shown the time evolution of the exponents $\alpha_{i}$ with $i=1,2$, obtained from the fit of the $E^{\rm ic}_{{\rm kin},p}(k)$ at low and high wave number for ($e$) species A and ($f$) species B. The vertical lines in ($e$) and ($f$) indicate the $-5/3$ and $-4$ scalings. The species A and species B comprises of $^{87}$Rb and $^{133}$Cs atoms, respectively and are confined within a two dimensional box geometry. The dynamics is triggered by applying a pair of LG laser pulses to the $^{87}$Rb atoms.}
	\label{kinetic_energy_spectra}
\end{figure}
To shed further light on the kinematics of the system, we perform the spectral analysis of the compressible and incompressible kinetic energy of each component according to the Eq~\eqref{kin_spec}~\cite{Nore1997, Horng2009}. Recall that the incompressible part being the divergence-free component of the condensate velocity, is associated with the quantum vortices. In contrast, the compressible part represents the energy associated with the acoustic waves.

The log-log plot $E^{\rm ic}_{{\rm kin}, p}(k)$ vs $k$ with $p=A,B$ for species A and species B are shown in Fig.~\ref{kinetic_energy_spectra}(a) and \ref{kinetic_energy_spectra}(b), respectively, corresponding to $\abs{l}=1$ and $a_{AB}=100a_{0}$. In particular, we extract the power-law exponents $\alpha_1$ and $\alpha_2$ ($E^{\rm ic}_{{\rm kin}, p}(k) \propto k^{-\alpha_{i}}$ with $i=1,2$) at low and high wave numbers, respectively, of the incompressible kinetic energy. These exponents provide information regarding the development of quantum turbulence in the system~\cite{Reeves2012}. As time progresses, $E^{\rm ic}_{{\rm kin}, p}(k)$ evolves into a stationary form. For species A, we find that the spectrum at $t=600$ms does not follow any power-law distribution  [Fig.~\ref{kinetic_energy_spectra}(a)]. However, in the long-time dynamics, when $E^{\rm ic}_{{\rm kin}, A}(k)$ becomes stationary, we notice the appearance of two ranges of spectral scaling within the $E^{\rm ic}_{{\rm kin}, A}(k)$ [Fig.~\ref{kinetic_energy_spectra}(a)]. The critical finding here is the conformation of these scalings to the power-law form. For species A, the first scaling appears over the range $0.3 < k <3$, where the spectrum resembles the well-known $k^{-5/3}$ Kolmogorov spectra; see Figs.~\ref{kinetic_energy_spectra}(a). In the second one, $E^{\rm ic}_{{\rm kin}, A}(k)$ closely conform to the $k^{-4}$ power law over the range $12 < k < 35$  [Fig~\ref{kinetic_energy_spectra}(a)]. Additionally, a range of wavenumber exists, namely $3 \le k \le 12$, where $E^{\rm ic}_{{\rm kin}, A}(k)$ does not exhibit the afore-mentioned power-law scaling behavior  [Fig.~\ref{kinetic_energy_spectra}(a)]. Furthermore, $E^{\rm ic}_{{\rm kin}, A}(k)$ located at $k=1$ increases, while that located at $k=10 $ decreases with time  [Fig.~\ref{kinetic_energy_spectra}(a)], indicating that the incompressible energy propagates towards larger $k$(small length scale) from small $k$(large length scale), which is the signature of the so-called inverse energy cascade~\cite{Kraichnan1967, kraichnan_1975}.

In classical 2D turbulence, the inverse energy cascade occurs due to the conservation of enstrophy(defined as mean-squared-vorticity), which is a fundamental feature~\cite{Kraichnan_1980, Sreenivasan1999}. Besides, in classical 2D turbulence, there exists a direct cascade of enstrophy from large to small length scale.  According to Kraichnan-Batchelor theory~\cite{Kraichnan1967, Batchelor1969}, this dual cascade of energy and entropy leads to the logarithmically bilinear spectrum in the incompressible kinetic energy, with the $k^{-5/3}$ dependence for large $k$ and $k^{-3}$ dependence for small $k$, for the 2D turbulence. Since vortex-antivortex pairs in the quantum fluid can annihilate each other, the enstrophy is not a conserved quantity here. This non-conservation of enstrophy makes the inverse energy cascade a debatable issue in 2D quantum turbulence~\cite{Parker2005, Numasato2010, White2012}.

Notably, the $k^{-3}$ power-law fails to develop in the incompressible sector of our system. The value of the scaling, instead, is close to $-4$   [Fig.~\ref{kinetic_energy_spectra}(a)]. Note that observation of the $k^{-4}$ spectrum, also known as the Saffman $-4$  power law~\cite{Saffman_1971}, has been reported in the previous works~\cite{Horng2009, Reeves2017}. This $k^{-4}$ scaling behavior is associated with randomly distributed vortex discontinuities, as shown in Fig.~\ref{energy_1.eps}$(b_3)$-$(b_6)$. Moreover, it has been shown that, $\alpha_2$, the exponent of the incompressible energy spectrum at high wavenumber can undergo a transition from $\alpha_2 =-4$ early to $\alpha_2 =-3$ at late time dynamics~\cite{Reeves2017, brachet1988}. We remark that our system does not display such a transition. Furthermore, in Saffman's theory~\cite{Saffman_1971}, the $k^{-4}$ spectrum is known as the 'dissipation spectrum.' However, we do not include any dissipative term in GP simulation. Therefore, the question regarding whether this $k^{-4}$ scaling appears in our system due to the so-called dissipation has to be addressed cautiously. Recall that our system does not include any vortical structure at the initial state. These vortices have been created by the 'chunk' of acoustic energy generated by the collision of segments of species A. Moreover, acoustic waves are emitted when a pair of vortex and antivortex annihilate each other, and the energy of the vortices is dissipated to the sound waves. Thus, the large amount of compressible energy can be regarded as both source and sink of the incompressible energy.  In this sense, our work closely follows what reported in Ref.~\cite{Horng2009}, where the compressible field, in a similar manner, is held responsible for the appearance of $k^{-4}$ scaling.

Surprisingly, we notice that $k^{-5/3}$ scaling in $E^{\rm ic}_{{\rm kin}, B}(k)$ is not well developed for species B, and it comes into sight over the range $0.8 <k <3$  [Fig.~\ref{kinetic_energy_spectra}(b)]. It is also observed that, unlike species A, the $k^{-4}$ scaling for species B is not pronounced at the high wavenumber region; see the range $0.3 < k< 10$, for instance in Fig.~\ref{kinetic_energy_spectra}(b). Furthermore, this indicates that both scaling regimes are shifted towards higher $k$ for species B when compared to those of species A.

Figures~\ref{kinetic_energy_spectra}(c)-(d) presents the compressible energy spectra, $E^{c}_{{\rm kin}, p}(k)$ (with $p=A,B$) for the same parameter set like that for Figs.~~\ref{kinetic_energy_spectra}(a)-(b). As time progresses, a power-law region with $E^{c}_{{\rm kin}, p}(k) \propto k$ develops on the low-$k$ side of the peak located at $k \approx 21$  [Figs.~\ref{kinetic_energy_spectra}(c)-(d)]. The power-law relation $E^{c}_{{\rm kin}, p}(k) \propto k$ shows excellent agreement with the relation that expresses the frequencies of Bogoliubov’s elementary excitations at low-wave number. This resemblance suggests that the $k<20$ region in the stationary state corresponds to the equilibration of the phonons.  The $E^{ c}_{{\rm kin},p}(k)$ spectrum falls steeply after the peak position  [Figs.~\ref{kinetic_energy_spectra}(c)-(d)].

Another important aspect of the above-mentioned scaling exponents involves their development in time for varying interspecies interaction Figs.~\ref{kinetic_energy_spectra}(e)-(f). For species A, $\alpha_1(t)$ and $\alpha_2(t)$ reside close to the values $-5/3$ and $-4$, respectively, from $t=0.8$s onward, irrespective of the interspecies interaction considered. The corresponding results for species B are shown in Fig.~\ref{kinetic_energy_spectra}(f). As previously noted, the exponents significantly deviate from the values $-5/3$ and $-4$ for most of the simulation.	This deviation is particularly dependent on $a_{AB}$. For example, for larger $a_{AB}$, $\alpha_1(t)$ seems to approach the value $-5/3$. However, a similar conclusion can not be drawn for the $\alpha_2(t)$ since it fails to exhibit any systematic behavior for varying $a_{AB}$.

\section{Conclusions}\label{Conclusion}
We have investigated LG beams-induced dynamics in a homogeneous binary BEC considering a mass difference between the two species.  A pair of LG laser pulses possessing equal magnitude but oppositely oriented winding are applied to species A, coupling its two internal states, while the pulses do not directly impact species B. Notably, the interaction of the pulses with the laser creates in the latter an azimuthally dependent inhomogeneous potential that leads an event involving dynamical segregation and collision of the segments of species A. We have demonstrated the outcome of this event, which results in the generation of vortices in species A, and their effects on species B. A particular focus of this work has been to describe the role played by different windings of the laser light and that of interspecies interaction. To gain insight into the  dynamical response of the condensates, we have utilized several diagnostics such as density, vorticity, and the time evolution of compressible and incompressible parts of the kinetic energy in real and momentum space.

In the first part of our work, we have focused on the density and vortices to exemplify various events during and after the light-matter interaction. The density of species A is segregated into two segments for unit-charged LG laser pulses. After merging two segments, we observe the generation of the soliton stripes and their ultimate relaxation into vortex-antivortex pairs only within species A for zero interspecies interaction.   Focusing on the non-zero interspecies interaction, the density disturbances in species A leave their foot-prints on species B, producing complementary density patterns and, ultimately, creating vortex-antivortex structures in the latter.  In contrast, the doubly charged pulses initially separate species A into four parts but produce phenomenologically similar dynamical events that are less dramatic than their unit-charged counterparts.

As the next step, we examined the dynamics of the system predominantly by invoking the compressible and incompressible parts of the kinetic energy. This description provides further supports to the phenomena already depicted through the density and vorticity profiles. For example, we characterise the transfer of the compressible and incompressible kinetic energies from species A to species B, and essentially, by calculating the time average of these energy components, we have found out the value of interspecies scattering lengths beyond which swirling and acoustic energies are more dominant in species B than species A. However, the most novel and highlighting part of this work addresses different power-law scaling of the kinetic energy spectra. We have observed $-5/3$ and $-4$ power-law scaling in the low and high wavenumber regime of the incompressible energy spectra of species A for any interspecies scattering length in the miscible regime. However, similar scaling is not properly manifested in species B. Since these scalings substantiate the emergence of quantum turbulence, we can claim that species A has been projected onto a turbulent regime due to its interaction with LG beams. Additionally, both species exhibit equilibration of sound waves in the long-time dynamics.

There are many promising research directions ushered by the present work to be pursued in future endeavors. A straightforward one is to consider the asymmetrical LG beam~\cite{Kovalev2016, Das_2020} and investigate the emergent density pattern and kinematics during light-matter interaction. More systematic studies of the dynamics involving the finite temperature effect would be interesting in its own right~\cite{Proukakis_2008}. Another critical prospect would be to employ dipolar BECs~\cite{Aikawa2012}, where one could further unravel the role of the long-range interaction~\cite{Lahaye2009}. In that direction, the additional long-range interaction would induce additional terms in the interaction Hamiltonian that might lead to novel density patterns and scaling law in the different kinetic energy spectra, not captured by the contact interaction. Furthermore, these calculations can also be performed beyond the paraxial limit~\cite{Bhowmik_2018, Bhowmik2016}. Finally, the studies mentioned above would also be equally fascinating at the beyond-mean field level, where the correlations among particles become significant~\cite{Cao2017}.

M.G.D, S.D. and K.M. have contributed equally to this work.
\begin{acknowledgments}
	K.M. gratefully acknowledges S.  I.  Mistakidis  for  insightful  discussions. We also acknowledge useful discussions with K. Seshasayanan.
\end{acknowledgments}


\appendix

\section{Electric field of vortex beams}\label{field}
The expression of electric field vectors involved here can be written as (for $\sigma=1$ and $2$)

\begin{equation}
	\label{elect_field}
	\begin{split}
		E_{\sigma}(x,y,t)=& \frac{\hat{\epsilon_{\sigma}}\mathcal{E}_{\sigma}(t)}{\sqrt{|l_{\sigma}|!}}\bigg(\frac{2}{w}\bigg)^{\frac{|l_{\sigma}|}{2}}(x^2  +y^2)^{\frac{|l_{\sigma}|}{2}} \\ & \times e^{-\frac{(x^2+y^2)}{w_{\sigma}^2}}e^{-i(k_{\sigma}z-\omega_\sigma t)},
	\end{split}
\end{equation}

where, $r=\sqrt{(x^2 + y^2)}$.  Moreover, $\mathcal{E}_{\sigma}(t), \hat{\epsilon}_{\sigma}, k_{\sigma}, \omega_{\sigma}$ and $w_{\sigma}$ are the corresponding amplitude profile, polarization vector, wave number, frequency, and beam waist of the $\sigma$-th laser pulse. The spatial amplitude profile arises due to LG beams propagating along $z$ direction, and the $\sigma$-th pulse is having topological charge of $l_{\sigma}$. If the temporal amplitude profiles of the laser pulses having the same form are considered, then,
\begin{align}\label{temp_profile}
	\mathcal{E}_{\sigma}(t) = \mathcal{E}_{\rm max}e^{-(\frac{t-\tau_{\sigma}}{T})^2}.
\end{align}
We consider both LG laser pulses to have the same maximum intensity $\varepsilon_{max}$, width $w_1= w_2 =w$, and centered temporally at $\tau_1 = \tau_2 = \tau$ and having pulse duration $T$.

Now, we can derive the Rabi frequency for the transition involving the above electric field as \cite{Mukherjee2021},
\begin{equation}
	\label{rabi_exp}
	|\Omega(\vb{r},t)|^2 = \bigg(\frac{\mathcal{E}_{\rm max}d}{\hbar}\bigg)^2 e^{-\frac{(t-\tau)^2}{T^2}}r^{2|l_{\sigma}|}e^{-2\frac{(x^2+y^2)}{w^2}},
\end{equation}
where $d$ is the transition dipole moment.

\section{Derivation of equations of motion}\label{ap:EOM}
We consider the condensate prepared with two different atomic elements. One species, named species A, has two internal states $\ket{1}$ and $\ket{2}$. Initially, all the atoms of species A are in the state $\ket{1}$. The intermediate energy level is denoted by $\ket{2}$ to which atoms of species A are transferred when the LG pulses are simultaneously applied to the condensate. The internal states of the atoms of  species B is irrelevant here since they does not take part directly in light-matter interaction.   We use bosonic field operators to represent the atoms in a particular state. Now, the Hamiltonian (refer to Eq. \eqref{ham_two_com}) for such a system, in the rotating wave approximation, is shown  in the main text.

The creation and annihilation field operators for species A and B commute with each other. Commutation relations obeyed by the bosonic field operators for species A are:
\begin{align}
	\label{com_1}
	 & \qty[\hat{\Psi}_{A,j}(\vb{r}), \hat{\Psi}_{A,k}^\dagger(\vb{r'})]=\delta\qty(\vb{r}-\vb{r'})\delta_{jk}, \\
	\label{com_2}
	 & \qty[\hat{\Psi}_{A,j}(\vb{r}), \hat{\Psi}_{A,k}(\vb{r'})]=0,                                             \\
	\label{com_3}
	 & \qty[\hat{\Psi}_{A,j}^\dagger(\vb{r}), \hat{\Psi}_{A,k}^\dagger(\vb{r'})]=0.
\end{align}

For the above-defined field operators [$\hat{\Psi}_{A,k} ($k=1,2$)$ and $\hat{\Psi}_{B})$], we may define the Heisenberg equations of motion as

\begin{equation}
	\label{heisen_eq_1}
	\im\hbar\pdv{\hat{\Psi}_{A,k}(\vb{r},t)}{t}=[\hat{\Psi}_{A,k}(\vb{r},t),\hat{H}],
\end{equation}
\begin{equation}
	\label{heisen_eq_2}
	\im\hbar\pdv{\hat{\Psi}_B (\vb{r},t)}{t}=[\hat{\Psi}_B(\vb{r},t),\hat{H}].
\end{equation}

Now, using the commutation relations shown above, we find the explicit forms of the Heisenberg equations of motion for $\hat{\Psi}_{A,1}$, $\hat{\Psi}_{A,2}$ and $\hat{\Psi}_B$, respectively, under the action of the Hamiltonian given by Eq.~\ref{ham_two_com} as,

\begin{align}
	\label{heis_A1}
	\im\hbar\pdv{\hat{\Psi}_{A,1}(\vb{r},t)}{t} & =\hat{h}_A \hat{\Psi}_{A,1}(\vb{r},t) + U_{AA} \hat{\Psi}^{\dagger}_{A,1}(\vb{r},t) \hat{\Psi}_{A,1}(\vb{r},t) \nonumber \\ \times \hat{\Psi}_{A,1}(\vb{r},t)
	                                            & + U_{AB} \hat{\Psi}^{\dagger}_B(\vb{r},t) \hat{\Psi}_B(\vb{r},t)\hat{\Psi}_{A,1}(\vb{r},t) \nonumber                     \\
	                                            & + \hbar\Omega(\vb{r},t)\hat{\Psi}_{A,2}(\vb{r},t)(e^{-il_1\phi}+e^{-il_2\phi}),
\end{align}

\begin{align}
	\label{heis_A2}
	\im\hbar\pdv{\hat{\Psi}_{A,2}(\vb{r},t)}{t} = & \hbar\Delta\hat{\Psi}_{A,2}(\vb{r},t) + \hbar\Omega(\vb{r},t)\hat{\Psi}_{A,1}(\vb{r},t) \nonumber \\&  \times (e^{il_1\phi}+e^{il_2\phi}),
\end{align}
and
\begin{align}
	\label{heis_B}
	 & \im\hbar\pdv{\hat{\Psi}_B(\vb{r},t)}{t}  =\hat{h}_B \hat{\Psi}_B(\vb{r},t) + U_{BB} \hat{\Psi}_B(\vb{r},t)\hat{\Psi}^{\dagger}_B(\vb{r},t) \nonumber \\ & \times \hat{\Psi}_B(\vb{r},t)
	+ U_{AB} \hat{\Psi}^{\dagger}_{A,1}(\vb{r},t)\hat{\Psi}_{A,1}(\vb{r},t) \hat{\Psi}_B(\vb{r},t).
\end{align}

Equation \eqref{heis_B} gives us the governing equation for species B.	We adiabatically eliminate the field operator $\hat{\Psi}_{A,2}(\vb{r},t)$ (as it is the intermediate state in our case), by setting,

\begin{align}
	\im\hbar\pdv{\hat{\Psi}_{A,2}(\vb{r},t)}{t} = 0,
\end{align}

and get,

\begin{align}
	\hat{\Psi}_{A,2}(\vb{r},t) = & -\frac{\Omega(\vb{r},t)}{\Delta}(e^{il_1\phi}+e^{il_2\phi})\hat{\Psi}_{A,1}(\vb{r},t).
\end{align}

On substituting the above in \eqref{heis_A1} and using trigonometric identities, we arrive at the result:

\begin{align}
	\label{dynam_eq_A1}
	\im\hbar\pdv{\hat{\Psi}_{A,1}(\vb{r},t)}{t} = & \hat{h}_{A} \hat{\Psi}_{A,1}(\vb{r},t) +  U_{AA} \hat{\Psi}^{\dagger}_{A,1}(\vb{r},t) \hat{\Psi}_{A,1}(\vb{r},t) \nonumber \\ \times \hat{\Psi}_{A,1}(\vb{r},t)
	                                              & + U_{AB} \hat{\Psi}^{\dagger}_B(\vb{r},t) \hat{\Psi}_B(\vb{r},t)\hat{\Psi}_{A,1}(\vb{r},t)  \nonumber                      \\
	                                              & - \frac{4\hbar}{\Delta}\abs{\Omega(\vb{r},t)}^2 \cos^2{\frac{(l_1-l_2)\phi}{2}} \hat{\Psi}_{A,1} (\vb{r},t).
\end{align}

Therefore, the two Eqs. \eqref{dynam_eq_A1} and \eqref{heis_B} govern the time evolution of the bosonic field operators for species A and B, respectively. At $t=0$, if we assume the low-energy $s$-wave scattering, neglect quantum fluctuations and suppress thermal fluctuations, we can replace the field operators by their corresponding wave functions. Thus, the Eqs. \eqref{dynam_eq_A1} and \eqref{heis_B} become:

\begin{align}
	\label{GP_eq_A1}
	\im\hbar\pdv{\Psi_{A,1}(\vb{r},t)}{t} = & \big[ -\frac{\hbar^2}{2m_A}\laplacian_\perp + V(\vb{r}) \big] \Psi_{A,1}(\vb{r},t) \nonumber           \\
	                                        & + U_{AA} \abs{\Psi_{A,1}(\vb{r},t)}^2 \Psi_{A,1}(\vb{r},t) \nonumber                                   \\
	                                        & + U_{AB} \abs{\Psi_{B}(\vb{r},t)}^2 \Psi_{A,1}(\vb{r},t) \nonumber                                     \\
	                                        & - \frac{4\hbar}{\Delta}\abs{\Omega(\vb{r},t)}^2 \cos^2{\frac{(l_1-l_2)\phi}{2}} \Psi_{A,1} (\vb{r},t),
\end{align}
and,
\begin{align}
	\label{GP_eq_B}
	\im\hbar\pdv{\Psi_B(\vb{r},t)}{t} = & \big[ -\frac{\hbar^2}{2m_B}\laplacian_\perp + V(\vb{r}) \big] \Psi_B(\vb{r},t) \nonumber \\
	                                    & + U_{BB} \abs{\Psi_B(\vb{r},t)}^2 \Psi_B(\vb{r},t) \nonumber                             \\
	                                    & + U_{AB} \abs{\Psi_{A,1}(\vb{r},t)}^2 \Psi_B(\vb{r},t),
\end{align}

where $m_A$ and $m_B$ are the atomic masses of species A and B, respectively, and $V(\vb{r})$ is the trapping potential of the BEC. For our problem, assuming the trapping frequency to be very weak, we take $V(\vb{r}) \approx 0$, i.e., the condensate is present in a homogeneous potential box in the x-y plane. In particular, we assume that z-direction is tightly confined compared to the x and y directions. As a result the dynamics along the $z$-direction is frozen and the effective dynamics of the condensates can be described in the 2D x-y plane by integrating the wavefunction along the z-directions; see Ref.~\cite{Mukherjee_2020, Soumik2017} for the relevant methodology. Thereafter, substituting the expression for the Rabi frequency as given by Eq. \eqref{rabi_exp} and non-dimensionalizing the above Eqs. \eqref{GP_eq_A1} and \eqref{GP_eq_B}, we can arrive at the coupled Gross-Pitaevskii Eqs. \eqref{GP_eqn1} and \eqref{GP_eqn2}, respectively. 

\bibliographystyle{apsrev4-1}
\bibliography{references}

\begin{thebibliography}{115}%
\makeatletter
\providecommand \@ifxundefined [1]{%
 \@ifx{#1\undefined}
}%
\providecommand \@ifnum [1]{%
 \ifnum #1\expandafter \@firstoftwo
 \else \expandafter \@secondoftwo
 \fi
}%
\providecommand \@ifx [1]{%
 \ifx #1\expandafter \@firstoftwo
 \else \expandafter \@secondoftwo
 \fi
}%
\providecommand \natexlab [1]{#1}%
\providecommand \enquote  [1]{``#1''}%
\providecommand \bibnamefont  [1]{#1}%
\providecommand \bibfnamefont [1]{#1}%
\providecommand \citenamefont [1]{#1}%
\providecommand \href@noop [0]{\@secondoftwo}%
\providecommand \href [0]{\begingroup \@sanitize@url \@href}%
\providecommand \@href[1]{\@@startlink{#1}\@@href}%
\providecommand \@@href[1]{\endgroup#1\@@endlink}%
\providecommand \@sanitize@url [0]{\catcode `\\12\catcode `\$12\catcode
  `\&12\catcode `\#12\catcode `\^12\catcode `\_12\catcode `\%12\relax}%
\providecommand \@@startlink[1]{}%
\providecommand \@@endlink[0]{}%
\providecommand \url  [0]{\begingroup\@sanitize@url \@url }%
\providecommand \@url [1]{\endgroup\@href {#1}{\urlprefix }}%
\providecommand \urlprefix  [0]{URL }%
\providecommand \Eprint [0]{\href }%
\providecommand \doibase [0]{http://dx.doi.org/}%
\providecommand \selectlanguage [0]{\@gobble}%
\providecommand \bibinfo  [0]{\@secondoftwo}%
\providecommand \bibfield  [0]{\@secondoftwo}%
\providecommand \translation [1]{[#1]}%
\providecommand \BibitemOpen [0]{}%
\providecommand \bibitemStop [0]{}%
\providecommand \bibitemNoStop [0]{.\EOS\space}%
\providecommand \EOS [0]{\spacefactor3000\relax}%
\providecommand \BibitemShut  [1]{\csname bibitem#1\endcsname}%
\let\auto@bib@innerbib\@empty
\bibitem [{\citenamefont {Chu}(1998)}]{Chu1998}%
  \BibitemOpen
  \bibfield  {author} {\bibinfo {author} {\bibfnamefont {S.}~\bibnamefont
  {Chu}},\ }\href {\doibase 10.1103/RevModPhys.70.685} {\bibfield  {journal}
  {\bibinfo  {journal} {Rev. Mod. Phys.}\ }\textbf {\bibinfo {volume} {70}},\
  \bibinfo {pages} {685} (\bibinfo {year} {1998})}\BibitemShut {NoStop}%
\bibitem [{\citenamefont {Cohen-Tannoudji}(1998)}]{Cohen1998}%
  \BibitemOpen
  \bibfield  {author} {\bibinfo {author} {\bibfnamefont {C.~N.}\ \bibnamefont
  {Cohen-Tannoudji}},\ }\href {\doibase 10.1103/RevModPhys.70.707} {\bibfield
  {journal} {\bibinfo  {journal} {Rev. Mod. Phys.}\ }\textbf {\bibinfo {volume}
  {70}},\ \bibinfo {pages} {707} (\bibinfo {year} {1998})}\BibitemShut
  {NoStop}%
\bibitem [{\citenamefont {Phillips}(1998)}]{Phillips1998}%
  \BibitemOpen
  \bibfield  {author} {\bibinfo {author} {\bibfnamefont {W.~D.}\ \bibnamefont
  {Phillips}},\ }\href {\doibase 10.1103/RevModPhys.70.721} {\bibfield
  {journal} {\bibinfo  {journal} {Rev. Mod. Phys.}\ }\textbf {\bibinfo {volume}
  {70}},\ \bibinfo {pages} {721} (\bibinfo {year} {1998})}\BibitemShut
  {NoStop}%
\bibitem [{\citenamefont {Ritsch}\ \emph {et~al.}(2013)\citenamefont {Ritsch},
  \citenamefont {Domokos}, \citenamefont {Brennecke},\ and\ \citenamefont
  {Esslinger}}]{Ritsch2013}%
  \BibitemOpen
  \bibfield  {author} {\bibinfo {author} {\bibfnamefont {H.}~\bibnamefont
  {Ritsch}}, \bibinfo {author} {\bibfnamefont {P.}~\bibnamefont {Domokos}},
  \bibinfo {author} {\bibfnamefont {F.}~\bibnamefont {Brennecke}}, \ and\
  \bibinfo {author} {\bibfnamefont {T.}~\bibnamefont {Esslinger}},\ }\href
  {\doibase 10.1103/RevModPhys.85.553} {\bibfield  {journal} {\bibinfo
  {journal} {Rev. Mod. Phys.}\ }\textbf {\bibinfo {volume} {85}},\ \bibinfo
  {pages} {553} (\bibinfo {year} {2013})}\BibitemShut {NoStop}%
\bibitem [{\citenamefont {Slama}\ \emph {et~al.}(2007)\citenamefont {Slama},
  \citenamefont {Bux}, \citenamefont {Krenz}, \citenamefont {Zimmermann},\ and\
  \citenamefont {Courteille}}]{Slama2007}%
  \BibitemOpen
  \bibfield  {author} {\bibinfo {author} {\bibfnamefont {S.}~\bibnamefont
  {Slama}}, \bibinfo {author} {\bibfnamefont {S.}~\bibnamefont {Bux}}, \bibinfo
  {author} {\bibfnamefont {G.}~\bibnamefont {Krenz}}, \bibinfo {author}
  {\bibfnamefont {C.}~\bibnamefont {Zimmermann}}, \ and\ \bibinfo {author}
  {\bibfnamefont {P.~W.}\ \bibnamefont {Courteille}},\ }\href {\doibase
  10.1103/PhysRevLett.98.053603} {\bibfield  {journal} {\bibinfo  {journal}
  {Phys. Rev. Lett.}\ }\textbf {\bibinfo {volume} {98}},\ \bibinfo {pages}
  {053603} (\bibinfo {year} {2007})}\BibitemShut {NoStop}%
\bibitem [{\citenamefont {Walther}\ \emph {et~al.}(2006)\citenamefont
  {Walther}, \citenamefont {Varcoe}, \citenamefont {Englert},\ and\
  \citenamefont {Becker}}]{Walther_2006}%
  \BibitemOpen
  \bibfield  {author} {\bibinfo {author} {\bibfnamefont {H.}~\bibnamefont
  {Walther}}, \bibinfo {author} {\bibfnamefont {B.~T.~H.}\ \bibnamefont
  {Varcoe}}, \bibinfo {author} {\bibfnamefont {B.-G.}\ \bibnamefont {Englert}},
  \ and\ \bibinfo {author} {\bibfnamefont {T.}~\bibnamefont {Becker}},\ }\href
  {\doibase 10.1088/0034-4885/69/5/r02} {\bibfield  {journal} {\bibinfo
  {journal} {Rep. Prog. Phys.}\ }\textbf {\bibinfo {volume} {69}},\ \bibinfo
  {pages} {1325} (\bibinfo {year} {2006})}\BibitemShut {NoStop}%
\bibitem [{\citenamefont {Caballero-Benitez}\ and\ \citenamefont
  {Mekhov}(2015)}]{Mekhov2015}%
  \BibitemOpen
  \bibfield  {author} {\bibinfo {author} {\bibfnamefont {S.~F.}\ \bibnamefont
  {Caballero-Benitez}}\ and\ \bibinfo {author} {\bibfnamefont {I.~B.}\
  \bibnamefont {Mekhov}},\ }\href {\doibase 10.1103/PhysRevLett.115.243604}
  {\bibfield  {journal} {\bibinfo  {journal} {Phys. Rev. Lett.}\ }\textbf
  {\bibinfo {volume} {115}},\ \bibinfo {pages} {243604} (\bibinfo {year}
  {2015})}\BibitemShut {NoStop}%
\bibitem [{\citenamefont {Dutra}(2005)}]{Dutra2005}%
  \BibitemOpen
  \bibfield  {author} {\bibinfo {author} {\bibfnamefont {S.~M.}\ \bibnamefont
  {Dutra}},\ }\href@noop {} {\emph {\bibinfo {title} {Cavity Quantum
  Electrodynamics: The Strange Theory of Light in a Box}}}\ (\bibinfo
  {publisher} {John Wiley \& Sons},\ \bibinfo {year} {2005})\BibitemShut
  {NoStop}%
\bibitem [{\citenamefont {Cooper}\ \emph {et~al.}(2019)\citenamefont {Cooper},
  \citenamefont {Dalibard},\ and\ \citenamefont {Spielman}}]{Cooper2019}%
  \BibitemOpen
  \bibfield  {author} {\bibinfo {author} {\bibfnamefont {N.~R.}\ \bibnamefont
  {Cooper}}, \bibinfo {author} {\bibfnamefont {J.}~\bibnamefont {Dalibard}}, \
  and\ \bibinfo {author} {\bibfnamefont {I.~B.}\ \bibnamefont {Spielman}},\
  }\href {\doibase 10.1103/RevModPhys.91.015005} {\bibfield  {journal}
  {\bibinfo  {journal} {Rev. Mod. Phys.}\ }\textbf {\bibinfo {volume} {91}},\
  \bibinfo {pages} {015005} (\bibinfo {year} {2019})}\BibitemShut {NoStop}%
\bibitem [{\citenamefont {Mivehvar}\ \emph {et~al.}(2021)\citenamefont
  {Mivehvar}, \citenamefont {Piazza}, \citenamefont {Donner},\ and\
  \citenamefont {Ritsch}}]{Mivehvar2021}%
  \BibitemOpen
  \bibfield  {author} {\bibinfo {author} {\bibfnamefont {F.}~\bibnamefont
  {Mivehvar}}, \bibinfo {author} {\bibfnamefont {F.}~\bibnamefont {Piazza}},
  \bibinfo {author} {\bibfnamefont {T.}~\bibnamefont {Donner}}, \ and\ \bibinfo
  {author} {\bibfnamefont {H.}~\bibnamefont {Ritsch}},\ }\href@noop {}
  {\enquote {\bibinfo {title} {Cavity qed with quantum gases: New paradigms in
  many-body physics},}\ } (\bibinfo {year} {2021}),\ \Eprint
  {http://arxiv.org/abs/2102.04473} {arXiv:2102.04473} \BibitemShut {NoStop}%
\bibitem [{\citenamefont {Grimm}\ \emph {et~al.}(2000)\citenamefont {Grimm},
  \citenamefont {Weidem\"uller},\ and\ \citenamefont
  {Ovchinnikov}}]{GRIMM200095}%
  \BibitemOpen
  \bibfield  {author} {\bibinfo {author} {\bibfnamefont {R.}~\bibnamefont
  {Grimm}}, \bibinfo {author} {\bibfnamefont {M.}~\bibnamefont
  {Weidem\"uller}}, \ and\ \bibinfo {author} {\bibfnamefont {Y.}~\bibnamefont
  {Ovchinnikov}},\ }\href {\doibase
  https://doi.org/10.1016/S1049-250X(08)60186-X} {\bibfield  {journal}
  {\bibinfo  {journal} {Adv. At. Mol .Opt. Phys.}\ }\textbf {\bibinfo {volume}
  {42}},\ \bibinfo {pages} {95} (\bibinfo {year} {2000})}\BibitemShut {NoStop}%
\bibitem [{\citenamefont {Garraway}\ and\ \citenamefont
  {Minogin}(2000)}]{Garraway2000}%
  \BibitemOpen
  \bibfield  {author} {\bibinfo {author} {\bibfnamefont {B.~M.}\ \bibnamefont
  {Garraway}}\ and\ \bibinfo {author} {\bibfnamefont {V.~G.}\ \bibnamefont
  {Minogin}},\ }\href {\doibase 10.1103/PhysRevA.62.043406} {\bibfield
  {journal} {\bibinfo  {journal} {Phys. Rev. A}\ }\textbf {\bibinfo {volume}
  {62}},\ \bibinfo {pages} {043406} (\bibinfo {year} {2000})}\BibitemShut
  {NoStop}%
\bibitem [{\citenamefont {Frese}\ \emph {et~al.}(2000)\citenamefont {Frese},
  \citenamefont {Ueberholz}, \citenamefont {Kuhr}, \citenamefont {Alt},
  \citenamefont {Schrader}, \citenamefont {Gomer},\ and\ \citenamefont
  {Meschede}}]{Frese2000}%
  \BibitemOpen
  \bibfield  {author} {\bibinfo {author} {\bibfnamefont {D.}~\bibnamefont
  {Frese}}, \bibinfo {author} {\bibfnamefont {B.}~\bibnamefont {Ueberholz}},
  \bibinfo {author} {\bibfnamefont {S.}~\bibnamefont {Kuhr}}, \bibinfo {author}
  {\bibfnamefont {W.}~\bibnamefont {Alt}}, \bibinfo {author} {\bibfnamefont
  {D.}~\bibnamefont {Schrader}}, \bibinfo {author} {\bibfnamefont
  {V.}~\bibnamefont {Gomer}}, \ and\ \bibinfo {author} {\bibfnamefont
  {D.}~\bibnamefont {Meschede}},\ }\href {\doibase 10.1103/PhysRevLett.85.3777}
  {\bibfield  {journal} {\bibinfo  {journal} {Phys. Rev. Lett.}\ }\textbf
  {\bibinfo {volume} {85}},\ \bibinfo {pages} {3777} (\bibinfo {year}
  {2000})}\BibitemShut {NoStop}%
\bibitem [{\citenamefont {Cornell}\ and\ \citenamefont
  {Wieman}(2002)}]{Cornell2002}%
  \BibitemOpen
  \bibfield  {author} {\bibinfo {author} {\bibfnamefont {E.~A.}\ \bibnamefont
  {Cornell}}\ and\ \bibinfo {author} {\bibfnamefont {C.~E.}\ \bibnamefont
  {Wieman}},\ }\href {\doibase 10.1103/RevModPhys.74.875} {\bibfield  {journal}
  {\bibinfo  {journal} {Rev. Mod. Phys.}\ }\textbf {\bibinfo {volume} {74}},\
  \bibinfo {pages} {875} (\bibinfo {year} {2002})}\BibitemShut {NoStop}%
\bibitem [{\citenamefont {Ketterle}(2002)}]{Ketterle2002}%
  \BibitemOpen
  \bibfield  {author} {\bibinfo {author} {\bibfnamefont {W.}~\bibnamefont
  {Ketterle}},\ }\href {\doibase 10.1103/RevModPhys.74.1131} {\bibfield
  {journal} {\bibinfo  {journal} {Rev. Mod. Phys.}\ }\textbf {\bibinfo {volume}
  {74}},\ \bibinfo {pages} {1131} (\bibinfo {year} {2002})}\BibitemShut
  {NoStop}%
\bibitem [{\citenamefont {Bloch}\ \emph {et~al.}(2008)\citenamefont {Bloch},
  \citenamefont {Dalibard},\ and\ \citenamefont {Zwerger}}]{Bloch2008}%
  \BibitemOpen
  \bibfield  {author} {\bibinfo {author} {\bibfnamefont {I.}~\bibnamefont
  {Bloch}}, \bibinfo {author} {\bibfnamefont {J.}~\bibnamefont {Dalibard}}, \
  and\ \bibinfo {author} {\bibfnamefont {W.}~\bibnamefont {Zwerger}},\ }\href
  {\doibase 10.1103/RevModPhys.80.885} {\bibfield  {journal} {\bibinfo
  {journal} {Rev. Mod. Phys.}\ }\textbf {\bibinfo {volume} {80}},\ \bibinfo
  {pages} {885} (\bibinfo {year} {2008})}\BibitemShut {NoStop}%
\bibitem [{\citenamefont {Giorgini}\ \emph {et~al.}(2008)\citenamefont
  {Giorgini}, \citenamefont {Pitaevskii},\ and\ \citenamefont
  {Stringari}}]{Giorgini2008}%
  \BibitemOpen
  \bibfield  {author} {\bibinfo {author} {\bibfnamefont {S.}~\bibnamefont
  {Giorgini}}, \bibinfo {author} {\bibfnamefont {L.~P.}\ \bibnamefont
  {Pitaevskii}}, \ and\ \bibinfo {author} {\bibfnamefont {S.}~\bibnamefont
  {Stringari}},\ }\href {\doibase 10.1103/RevModPhys.80.1215} {\bibfield
  {journal} {\bibinfo  {journal} {Rev. Mod. Phys.}\ }\textbf {\bibinfo {volume}
  {80}},\ \bibinfo {pages} {1215} (\bibinfo {year} {2008})}\BibitemShut
  {NoStop}%
\bibitem [{\citenamefont {Gordon}\ and\ \citenamefont
  {Ashkin}(1980)}]{Gordon1980}%
  \BibitemOpen
  \bibfield  {author} {\bibinfo {author} {\bibfnamefont {J.~P.}\ \bibnamefont
  {Gordon}}\ and\ \bibinfo {author} {\bibfnamefont {A.}~\bibnamefont
  {Ashkin}},\ }\href {\doibase 10.1103/PhysRevA.21.1606} {\bibfield  {journal}
  {\bibinfo  {journal} {Phys. Rev. A}\ }\textbf {\bibinfo {volume} {21}},\
  \bibinfo {pages} {1606} (\bibinfo {year} {1980})}\BibitemShut {NoStop}%
\bibitem [{\citenamefont {Allen}\ \emph {et~al.}(1992)\citenamefont {Allen},
  \citenamefont {Beijersbergen}, \citenamefont {Spreeuw},\ and\ \citenamefont
  {Woerdman}}]{Allen1992}%
  \BibitemOpen
  \bibfield  {author} {\bibinfo {author} {\bibfnamefont {L.}~\bibnamefont
  {Allen}}, \bibinfo {author} {\bibfnamefont {M.~W.}\ \bibnamefont
  {Beijersbergen}}, \bibinfo {author} {\bibfnamefont {R.~J.~C.}\ \bibnamefont
  {Spreeuw}}, \ and\ \bibinfo {author} {\bibfnamefont {J.~P.}\ \bibnamefont
  {Woerdman}},\ }\href {\doibase 10.1103/PhysRevA.45.8185} {\bibfield
  {journal} {\bibinfo  {journal} {Phys. Rev. A}\ }\textbf {\bibinfo {volume}
  {45}},\ \bibinfo {pages} {8185} (\bibinfo {year} {1992})}\BibitemShut
  {NoStop}%
\bibitem [{\citenamefont {Yao}\ and\ \citenamefont
  {Padgett}(2011{\natexlab{a}})}]{Alison2011}%
  \BibitemOpen
  \bibfield  {author} {\bibinfo {author} {\bibfnamefont {A.~M.}\ \bibnamefont
  {Yao}}\ and\ \bibinfo {author} {\bibfnamefont {M.~J.}\ \bibnamefont
  {Padgett}},\ }\href {\doibase 10.1364/AOP.3.000161} {\bibfield  {journal}
  {\bibinfo  {journal} {Adv. Opt. Photon.}\ }\textbf {\bibinfo {volume} {3}},\
  \bibinfo {pages} {161} (\bibinfo {year} {2011}{\natexlab{a}})}\BibitemShut
  {NoStop}%
\bibitem [{\citenamefont {Padgett}(2017{\natexlab{a}})}]{Miles2017}%
  \BibitemOpen
  \bibfield  {author} {\bibinfo {author} {\bibfnamefont {M.~J.}\ \bibnamefont
  {Padgett}},\ }\href {\doibase 10.1364/OE.25.011265} {\bibfield  {journal}
  {\bibinfo  {journal} {Opt. Express}\ }\textbf {\bibinfo {volume} {25}},\
  \bibinfo {pages} {11265} (\bibinfo {year} {2017}{\natexlab{a}})}\BibitemShut
  {NoStop}%
\bibitem [{\citenamefont {Dennis}\ \emph {et~al.}(2009)\citenamefont {Dennis},
  \citenamefont {O'Holleran},\ and\ \citenamefont {Padgett}}]{DENNIS2009}%
  \BibitemOpen
  \bibfield  {author} {\bibinfo {author} {\bibfnamefont {M.~R.}\ \bibnamefont
  {Dennis}}, \bibinfo {author} {\bibfnamefont {K.}~\bibnamefont {O'Holleran}},
  \ and\ \bibinfo {author} {\bibfnamefont {M.~J.}\ \bibnamefont {Padgett}},\
  }\href {\doibase https://doi.org/10.1016/S0079-6638(08)00205-9} {\bibfield
  {journal} {\bibinfo  {journal} {Prog. Opt.}\ }\textbf {\bibinfo {volume}
  {53}},\ \bibinfo {pages} {293} (\bibinfo {year} {2009})}\BibitemShut
  {NoStop}%
\bibitem [{\citenamefont {Ren}\ \emph {et~al.}(2010)\citenamefont {Ren},
  \citenamefont {Li}, \citenamefont {Huang}, \citenamefont {Wu}, \citenamefont
  {Gao}, \citenamefont {Wang},\ and\ \citenamefont {Li}}]{Ren2010}%
  \BibitemOpen
  \bibfield  {author} {\bibinfo {author} {\bibfnamefont {Y.-X.}\ \bibnamefont
  {Ren}}, \bibinfo {author} {\bibfnamefont {M.}~\bibnamefont {Li}}, \bibinfo
  {author} {\bibfnamefont {K.}~\bibnamefont {Huang}}, \bibinfo {author}
  {\bibfnamefont {J.-G.}\ \bibnamefont {Wu}}, \bibinfo {author} {\bibfnamefont
  {H.-F.}\ \bibnamefont {Gao}}, \bibinfo {author} {\bibfnamefont {Z.-Q.}\
  \bibnamefont {Wang}}, \ and\ \bibinfo {author} {\bibfnamefont {Y.-M.}\
  \bibnamefont {Li}},\ }\href {\doibase 10.1364/AO.49.001838} {\bibfield
  {journal} {\bibinfo  {journal} {Appl. Opt.}\ }\textbf {\bibinfo {volume}
  {49}},\ \bibinfo {pages} {1838} (\bibinfo {year} {2010})}\BibitemShut
  {NoStop}%
\bibitem [{\citenamefont {Ruffato}\ \emph {et~al.}(2014)\citenamefont
  {Ruffato}, \citenamefont {Massari},\ and\ \citenamefont
  {Romanato}}]{Ruffato2014}%
  \BibitemOpen
  \bibfield  {author} {\bibinfo {author} {\bibfnamefont {G.}~\bibnamefont
  {Ruffato}}, \bibinfo {author} {\bibfnamefont {M.}~\bibnamefont {Massari}}, \
  and\ \bibinfo {author} {\bibfnamefont {F.}~\bibnamefont {Romanato}},\ }\href
  {\doibase 10.1364/OL.39.005094} {\bibfield  {journal} {\bibinfo  {journal}
  {Opt. Lett.}\ }\textbf {\bibinfo {volume} {39}},\ \bibinfo {pages} {5094}
  (\bibinfo {year} {2014})}\BibitemShut {NoStop}%
\bibitem [{\citenamefont {Sueda}\ \emph {et~al.}(2004)\citenamefont {Sueda},
  \citenamefont {Miyaji}, \citenamefont {Miyanaga},\ and\ \citenamefont
  {Nakatsuka}}]{Sueda2004}%
  \BibitemOpen
  \bibfield  {author} {\bibinfo {author} {\bibfnamefont {K.}~\bibnamefont
  {Sueda}}, \bibinfo {author} {\bibfnamefont {G.}~\bibnamefont {Miyaji}},
  \bibinfo {author} {\bibfnamefont {N.}~\bibnamefont {Miyanaga}}, \ and\
  \bibinfo {author} {\bibfnamefont {M.}~\bibnamefont {Nakatsuka}},\ }\href
  {\doibase 10.1364/OPEX.12.003548} {\bibfield  {journal} {\bibinfo  {journal}
  {Opt. Express}\ }\textbf {\bibinfo {volume} {12}},\ \bibinfo {pages} {3548}
  (\bibinfo {year} {2004})}\BibitemShut {NoStop}%
\bibitem [{\citenamefont {Beijersbergen}\ \emph {et~al.}(1993)\citenamefont
  {Beijersbergen}, \citenamefont {Allen}, \citenamefont {{van der Veen}},\ and\
  \citenamefont {Woerdman}}]{Beijersbergen1993}%
  \BibitemOpen
  \bibfield  {author} {\bibinfo {author} {\bibfnamefont {M.}~\bibnamefont
  {Beijersbergen}}, \bibinfo {author} {\bibfnamefont {L.}~\bibnamefont
  {Allen}}, \bibinfo {author} {\bibfnamefont {H.}~\bibnamefont {{van der
  Veen}}}, \ and\ \bibinfo {author} {\bibfnamefont {J.}~\bibnamefont
  {Woerdman}},\ }\href {\doibase https://doi.org/10.1016/0030-4018(93)90535-D}
  {\bibfield  {journal} {\bibinfo  {journal} {Opt. Commun.}\ }\textbf {\bibinfo
  {volume} {96}},\ \bibinfo {pages} {123} (\bibinfo {year} {1993})}\BibitemShut
  {NoStop}%
\bibitem [{\citenamefont {Beijersbergen}\ \emph {et~al.}(1994)\citenamefont
  {Beijersbergen}, \citenamefont {Coerwinkel}, \citenamefont {Kristensen},\
  and\ \citenamefont {Woerdman}}]{Beijersbergen1994}%
  \BibitemOpen
  \bibfield  {author} {\bibinfo {author} {\bibfnamefont {M.}~\bibnamefont
  {Beijersbergen}}, \bibinfo {author} {\bibfnamefont {R.}~\bibnamefont
  {Coerwinkel}}, \bibinfo {author} {\bibfnamefont {M.}~\bibnamefont
  {Kristensen}}, \ and\ \bibinfo {author} {\bibfnamefont {J.}~\bibnamefont
  {Woerdman}},\ }\href {\doibase https://doi.org/10.1016/0030-4018(94)90638-6}
  {\bibfield  {journal} {\bibinfo  {journal} {Opt. Commun.}\ }\textbf {\bibinfo
  {volume} {112}},\ \bibinfo {pages} {321} (\bibinfo {year}
  {1994})}\BibitemShut {NoStop}%
\bibitem [{\citenamefont {Heckenberg}\ \emph {et~al.}(1992)\citenamefont
  {Heckenberg}, \citenamefont {McDuff}, \citenamefont {Smith},\ and\
  \citenamefont {White}}]{Heckenberg1992}%
  \BibitemOpen
  \bibfield  {author} {\bibinfo {author} {\bibfnamefont {N.~R.}\ \bibnamefont
  {Heckenberg}}, \bibinfo {author} {\bibfnamefont {R.}~\bibnamefont {McDuff}},
  \bibinfo {author} {\bibfnamefont {C.~P.}\ \bibnamefont {Smith}}, \ and\
  \bibinfo {author} {\bibfnamefont {A.~G.}\ \bibnamefont {White}},\ }\href
  {\doibase 10.1364/OL.17.000221} {\bibfield  {journal} {\bibinfo  {journal}
  {Opt. Lett.}\ }\textbf {\bibinfo {volume} {17}},\ \bibinfo {pages} {221}
  (\bibinfo {year} {1992})}\BibitemShut {NoStop}%
\bibitem [{\citenamefont {Mair}\ \emph {et~al.}(2001)\citenamefont {Mair},
  \citenamefont {Vaziri}, \citenamefont {Weihs},\ and\ \citenamefont
  {Zeilinger}}]{Mair2001}%
  \BibitemOpen
  \bibfield  {author} {\bibinfo {author} {\bibfnamefont {A.}~\bibnamefont
  {Mair}}, \bibinfo {author} {\bibfnamefont {A.}~\bibnamefont {Vaziri}},
  \bibinfo {author} {\bibfnamefont {G.}~\bibnamefont {Weihs}}, \ and\ \bibinfo
  {author} {\bibfnamefont {A.}~\bibnamefont {Zeilinger}},\ }\href {\doibase
  10.1038/35085529} {\bibfield  {journal} {\bibinfo  {journal} {Nature}\
  }\textbf {\bibinfo {volume} {412}},\ \bibinfo {pages} {313} (\bibinfo {year}
  {2001})}\BibitemShut {NoStop}%
\bibitem [{\citenamefont {Allen}\ \emph {et~al.}(1999)\citenamefont {Allen},
  \citenamefont {Padgett},\ and\ \citenamefont {Babiker}}]{ALLEN1999}%
  \BibitemOpen
  \bibfield  {author} {\bibinfo {author} {\bibfnamefont {L.}~\bibnamefont
  {Allen}}, \bibinfo {author} {\bibfnamefont {M.}~\bibnamefont {Padgett}}, \
  and\ \bibinfo {author} {\bibfnamefont {M.}~\bibnamefont {Babiker}},\ }\href
  {\doibase https://doi.org/10.1016/S0079-6638(08)70391-3} {\bibfield
  {journal} {\bibinfo  {journal} {Prog. Opt.}\ }\textbf {\bibinfo {volume}
  {39}},\ \bibinfo {pages} {291} (\bibinfo {year} {1999})}\BibitemShut
  {NoStop}%
\bibitem [{\citenamefont {Yao}\ and\ \citenamefont
  {Padgett}(2011{\natexlab{b}})}]{Yao_2011}%
  \BibitemOpen
  \bibfield  {author} {\bibinfo {author} {\bibfnamefont {A.~M.}\ \bibnamefont
  {Yao}}\ and\ \bibinfo {author} {\bibfnamefont {M.~J.}\ \bibnamefont
  {Padgett}},\ }\href {\doibase 10.1364/AOP.3.000161} {\bibfield  {journal}
  {\bibinfo  {journal} {Adv. Opt. Photon.}\ }\textbf {\bibinfo {volume} {3}},\
  \bibinfo {pages} {161} (\bibinfo {year} {2011}{\natexlab{b}})}\BibitemShut
  {NoStop}%
\bibitem [{\citenamefont {Franke-Arnold}\ \emph {et~al.}(2008)\citenamefont
  {Franke-Arnold}, \citenamefont {Allen},\ and\ \citenamefont
  {Padgett}}]{Franke-Arnold2008}%
  \BibitemOpen
  \bibfield  {author} {\bibinfo {author} {\bibfnamefont {S.}~\bibnamefont
  {Franke-Arnold}}, \bibinfo {author} {\bibfnamefont {L.}~\bibnamefont
  {Allen}}, \ and\ \bibinfo {author} {\bibfnamefont {M.}~\bibnamefont
  {Padgett}},\ }\href {\doibase https://doi.org/10.1002/lpor.200810007}
  {\bibfield  {journal} {\bibinfo  {journal} {Laser Photonics Rev.}\ }\textbf
  {\bibinfo {volume} {2}},\ \bibinfo {pages} {299} (\bibinfo {year}
  {2008})}\BibitemShut {NoStop}%
\bibitem [{\citenamefont {Shen}\ \emph {et~al.}(2019)\citenamefont {Shen},
  \citenamefont {Wang}, \citenamefont {Xie}, \citenamefont {Min}, \citenamefont
  {Fu}, \citenamefont {Liu}, \citenamefont {Gong},\ and\ \citenamefont
  {Yuan}}]{Shen2019}%
  \BibitemOpen
  \bibfield  {author} {\bibinfo {author} {\bibfnamefont {Y.}~\bibnamefont
  {Shen}}, \bibinfo {author} {\bibfnamefont {X.}~\bibnamefont {Wang}}, \bibinfo
  {author} {\bibfnamefont {Z.}~\bibnamefont {Xie}}, \bibinfo {author}
  {\bibfnamefont {C.}~\bibnamefont {Min}}, \bibinfo {author} {\bibfnamefont
  {X.}~\bibnamefont {Fu}}, \bibinfo {author} {\bibfnamefont {Q.}~\bibnamefont
  {Liu}}, \bibinfo {author} {\bibfnamefont {M.}~\bibnamefont {Gong}}, \ and\
  \bibinfo {author} {\bibfnamefont {X.}~\bibnamefont {Yuan}},\ }\href {\doibase
  10.1038/s41377-019-0194-2} {\bibfield  {journal} {\bibinfo  {journal} {Light
  Sci. Appl.}\ }\textbf {\bibinfo {volume} {8}},\ \bibinfo {pages} {90}
  (\bibinfo {year} {2019})}\BibitemShut {NoStop}%
\bibitem [{\citenamefont {Padgett}(2017{\natexlab{b}})}]{Padgett2017}%
  \BibitemOpen
  \bibfield  {author} {\bibinfo {author} {\bibfnamefont {M.~J.}\ \bibnamefont
  {Padgett}},\ }\href {\doibase 10.1364/OE.25.011265} {\bibfield  {journal}
  {\bibinfo  {journal} {Opt. Express}\ }\textbf {\bibinfo {volume} {25}},\
  \bibinfo {pages} {11265} (\bibinfo {year} {2017}{\natexlab{b}})}\BibitemShut
  {NoStop}%
\bibitem [{\citenamefont {Allen}\ \emph {et~al.}(1996)\citenamefont {Allen},
  \citenamefont {Babiker}, \citenamefont {Lai},\ and\ \citenamefont
  {Lembessis}}]{Allen1996}%
  \BibitemOpen
  \bibfield  {author} {\bibinfo {author} {\bibfnamefont {L.}~\bibnamefont
  {Allen}}, \bibinfo {author} {\bibfnamefont {M.}~\bibnamefont {Babiker}},
  \bibinfo {author} {\bibfnamefont {W.~K.}\ \bibnamefont {Lai}}, \ and\
  \bibinfo {author} {\bibfnamefont {V.~E.}\ \bibnamefont {Lembessis}},\ }\href
  {\doibase 10.1103/PhysRevA.54.4259} {\bibfield  {journal} {\bibinfo
  {journal} {Phys. Rev. A}\ }\textbf {\bibinfo {volume} {54}},\ \bibinfo
  {pages} {4259} (\bibinfo {year} {1996})}\BibitemShut {NoStop}%
\bibitem [{\citenamefont {Babiker}\ \emph {et~al.}(2002)\citenamefont
  {Babiker}, \citenamefont {Bennett}, \citenamefont {Andrews},\ and\
  \citenamefont {D\'avila~Romero}}]{Babiker2002}%
  \BibitemOpen
  \bibfield  {author} {\bibinfo {author} {\bibfnamefont {M.}~\bibnamefont
  {Babiker}}, \bibinfo {author} {\bibfnamefont {C.~R.}\ \bibnamefont
  {Bennett}}, \bibinfo {author} {\bibfnamefont {D.~L.}\ \bibnamefont
  {Andrews}}, \ and\ \bibinfo {author} {\bibfnamefont {L.~C.}\ \bibnamefont
  {D\'avila~Romero}},\ }\href {\doibase 10.1103/PhysRevLett.89.143601}
  {\bibfield  {journal} {\bibinfo  {journal} {Phys. Rev. Lett.}\ }\textbf
  {\bibinfo {volume} {89}},\ \bibinfo {pages} {143601} (\bibinfo {year}
  {2002})}\BibitemShut {NoStop}%
\bibitem [{\citenamefont {Lloyd}\ \emph {et~al.}(2012)\citenamefont {Lloyd},
  \citenamefont {Babiker},\ and\ \citenamefont {Yuan}}]{Lloyd2012}%
  \BibitemOpen
  \bibfield  {author} {\bibinfo {author} {\bibfnamefont {S.}~\bibnamefont
  {Lloyd}}, \bibinfo {author} {\bibfnamefont {M.}~\bibnamefont {Babiker}}, \
  and\ \bibinfo {author} {\bibfnamefont {J.}~\bibnamefont {Yuan}},\ }\href
  {\doibase 10.1103/PhysRevLett.108.074802} {\bibfield  {journal} {\bibinfo
  {journal} {Phys. Rev. Lett.}\ }\textbf {\bibinfo {volume} {108}},\ \bibinfo
  {pages} {074802} (\bibinfo {year} {2012})}\BibitemShut {NoStop}%
\bibitem [{\citenamefont {Power}\ \emph {et~al.}(1995)\citenamefont {Power},
  \citenamefont {Allen}, \citenamefont {Babiker},\ and\ \citenamefont
  {Lembessis}}]{Power1995}%
  \BibitemOpen
  \bibfield  {author} {\bibinfo {author} {\bibfnamefont {W.~L.}\ \bibnamefont
  {Power}}, \bibinfo {author} {\bibfnamefont {L.}~\bibnamefont {Allen}},
  \bibinfo {author} {\bibfnamefont {M.}~\bibnamefont {Babiker}}, \ and\
  \bibinfo {author} {\bibfnamefont {V.~E.}\ \bibnamefont {Lembessis}},\ }\href
  {\doibase 10.1103/PhysRevA.52.479} {\bibfield  {journal} {\bibinfo  {journal}
  {Phys. Rev. A}\ }\textbf {\bibinfo {volume} {52}},\ \bibinfo {pages} {479}
  (\bibinfo {year} {1995})}\BibitemShut {NoStop}%
\bibitem [{\citenamefont {Araoka}\ \emph {et~al.}(2005)\citenamefont {Araoka},
  \citenamefont {Verbiest}, \citenamefont {Clays},\ and\ \citenamefont
  {Persoons}}]{Araoka2005}%
  \BibitemOpen
  \bibfield  {author} {\bibinfo {author} {\bibfnamefont {F.}~\bibnamefont
  {Araoka}}, \bibinfo {author} {\bibfnamefont {T.}~\bibnamefont {Verbiest}},
  \bibinfo {author} {\bibfnamefont {K.}~\bibnamefont {Clays}}, \ and\ \bibinfo
  {author} {\bibfnamefont {A.}~\bibnamefont {Persoons}},\ }\href {\doibase
  10.1103/PhysRevA.71.055401} {\bibfield  {journal} {\bibinfo  {journal} {Phys.
  Rev. A}\ }\textbf {\bibinfo {volume} {71}},\ \bibinfo {pages} {055401}
  (\bibinfo {year} {2005})}\BibitemShut {NoStop}%
\bibitem [{\citenamefont {Bougouffa}\ and\ \citenamefont
  {Babiker}(2020)}]{Bougouffa2020}%
  \BibitemOpen
  \bibfield  {author} {\bibinfo {author} {\bibfnamefont {S.}~\bibnamefont
  {Bougouffa}}\ and\ \bibinfo {author} {\bibfnamefont {M.}~\bibnamefont
  {Babiker}},\ }\href {\doibase 10.1103/PhysRevA.102.063706} {\bibfield
  {journal} {\bibinfo  {journal} {Phys. Rev. A}\ }\textbf {\bibinfo {volume}
  {102}},\ \bibinfo {pages} {063706} (\bibinfo {year} {2020})}\BibitemShut
  {NoStop}%
\bibitem [{\citenamefont {Mashhadi}(2017)}]{Mashhadi_2017}%
  \BibitemOpen
  \bibfield  {author} {\bibinfo {author} {\bibfnamefont {L.}~\bibnamefont
  {Mashhadi}},\ }\href {\doibase 10.1088/1361-6455/aa93cb} {\bibfield
  {journal} {\bibinfo  {journal} {J. Phys. B: At. Mol. Opt. Phys.}\ }\textbf
  {\bibinfo {volume} {50}},\ \bibinfo {pages} {245201} (\bibinfo {year}
  {2017})}\BibitemShut {NoStop}%
\bibitem [{\citenamefont {Quinteiro}\ \emph {et~al.}(2017)\citenamefont
  {Quinteiro}, \citenamefont {Reiter},\ and\ \citenamefont
  {Kuhn}}]{Quinteiro2017}%
  \BibitemOpen
  \bibfield  {author} {\bibinfo {author} {\bibfnamefont {G.~F.}\ \bibnamefont
  {Quinteiro}}, \bibinfo {author} {\bibfnamefont {D.~E.}\ \bibnamefont
  {Reiter}}, \ and\ \bibinfo {author} {\bibfnamefont {T.}~\bibnamefont
  {Kuhn}},\ }\href {\doibase 10.1103/PhysRevA.95.012106} {\bibfield  {journal}
  {\bibinfo  {journal} {Phys. Rev. A}\ }\textbf {\bibinfo {volume} {95}},\
  \bibinfo {pages} {012106} (\bibinfo {year} {2017})}\BibitemShut {NoStop}%
\bibitem [{\citenamefont {Mukherjee}\ \emph {et~al.}(2017)\citenamefont
  {Mukherjee}, \citenamefont {Majumder}, \citenamefont {Mondal},\ and\
  \citenamefont {Deb}}]{Mukherjee_2017}%
  \BibitemOpen
  \bibfield  {author} {\bibinfo {author} {\bibfnamefont {K.}~\bibnamefont
  {Mukherjee}}, \bibinfo {author} {\bibfnamefont {S.}~\bibnamefont {Majumder}},
  \bibinfo {author} {\bibfnamefont {P.~K.}\ \bibnamefont {Mondal}}, \ and\
  \bibinfo {author} {\bibfnamefont {B.}~\bibnamefont {Deb}},\ }\href {\doibase
  10.1088/1361-6455/aa90d3} {\bibfield  {journal} {\bibinfo  {journal} {J.
  Phys. B: At. Mol. and Opt. Phys.}\ }\textbf {\bibinfo {volume} {51}},\
  \bibinfo {pages} {015004} (\bibinfo {year} {2017})}\BibitemShut {NoStop}%
\bibitem [{\citenamefont {Andersen}\ \emph {et~al.}(2006)\citenamefont
  {Andersen}, \citenamefont {Ryu}, \citenamefont {Clad\'e}, \citenamefont
  {Natarajan}, \citenamefont {Vaziri}, \citenamefont {Helmerson},\ and\
  \citenamefont {Phillips}}]{Andersen2006}%
  \BibitemOpen
  \bibfield  {author} {\bibinfo {author} {\bibfnamefont {M.~F.}\ \bibnamefont
  {Andersen}}, \bibinfo {author} {\bibfnamefont {C.}~\bibnamefont {Ryu}},
  \bibinfo {author} {\bibfnamefont {P.}~\bibnamefont {Clad\'e}}, \bibinfo
  {author} {\bibfnamefont {V.}~\bibnamefont {Natarajan}}, \bibinfo {author}
  {\bibfnamefont {A.}~\bibnamefont {Vaziri}}, \bibinfo {author} {\bibfnamefont
  {K.}~\bibnamefont {Helmerson}}, \ and\ \bibinfo {author} {\bibfnamefont
  {W.~D.}\ \bibnamefont {Phillips}},\ }\href {\doibase
  10.1103/PhysRevLett.97.170406} {\bibfield  {journal} {\bibinfo  {journal}
  {Phys. Rev. Lett.}\ }\textbf {\bibinfo {volume} {97}},\ \bibinfo {pages}
  {170406} (\bibinfo {year} {2006})}\BibitemShut {NoStop}%
\bibitem [{\citenamefont {Mondal}\ \emph {et~al.}(2014)\citenamefont {Mondal},
  \citenamefont {Deb},\ and\ \citenamefont {Majumder}}]{Mondal2014}%
  \BibitemOpen
  \bibfield  {author} {\bibinfo {author} {\bibfnamefont {P.~K.}\ \bibnamefont
  {Mondal}}, \bibinfo {author} {\bibfnamefont {B.}~\bibnamefont {Deb}}, \ and\
  \bibinfo {author} {\bibfnamefont {S.}~\bibnamefont {Majumder}},\ }\href
  {\doibase 10.1103/PhysRevA.89.063418} {\bibfield  {journal} {\bibinfo
  {journal} {Phys. Rev. A}\ }\textbf {\bibinfo {volume} {89}},\ \bibinfo
  {pages} {063418} (\bibinfo {year} {2014})}\BibitemShut {NoStop}%
\bibitem [{\citenamefont {Mondal}\ \emph {et~al.}(2015)\citenamefont {Mondal},
  \citenamefont {Deb},\ and\ \citenamefont {Majumder}}]{Mondal2015}%
  \BibitemOpen
  \bibfield  {author} {\bibinfo {author} {\bibfnamefont {P.~K.}\ \bibnamefont
  {Mondal}}, \bibinfo {author} {\bibfnamefont {B.}~\bibnamefont {Deb}}, \ and\
  \bibinfo {author} {\bibfnamefont {S.}~\bibnamefont {Majumder}},\ }\href
  {\doibase 10.1103/PhysRevA.92.043603} {\bibfield  {journal} {\bibinfo
  {journal} {Phys. Rev. A}\ }\textbf {\bibinfo {volume} {92}},\ \bibinfo
  {pages} {043603} (\bibinfo {year} {2015})}\BibitemShut {NoStop}%
\bibitem [{\citenamefont {Bhowmik}\ \emph {et~al.}(2016)\citenamefont
  {Bhowmik}, \citenamefont {Mondal}, \citenamefont {Majumder},\ and\
  \citenamefont {Deb}}]{Bhowmik2016}%
  \BibitemOpen
  \bibfield  {author} {\bibinfo {author} {\bibfnamefont {A.}~\bibnamefont
  {Bhowmik}}, \bibinfo {author} {\bibfnamefont {P.~K.}\ \bibnamefont {Mondal}},
  \bibinfo {author} {\bibfnamefont {S.}~\bibnamefont {Majumder}}, \ and\
  \bibinfo {author} {\bibfnamefont {B.}~\bibnamefont {Deb}},\ }\href {\doibase
  10.1103/PhysRevA.93.063852} {\bibfield  {journal} {\bibinfo  {journal} {Phys.
  Rev. A}\ }\textbf {\bibinfo {volume} {93}},\ \bibinfo {pages} {063852}
  (\bibinfo {year} {2016})}\BibitemShut {NoStop}%
\bibitem [{\citenamefont {Kanamoto}\ \emph {et~al.}(2007)\citenamefont
  {Kanamoto}, \citenamefont {Wright},\ and\ \citenamefont
  {Meystre}}]{Kanamoto2007}%
  \BibitemOpen
  \bibfield  {author} {\bibinfo {author} {\bibfnamefont {R.}~\bibnamefont
  {Kanamoto}}, \bibinfo {author} {\bibfnamefont {E.~M.}\ \bibnamefont
  {Wright}}, \ and\ \bibinfo {author} {\bibfnamefont {P.}~\bibnamefont
  {Meystre}},\ }\href {\doibase 10.1103/PhysRevA.75.063623} {\bibfield
  {journal} {\bibinfo  {journal} {Phys. Rev. A}\ }\textbf {\bibinfo {volume}
  {75}},\ \bibinfo {pages} {063623} (\bibinfo {year} {2007})}\BibitemShut
  {NoStop}%
\bibitem [{\citenamefont {Wright}\ \emph {et~al.}(2008)\citenamefont {Wright},
  \citenamefont {Leslie},\ and\ \citenamefont {Bigelow}}]{Wright2008}%
  \BibitemOpen
  \bibfield  {author} {\bibinfo {author} {\bibfnamefont {K.~C.}\ \bibnamefont
  {Wright}}, \bibinfo {author} {\bibfnamefont {L.~S.}\ \bibnamefont {Leslie}},
  \ and\ \bibinfo {author} {\bibfnamefont {N.~P.}\ \bibnamefont {Bigelow}},\
  }\href {\doibase 10.1103/PhysRevA.77.041601} {\bibfield  {journal} {\bibinfo
  {journal} {Phys. Rev. A}\ }\textbf {\bibinfo {volume} {77}},\ \bibinfo
  {pages} {041601} (\bibinfo {year} {2008})}\BibitemShut {NoStop}%
\bibitem [{\citenamefont {Bhowmik}\ \emph {et~al.}(2018)\citenamefont
  {Bhowmik}, \citenamefont {Mondal}, \citenamefont {Majumder},\ and\
  \citenamefont {Deb}}]{Bhowmik_2018_paraxial}%
  \BibitemOpen
  \bibfield  {author} {\bibinfo {author} {\bibfnamefont {A.}~\bibnamefont
  {Bhowmik}}, \bibinfo {author} {\bibfnamefont {P.~K.}\ \bibnamefont {Mondal}},
  \bibinfo {author} {\bibfnamefont {S.}~\bibnamefont {Majumder}}, \ and\
  \bibinfo {author} {\bibfnamefont {B.}~\bibnamefont {Deb}},\ }\href {\doibase
  10.1088/1361-6455/aac626} {\bibfield  {journal} {\bibinfo  {journal} {J.
  Phys. B: At. Mol. and Opt. Phys.}\ }\textbf {\bibinfo {volume} {51}},\
  \bibinfo {pages} {135003} (\bibinfo {year} {2018})}\BibitemShut {NoStop}%
\bibitem [{\citenamefont {Bhowmik}\ and\ \citenamefont
  {Majumder}(2018)}]{Bhowmik_2018}%
  \BibitemOpen
  \bibfield  {author} {\bibinfo {author} {\bibfnamefont {A.}~\bibnamefont
  {Bhowmik}}\ and\ \bibinfo {author} {\bibfnamefont {S.}~\bibnamefont
  {Majumder}},\ }\href {\doibase 10.1088/2399-6528/aaf189} {\bibfield
  {journal} {\bibinfo  {journal} {J. Phys. Commun.}\ }\textbf {\bibinfo
  {volume} {2}},\ \bibinfo {pages} {125001} (\bibinfo {year}
  {2018})}\BibitemShut {NoStop}%
\bibitem [{\citenamefont {Das}\ \emph {et~al.}(2020)\citenamefont {Das},
  \citenamefont {Bhowmik}, \citenamefont {Mukherjee},\ and\ \citenamefont
  {Majumder}}]{Das_2020}%
  \BibitemOpen
  \bibfield  {author} {\bibinfo {author} {\bibfnamefont {S.}~\bibnamefont
  {Das}}, \bibinfo {author} {\bibfnamefont {A.}~\bibnamefont {Bhowmik}},
  \bibinfo {author} {\bibfnamefont {K.}~\bibnamefont {Mukherjee}}, \ and\
  \bibinfo {author} {\bibfnamefont {S.}~\bibnamefont {Majumder}},\ }\href
  {\doibase 10.1088/1361-6455/ab42c4} {\bibfield  {journal} {\bibinfo
  {journal} {J. Phys. B: At. Mol. and Opt. Phys.}\ }\textbf {\bibinfo {volume}
  {53}},\ \bibinfo {pages} {025302} (\bibinfo {year} {2020})}\BibitemShut
  {NoStop}%
\bibitem [{\citenamefont {Tempere}\ \emph {et~al.}(2001)\citenamefont
  {Tempere}, \citenamefont {Devreese},\ and\ \citenamefont
  {Abraham}}]{Tempere2001}%
  \BibitemOpen
  \bibfield  {author} {\bibinfo {author} {\bibfnamefont {J.}~\bibnamefont
  {Tempere}}, \bibinfo {author} {\bibfnamefont {J.~T.}\ \bibnamefont
  {Devreese}}, \ and\ \bibinfo {author} {\bibfnamefont {E.~R.~I.}\ \bibnamefont
  {Abraham}},\ }\href {\doibase 10.1103/PhysRevA.64.023603} {\bibfield
  {journal} {\bibinfo  {journal} {Phys. Rev. A}\ }\textbf {\bibinfo {volume}
  {64}},\ \bibinfo {pages} {023603} (\bibinfo {year} {2001})}\BibitemShut
  {NoStop}%
\bibitem [{\citenamefont {Marzlin}\ \emph {et~al.}(1997)\citenamefont
  {Marzlin}, \citenamefont {Zhang},\ and\ \citenamefont
  {Wright}}]{Marzlin1997}%
  \BibitemOpen
  \bibfield  {author} {\bibinfo {author} {\bibfnamefont {K.-P.}\ \bibnamefont
  {Marzlin}}, \bibinfo {author} {\bibfnamefont {W.}~\bibnamefont {Zhang}}, \
  and\ \bibinfo {author} {\bibfnamefont {E.~M.}\ \bibnamefont {Wright}},\
  }\href {\doibase 10.1103/PhysRevLett.79.4728} {\bibfield  {journal} {\bibinfo
   {journal} {Phys. Rev. Lett.}\ }\textbf {\bibinfo {volume} {79}},\ \bibinfo
  {pages} {4728} (\bibinfo {year} {1997})}\BibitemShut {NoStop}%
\bibitem [{\citenamefont {Simula}\ \emph {et~al.}(2008)\citenamefont {Simula},
  \citenamefont {Nygaard}, \citenamefont {Hu}, \citenamefont {Collins},
  \citenamefont {Schneider},\ and\ \citenamefont {M\o{}lmer}}]{Simula2008}%
  \BibitemOpen
  \bibfield  {author} {\bibinfo {author} {\bibfnamefont {T.~P.}\ \bibnamefont
  {Simula}}, \bibinfo {author} {\bibfnamefont {N.}~\bibnamefont {Nygaard}},
  \bibinfo {author} {\bibfnamefont {S.~X.}\ \bibnamefont {Hu}}, \bibinfo
  {author} {\bibfnamefont {L.~A.}\ \bibnamefont {Collins}}, \bibinfo {author}
  {\bibfnamefont {B.~I.}\ \bibnamefont {Schneider}}, \ and\ \bibinfo {author}
  {\bibfnamefont {K.}~\bibnamefont {M\o{}lmer}},\ }\href {\doibase
  10.1103/PhysRevA.77.015401} {\bibfield  {journal} {\bibinfo  {journal} {Phys.
  Rev. A}\ }\textbf {\bibinfo {volume} {77}},\ \bibinfo {pages} {015401}
  (\bibinfo {year} {2008})}\BibitemShut {NoStop}%
\bibitem [{\citenamefont {Nandi}\ \emph {et~al.}(2004)\citenamefont {Nandi},
  \citenamefont {Walser},\ and\ \citenamefont {Schleich}}]{Nandi2004}%
  \BibitemOpen
  \bibfield  {author} {\bibinfo {author} {\bibfnamefont {G.}~\bibnamefont
  {Nandi}}, \bibinfo {author} {\bibfnamefont {R.}~\bibnamefont {Walser}}, \
  and\ \bibinfo {author} {\bibfnamefont {W.~P.}\ \bibnamefont {Schleich}},\
  }\href {\doibase 10.1103/PhysRevA.69.063606} {\bibfield  {journal} {\bibinfo
  {journal} {Phys. Rev. A}\ }\textbf {\bibinfo {volume} {69}},\ \bibinfo
  {pages} {063606} (\bibinfo {year} {2004})}\BibitemShut {NoStop}%
\bibitem [{\citenamefont {Mukherjee}\ \emph {et~al.}(2021)\citenamefont
  {Mukherjee}, \citenamefont {Bandyopadhyay}, \citenamefont {Angom},
  \citenamefont {Martin},\ and\ \citenamefont {Majumder}}]{Mukherjee2021}%
  \BibitemOpen
  \bibfield  {author} {\bibinfo {author} {\bibfnamefont {K.}~\bibnamefont
  {Mukherjee}}, \bibinfo {author} {\bibfnamefont {S.}~\bibnamefont
  {Bandyopadhyay}}, \bibinfo {author} {\bibfnamefont {D.}~\bibnamefont
  {Angom}}, \bibinfo {author} {\bibfnamefont {A.~M.}\ \bibnamefont {Martin}}, \
  and\ \bibinfo {author} {\bibfnamefont {S.}~\bibnamefont {Majumder}},\ }\href
  {\doibase 10.3390/atoms9010014} {\bibfield  {journal} {\bibinfo  {journal}
  {Atoms}\ }\textbf {\bibinfo {volume} {9}},\ \bibinfo {pages} {14} (\bibinfo
  {year} {2021})}\BibitemShut {NoStop}%
\bibitem [{\citenamefont {Huang}\ \emph {et~al.}(2016)\citenamefont {Huang},
  \citenamefont {Miao}, \citenamefont {He}, \citenamefont {Pang}, \citenamefont
  {Li},\ and\ \citenamefont {Wang}}]{HUANG2016132}%
  \BibitemOpen
  \bibfield  {author} {\bibinfo {author} {\bibfnamefont {S.}~\bibnamefont
  {Huang}}, \bibinfo {author} {\bibfnamefont {Z.}~\bibnamefont {Miao}},
  \bibinfo {author} {\bibfnamefont {C.}~\bibnamefont {He}}, \bibinfo {author}
  {\bibfnamefont {F.}~\bibnamefont {Pang}}, \bibinfo {author} {\bibfnamefont
  {Y.}~\bibnamefont {Li}}, \ and\ \bibinfo {author} {\bibfnamefont
  {T.}~\bibnamefont {Wang}},\ }\href {\doibase
  https://doi.org/10.1016/j.optlaseng.2015.10.008} {\bibfield  {journal}
  {\bibinfo  {journal} {Opt. Lasers Eng.}\ }\textbf {\bibinfo {volume} {78}},\
  \bibinfo {pages} {132} (\bibinfo {year} {2016})}\BibitemShut {NoStop}%
\bibitem [{\citenamefont {Tao}\ \emph {et~al.}(2006)\citenamefont {Tao},
  \citenamefont {Yuan}, \citenamefont {Lin},\ and\ \citenamefont
  {Burge}}]{Tao2006}%
  \BibitemOpen
  \bibfield  {author} {\bibinfo {author} {\bibfnamefont {S.~H.}\ \bibnamefont
  {Tao}}, \bibinfo {author} {\bibfnamefont {X.-C.}\ \bibnamefont {Yuan}},
  \bibinfo {author} {\bibfnamefont {J.}~\bibnamefont {Lin}}, \ and\ \bibinfo
  {author} {\bibfnamefont {R.~E.}\ \bibnamefont {Burge}},\ }\href {\doibase
  10.1364/OPEX.14.000535} {\bibfield  {journal} {\bibinfo  {journal} {Opt.
  Express}\ }\textbf {\bibinfo {volume} {14}},\ \bibinfo {pages} {535}
  (\bibinfo {year} {2006})}\BibitemShut {NoStop}%
\bibitem [{\citenamefont {Yang}\ \emph {et~al.}(2011)\citenamefont {Yang},
  \citenamefont {Zhao}, \citenamefont {Zhao},\ and\ \citenamefont
  {Kong}}]{YANG20113597}%
  \BibitemOpen
  \bibfield  {author} {\bibinfo {author} {\bibfnamefont {D.}~\bibnamefont
  {Yang}}, \bibinfo {author} {\bibfnamefont {J.}~\bibnamefont {Zhao}}, \bibinfo
  {author} {\bibfnamefont {T.}~\bibnamefont {Zhao}}, \ and\ \bibinfo {author}
  {\bibfnamefont {L.}~\bibnamefont {Kong}},\ }\href {\doibase
  https://doi.org/10.1016/j.optcom.2011.03.075} {\bibfield  {journal} {\bibinfo
   {journal} {Opt. Commun.}\ }\textbf {\bibinfo {volume} {284}},\ \bibinfo
  {pages} {3597} (\bibinfo {year} {2011})}\BibitemShut {NoStop}%
\bibitem [{\citenamefont {Franke-Arnold}\ \emph {et~al.}(2007)\citenamefont
  {Franke-Arnold}, \citenamefont {Leach}, \citenamefont {Padgett},
  \citenamefont {Lembessis}, \citenamefont {Ellinas}, \citenamefont {Wright},
  \citenamefont {Girkin}, \citenamefont {\"{O}hberg},\ and\ \citenamefont
  {Arnold}}]{Franke_2007}%
  \BibitemOpen
  \bibfield  {author} {\bibinfo {author} {\bibfnamefont {S.}~\bibnamefont
  {Franke-Arnold}}, \bibinfo {author} {\bibfnamefont {J.}~\bibnamefont
  {Leach}}, \bibinfo {author} {\bibfnamefont {M.~J.}\ \bibnamefont {Padgett}},
  \bibinfo {author} {\bibfnamefont {V.~E.}\ \bibnamefont {Lembessis}}, \bibinfo
  {author} {\bibfnamefont {D.}~\bibnamefont {Ellinas}}, \bibinfo {author}
  {\bibfnamefont {A.~J.}\ \bibnamefont {Wright}}, \bibinfo {author}
  {\bibfnamefont {J.~M.}\ \bibnamefont {Girkin}}, \bibinfo {author}
  {\bibfnamefont {P.}~\bibnamefont {\"{O}hberg}}, \ and\ \bibinfo {author}
  {\bibfnamefont {A.~S.}\ \bibnamefont {Arnold}},\ }\href {\doibase
  10.1364/OE.15.008619} {\bibfield  {journal} {\bibinfo  {journal} {Opt.
  Express}\ }\textbf {\bibinfo {volume} {15}},\ \bibinfo {pages} {8619}
  (\bibinfo {year} {2007})}\BibitemShut {NoStop}%
\bibitem [{\citenamefont {Chin}\ \emph {et~al.}(2010)\citenamefont {Chin},
  \citenamefont {Grimm}, \citenamefont {Julienne},\ and\ \citenamefont
  {Tiesinga}}]{chin2010}%
  \BibitemOpen
  \bibfield  {author} {\bibinfo {author} {\bibfnamefont {C.}~\bibnamefont
  {Chin}}, \bibinfo {author} {\bibfnamefont {R.}~\bibnamefont {Grimm}},
  \bibinfo {author} {\bibfnamefont {P.}~\bibnamefont {Julienne}}, \ and\
  \bibinfo {author} {\bibfnamefont {E.}~\bibnamefont {Tiesinga}},\ }\href
  {\doibase 10.1103/RevModPhys.82.1225} {\bibfield  {journal} {\bibinfo
  {journal} {Rev. Mod. Phys.}\ }\textbf {\bibinfo {volume} {82}},\ \bibinfo
  {pages} {1225} (\bibinfo {year} {2010})}\BibitemShut {NoStop}%
\bibitem [{\citenamefont {K\"ohler}\ \emph {et~al.}(2006)\citenamefont
  {K\"ohler}, \citenamefont {G\'oral},\ and\ \citenamefont
  {Julienne}}]{Thorsten2006}%
  \BibitemOpen
  \bibfield  {author} {\bibinfo {author} {\bibfnamefont {T.}~\bibnamefont
  {K\"ohler}}, \bibinfo {author} {\bibfnamefont {K.}~\bibnamefont {G\'oral}}, \
  and\ \bibinfo {author} {\bibfnamefont {P.~S.}\ \bibnamefont {Julienne}},\
  }\href {\doibase 10.1103/RevModPhys.78.1311} {\bibfield  {journal} {\bibinfo
  {journal} {Rev. Mod. Phys.}\ }\textbf {\bibinfo {volume} {78}},\ \bibinfo
  {pages} {1311} (\bibinfo {year} {2006})}\BibitemShut {NoStop}%
\bibitem [{\citenamefont {G\"orlitz}\ \emph {et~al.}(2001)\citenamefont
  {G\"orlitz}, \citenamefont {Vogels}, \citenamefont {Leanhardt}, \citenamefont
  {Raman}, \citenamefont {Gustavson}, \citenamefont {Abo-Shaeer}, \citenamefont
  {Chikkatur}, \citenamefont {Gupta}, \citenamefont {Inouye}, \citenamefont
  {Rosenband},\ and\ \citenamefont {Ketterle}}]{Vogels2001}%
  \BibitemOpen
  \bibfield  {author} {\bibinfo {author} {\bibfnamefont {A.}~\bibnamefont
  {G\"orlitz}}, \bibinfo {author} {\bibfnamefont {J.~M.}\ \bibnamefont
  {Vogels}}, \bibinfo {author} {\bibfnamefont {A.~E.}\ \bibnamefont
  {Leanhardt}}, \bibinfo {author} {\bibfnamefont {C.}~\bibnamefont {Raman}},
  \bibinfo {author} {\bibfnamefont {T.~L.}\ \bibnamefont {Gustavson}}, \bibinfo
  {author} {\bibfnamefont {J.~R.}\ \bibnamefont {Abo-Shaeer}}, \bibinfo
  {author} {\bibfnamefont {A.~P.}\ \bibnamefont {Chikkatur}}, \bibinfo {author}
  {\bibfnamefont {S.}~\bibnamefont {Gupta}}, \bibinfo {author} {\bibfnamefont
  {S.}~\bibnamefont {Inouye}}, \bibinfo {author} {\bibfnamefont
  {T.}~\bibnamefont {Rosenband}}, \ and\ \bibinfo {author} {\bibfnamefont
  {W.}~\bibnamefont {Ketterle}},\ }\href {\doibase
  10.1103/PhysRevLett.87.130402} {\bibfield  {journal} {\bibinfo  {journal}
  {Phys. Rev. Lett.}\ }\textbf {\bibinfo {volume} {87}},\ \bibinfo {pages}
  {130402} (\bibinfo {year} {2001})}\BibitemShut {NoStop}%
\bibitem [{\citenamefont {Nicolin}(2011)}]{Nicolin2011}%
  \BibitemOpen
  \bibfield  {author} {\bibinfo {author} {\bibfnamefont {A.~I.}\ \bibnamefont
  {Nicolin}},\ }\href {\doibase 10.1103/PhysRevE.84.056202} {\bibfield
  {journal} {\bibinfo  {journal} {Phys. Rev. E}\ }\textbf {\bibinfo {volume}
  {84}},\ \bibinfo {pages} {056202} (\bibinfo {year} {2011})}\BibitemShut
  {NoStop}%
\bibitem [{\citenamefont {Staliunas}\ \emph {et~al.}(2002)\citenamefont
  {Staliunas}, \citenamefont {Longhi},\ and\ \citenamefont
  {de~Valc\'arcel}}]{Staliunas2002}%
  \BibitemOpen
  \bibfield  {author} {\bibinfo {author} {\bibfnamefont {K.}~\bibnamefont
  {Staliunas}}, \bibinfo {author} {\bibfnamefont {S.}~\bibnamefont {Longhi}}, \
  and\ \bibinfo {author} {\bibfnamefont {G.~J.}\ \bibnamefont
  {de~Valc\'arcel}},\ }\href {\doibase 10.1103/PhysRevLett.89.210406}
  {\bibfield  {journal} {\bibinfo  {journal} {Phys. Rev. Lett.}\ }\textbf
  {\bibinfo {volume} {89}},\ \bibinfo {pages} {210406} (\bibinfo {year}
  {2002})}\BibitemShut {NoStop}%
\bibitem [{\citenamefont {Engels}\ \emph {et~al.}(2007)\citenamefont {Engels},
  \citenamefont {Atherton},\ and\ \citenamefont {Hoefer}}]{Engels2007}%
  \BibitemOpen
  \bibfield  {author} {\bibinfo {author} {\bibfnamefont {P.}~\bibnamefont
  {Engels}}, \bibinfo {author} {\bibfnamefont {C.}~\bibnamefont {Atherton}}, \
  and\ \bibinfo {author} {\bibfnamefont {M.~A.}\ \bibnamefont {Hoefer}},\
  }\href {\doibase 10.1103/PhysRevLett.98.095301} {\bibfield  {journal}
  {\bibinfo  {journal} {Phys. Rev. Lett.}\ }\textbf {\bibinfo {volume} {98}},\
  \bibinfo {pages} {095301} (\bibinfo {year} {2007})}\BibitemShut {NoStop}%
\bibitem [{\citenamefont {Maity}\ \emph {et~al.}(2020)\citenamefont {Maity},
  \citenamefont {Mukherjee}, \citenamefont {Mistakidis}, \citenamefont {Das},
  \citenamefont {Kevrekidis}, \citenamefont {Majumder},\ and\ \citenamefont
  {Schmelcher}}]{Maity2020}%
  \BibitemOpen
  \bibfield  {author} {\bibinfo {author} {\bibfnamefont {D.~K.}\ \bibnamefont
  {Maity}}, \bibinfo {author} {\bibfnamefont {K.}~\bibnamefont {Mukherjee}},
  \bibinfo {author} {\bibfnamefont {S.~I.}\ \bibnamefont {Mistakidis}},
  \bibinfo {author} {\bibfnamefont {S.}~\bibnamefont {Das}}, \bibinfo {author}
  {\bibfnamefont {P.~G.}\ \bibnamefont {Kevrekidis}}, \bibinfo {author}
  {\bibfnamefont {S.}~\bibnamefont {Majumder}}, \ and\ \bibinfo {author}
  {\bibfnamefont {P.}~\bibnamefont {Schmelcher}},\ }\href {\doibase
  10.1103/PhysRevA.102.033320} {\bibfield  {journal} {\bibinfo  {journal}
  {Phys. Rev. A}\ }\textbf {\bibinfo {volume} {102}},\ \bibinfo {pages}
  {033320} (\bibinfo {year} {2020})}\BibitemShut {NoStop}%
\bibitem [{\citenamefont {Zhang}\ \emph {et~al.}(2020)\citenamefont {Zhang},
  \citenamefont {Yao}, \citenamefont {Feng}, \citenamefont {Hu},\ and\
  \citenamefont {Chin}}]{Zhang2020}%
  \BibitemOpen
  \bibfield  {author} {\bibinfo {author} {\bibfnamefont {Z.}~\bibnamefont
  {Zhang}}, \bibinfo {author} {\bibfnamefont {K.-X.}\ \bibnamefont {Yao}},
  \bibinfo {author} {\bibfnamefont {L.}~\bibnamefont {Feng}}, \bibinfo {author}
  {\bibfnamefont {J.}~\bibnamefont {Hu}}, \ and\ \bibinfo {author}
  {\bibfnamefont {C.}~\bibnamefont {Chin}},\ }\href {\doibase
  10.1038/s41567-020-0839-3} {\bibfield  {journal} {\bibinfo  {journal} {Nat.
  Phys}\ }\textbf {\bibinfo {volume} {16}},\ \bibinfo {pages} {652} (\bibinfo
  {year} {2020})}\BibitemShut {NoStop}%
\bibitem [{\citenamefont {Anglin}\ and\ \citenamefont
  {Zurek}(1999)}]{Anglin1999}%
  \BibitemOpen
  \bibfield  {author} {\bibinfo {author} {\bibfnamefont {J.~R.}\ \bibnamefont
  {Anglin}}\ and\ \bibinfo {author} {\bibfnamefont {W.~H.}\ \bibnamefont
  {Zurek}},\ }\href {\doibase 10.1103/PhysRevLett.83.1707} {\bibfield
  {journal} {\bibinfo  {journal} {Phys. Rev. Lett.}\ }\textbf {\bibinfo
  {volume} {83}},\ \bibinfo {pages} {1707} (\bibinfo {year}
  {1999})}\BibitemShut {NoStop}%
\bibitem [{\citenamefont {Mukherjee}\ \emph {et~al.}(2020)\citenamefont
  {Mukherjee}, \citenamefont {Mistakidis}, \citenamefont {Kevrekidis},\ and\
  \citenamefont {Schmelcher}}]{Mukherjee_2020}%
  \BibitemOpen
  \bibfield  {author} {\bibinfo {author} {\bibfnamefont {K.}~\bibnamefont
  {Mukherjee}}, \bibinfo {author} {\bibfnamefont {S.}~\bibnamefont
  {Mistakidis}}, \bibinfo {author} {\bibfnamefont {P.~G.}\ \bibnamefont
  {Kevrekidis}}, \ and\ \bibinfo {author} {\bibfnamefont {P.}~\bibnamefont
  {Schmelcher}},\ }\href {https://doi.org/10.1088/1361-6455/ab678d} {\bibfield
  {journal} {\bibinfo  {journal} {J. Phys. B: At. Mol. Opt. Phys.}\ } (\bibinfo
  {year} {2020})}\BibitemShut {NoStop}%
\bibitem [{\citenamefont {Kevrekidis}\ \emph {et~al.}(2004)\citenamefont
  {Kevrekidis}, \citenamefont {Carretero-Gonz{\'a}lez}, \citenamefont
  {Frantzeskakis},\ and\ \citenamefont {Kevrekidis}}]{KEVREKIDIS_2004}%
  \BibitemOpen
  \bibfield  {author} {\bibinfo {author} {\bibfnamefont {P.~G.}\ \bibnamefont
  {Kevrekidis}}, \bibinfo {author} {\bibfnamefont {R.}~\bibnamefont
  {Carretero-Gonz{\'a}lez}}, \bibinfo {author} {\bibfnamefont {D.~J.}\
  \bibnamefont {Frantzeskakis}}, \ and\ \bibinfo {author} {\bibfnamefont
  {I.~G.}\ \bibnamefont {Kevrekidis}},\ }\href {\doibase
  10.1142/S0217984904007967} {\bibfield  {journal} {\bibinfo  {journal} {Mod.
  Phys. Lett. B.}\ }\textbf {\bibinfo {volume} {18}},\ \bibinfo {pages} {1481}
  (\bibinfo {year} {2004})}\BibitemShut {NoStop}%
\bibitem [{\citenamefont {Kevrekidis}\ and\ \citenamefont
  {Frantzeskakis}(2004)}]{KEVREKIDIS2004}%
  \BibitemOpen
  \bibfield  {author} {\bibinfo {author} {\bibfnamefont {P.~G.}\ \bibnamefont
  {Kevrekidis}}\ and\ \bibinfo {author} {\bibfnamefont {D.~J.}\ \bibnamefont
  {Frantzeskakis}},\ }\href {\doibase 10.1142/S0217984904006809} {\bibfield
  {journal} {\bibinfo  {journal} {Mod. Phys. Lett. B.}\ }\textbf {\bibinfo
  {volume} {18}},\ \bibinfo {pages} {173} (\bibinfo {year} {2004})}\BibitemShut
  {NoStop}%
\bibitem [{\citenamefont {Fetter}\ and\ \citenamefont
  {Svidzinsky}(2001)}]{Fetter_2001}%
  \BibitemOpen
  \bibfield  {author} {\bibinfo {author} {\bibfnamefont {A.~L.}\ \bibnamefont
  {Fetter}}\ and\ \bibinfo {author} {\bibfnamefont {A.~A.}\ \bibnamefont
  {Svidzinsky}},\ }\href {\doibase 10.1088/0953-8984/13/12/201} {\bibfield
  {journal} {\bibinfo  {journal} {J. Condens. Matter Phys.}\ }\textbf {\bibinfo
  {volume} {13}},\ \bibinfo {pages} {R135} (\bibinfo {year}
  {2001})}\BibitemShut {NoStop}%
\bibitem [{\citenamefont {Law}\ \emph {et~al.}(2010)\citenamefont {Law},
  \citenamefont {Kevrekidis},\ and\ \citenamefont {Tuckerman}}]{Law2010}%
  \BibitemOpen
  \bibfield  {author} {\bibinfo {author} {\bibfnamefont {K.~J.~H.}\
  \bibnamefont {Law}}, \bibinfo {author} {\bibfnamefont {P.~G.}\ \bibnamefont
  {Kevrekidis}}, \ and\ \bibinfo {author} {\bibfnamefont {L.~S.}\ \bibnamefont
  {Tuckerman}},\ }\href {\doibase 10.1103/PhysRevLett.105.160405} {\bibfield
  {journal} {\bibinfo  {journal} {Phys. Rev. Lett.}\ }\textbf {\bibinfo
  {volume} {105}},\ \bibinfo {pages} {160405} (\bibinfo {year}
  {2010})}\BibitemShut {NoStop}%
\bibitem [{\citenamefont {Allen}\ \emph {et~al.}(2014)\citenamefont {Allen},
  \citenamefont {Parker}, \citenamefont {Proukakis},\ and\ \citenamefont
  {Barenghi}}]{Allen_2014}%
  \BibitemOpen
  \bibfield  {author} {\bibinfo {author} {\bibfnamefont {A.~J.}\ \bibnamefont
  {Allen}}, \bibinfo {author} {\bibfnamefont {N.~G.}\ \bibnamefont {Parker}},
  \bibinfo {author} {\bibfnamefont {N.~P.}\ \bibnamefont {Proukakis}}, \ and\
  \bibinfo {author} {\bibfnamefont {C.~F.}\ \bibnamefont {Barenghi}},\ }\href
  {\doibase 10.1088/1742-6596/544/1/012023} {\bibfield  {journal} {\bibinfo
  {journal} {J. Phys. Conf. Ser.}\ }\textbf {\bibinfo {volume} {544}},\
  \bibinfo {pages} {012023} (\bibinfo {year} {2014})}\BibitemShut {NoStop}%
\bibitem [{\citenamefont {Tsatsos}\ \emph {et~al.}(2016)\citenamefont
  {Tsatsos}, \citenamefont {Tavares}, \citenamefont {Cidrim}, \citenamefont
  {Fritsch}, \citenamefont {Caracanhas}, \citenamefont {{dos Santos}},
  \citenamefont {Barenghi},\ and\ \citenamefont {Bagnato}}]{TSATSOS20161}%
  \BibitemOpen
  \bibfield  {author} {\bibinfo {author} {\bibfnamefont {M.~C.}\ \bibnamefont
  {Tsatsos}}, \bibinfo {author} {\bibfnamefont {P.~E.}\ \bibnamefont
  {Tavares}}, \bibinfo {author} {\bibfnamefont {A.}~\bibnamefont {Cidrim}},
  \bibinfo {author} {\bibfnamefont {A.~R.}\ \bibnamefont {Fritsch}}, \bibinfo
  {author} {\bibfnamefont {M.~A.}\ \bibnamefont {Caracanhas}}, \bibinfo
  {author} {\bibfnamefont {F.~E.~A.}\ \bibnamefont {{dos Santos}}}, \bibinfo
  {author} {\bibfnamefont {C.~F.}\ \bibnamefont {Barenghi}}, \ and\ \bibinfo
  {author} {\bibfnamefont {V.~S.}\ \bibnamefont {Bagnato}},\ }\href {\doibase
  https://doi.org/10.1016/j.physrep.2016.02.003} {\bibfield  {journal}
  {\bibinfo  {journal} {Phys. Rep.}\ }\textbf {\bibinfo {volume} {622}},\
  \bibinfo {pages} {1 } (\bibinfo {year} {2016})}\BibitemShut {NoStop}%
\bibitem [{\citenamefont {White}\ \emph {et~al.}(2012)\citenamefont {White},
  \citenamefont {Barenghi},\ and\ \citenamefont {Proukakis}}]{White2012}%
  \BibitemOpen
  \bibfield  {author} {\bibinfo {author} {\bibfnamefont {A.~C.}\ \bibnamefont
  {White}}, \bibinfo {author} {\bibfnamefont {C.~F.}\ \bibnamefont {Barenghi}},
  \ and\ \bibinfo {author} {\bibfnamefont {N.~P.}\ \bibnamefont {Proukakis}},\
  }\href {\doibase 10.1103/PhysRevA.86.013635} {\bibfield  {journal} {\bibinfo
  {journal} {Phys. Rev. A}\ }\textbf {\bibinfo {volume} {86}},\ \bibinfo
  {pages} {013635} (\bibinfo {year} {2012})}\BibitemShut {NoStop}%
\bibitem [{\citenamefont {White}\ \emph {et~al.}(2014)\citenamefont {White},
  \citenamefont {Anderson},\ and\ \citenamefont {Bagnato}}]{White4719}%
  \BibitemOpen
  \bibfield  {author} {\bibinfo {author} {\bibfnamefont {A.~C.}\ \bibnamefont
  {White}}, \bibinfo {author} {\bibfnamefont {B.~P.}\ \bibnamefont {Anderson}},
  \ and\ \bibinfo {author} {\bibfnamefont {V.~S.}\ \bibnamefont {Bagnato}},\
  }\href {\doibase 10.1073/pnas.1312737110} {\bibfield  {journal} {\bibinfo
  {journal} {Proc. Natl. Acad. Sci. U.S.A.}\ }\textbf {\bibinfo {volume}
  {111}},\ \bibinfo {pages} {4719} (\bibinfo {year} {2014})}\BibitemShut
  {NoStop}%
\bibitem [{\citenamefont {Madeira}\ \emph {et~al.}(2020)\citenamefont
  {Madeira}, \citenamefont {Cidrim}, \citenamefont {Hemmerling}, \citenamefont
  {Caracanhas}, \citenamefont {dos Santos},\ and\ \citenamefont
  {Bagnato}}]{Madeira2020}%
  \BibitemOpen
  \bibfield  {author} {\bibinfo {author} {\bibfnamefont {L.}~\bibnamefont
  {Madeira}}, \bibinfo {author} {\bibfnamefont {A.}~\bibnamefont {Cidrim}},
  \bibinfo {author} {\bibfnamefont {M.}~\bibnamefont {Hemmerling}}, \bibinfo
  {author} {\bibfnamefont {M.~A.}\ \bibnamefont {Caracanhas}}, \bibinfo
  {author} {\bibfnamefont {F.~E.~A.}\ \bibnamefont {dos Santos}}, \ and\
  \bibinfo {author} {\bibfnamefont {V.~S.}\ \bibnamefont {Bagnato}},\ }\href
  {\doibase 10.1116/5.0016751} {\bibfield  {journal} {\bibinfo  {journal} {AVS
  Quantum Sci.}\ }\textbf {\bibinfo {volume} {2}},\ \bibinfo {pages} {035901}
  (\bibinfo {year} {2020})}\BibitemShut {NoStop}%
\bibitem [{\citenamefont {Kraichnan}\ and\ \citenamefont
  {Montgomery}(1980)}]{Kraichnan_1980}%
  \BibitemOpen
  \bibfield  {author} {\bibinfo {author} {\bibfnamefont {R.~H.}\ \bibnamefont
  {Kraichnan}}\ and\ \bibinfo {author} {\bibfnamefont {D.}~\bibnamefont
  {Montgomery}},\ }\href {\doibase 10.1088/0034-4885/43/5/001} {\bibfield
  {journal} {\bibinfo  {journal} {Rep. Prog. Phys.}\ }\textbf {\bibinfo
  {volume} {43}},\ \bibinfo {pages} {547} (\bibinfo {year} {1980})}\BibitemShut
  {NoStop}%
\bibitem [{\citenamefont {Navon}\ \emph {et~al.}(2016)\citenamefont {Navon},
  \citenamefont {Gaunt}, \citenamefont {Smith},\ and\ \citenamefont
  {Hadzibabic}}]{Navon2016}%
  \BibitemOpen
  \bibfield  {author} {\bibinfo {author} {\bibfnamefont {N.}~\bibnamefont
  {Navon}}, \bibinfo {author} {\bibfnamefont {A.~L.}\ \bibnamefont {Gaunt}},
  \bibinfo {author} {\bibfnamefont {R.~P.}\ \bibnamefont {Smith}}, \ and\
  \bibinfo {author} {\bibfnamefont {Z.}~\bibnamefont {Hadzibabic}},\ }\href
  {\doibase 10.1038/nature20114} {\bibfield  {journal} {\bibinfo  {journal}
  {Nature}\ }\textbf {\bibinfo {volume} {539}},\ \bibinfo {pages} {72}
  (\bibinfo {year} {2016})}\BibitemShut {NoStop}%
\bibitem [{\citenamefont {Navon}\ \emph {et~al.}(2019)\citenamefont {Navon},
  \citenamefont {Eigen}, \citenamefont {Zhang}, \citenamefont {Lopes},
  \citenamefont {Gaunt}, \citenamefont {Fujimoto}, \citenamefont {Tsubota},
  \citenamefont {Smith},\ and\ \citenamefont {Hadzibabic}}]{Navon2019}%
  \BibitemOpen
  \bibfield  {author} {\bibinfo {author} {\bibfnamefont {N.}~\bibnamefont
  {Navon}}, \bibinfo {author} {\bibfnamefont {C.}~\bibnamefont {Eigen}},
  \bibinfo {author} {\bibfnamefont {J.}~\bibnamefont {Zhang}}, \bibinfo
  {author} {\bibfnamefont {R.}~\bibnamefont {Lopes}}, \bibinfo {author}
  {\bibfnamefont {A.~L.}\ \bibnamefont {Gaunt}}, \bibinfo {author}
  {\bibfnamefont {K.}~\bibnamefont {Fujimoto}}, \bibinfo {author}
  {\bibfnamefont {M.}~\bibnamefont {Tsubota}}, \bibinfo {author} {\bibfnamefont
  {R.~P.}\ \bibnamefont {Smith}}, \ and\ \bibinfo {author} {\bibfnamefont
  {Z.}~\bibnamefont {Hadzibabic}},\ }\href {\doibase 10.1126/science.aau6103}
  {\bibfield  {journal} {\bibinfo  {journal} {Science}\ }\textbf {\bibinfo
  {volume} {366}},\ \bibinfo {pages} {382} (\bibinfo {year}
  {2019})}\BibitemShut {NoStop}%
\bibitem [{\citenamefont {Horng}\ \emph {et~al.}(2009)\citenamefont {Horng},
  \citenamefont {Hsueh}, \citenamefont {Su}, \citenamefont {Kao},\ and\
  \citenamefont {Gou}}]{Horng2009}%
  \BibitemOpen
  \bibfield  {author} {\bibinfo {author} {\bibfnamefont {T.-L.}\ \bibnamefont
  {Horng}}, \bibinfo {author} {\bibfnamefont {C.-H.}\ \bibnamefont {Hsueh}},
  \bibinfo {author} {\bibfnamefont {S.-W.}\ \bibnamefont {Su}}, \bibinfo
  {author} {\bibfnamefont {Y.-M.}\ \bibnamefont {Kao}}, \ and\ \bibinfo
  {author} {\bibfnamefont {S.-C.}\ \bibnamefont {Gou}},\ }\href {\doibase
  10.1103/PhysRevA.80.023618} {\bibfield  {journal} {\bibinfo  {journal} {Phys.
  Rev. A}\ }\textbf {\bibinfo {volume} {80}},\ \bibinfo {pages} {023618}
  (\bibinfo {year} {2009})}\BibitemShut {NoStop}%
\bibitem [{\citenamefont {LeBlanc}\ and\ \citenamefont
  {Thywissen}(2007)}]{LeBlanc2007}%
  \BibitemOpen
  \bibfield  {author} {\bibinfo {author} {\bibfnamefont {L.~J.}\ \bibnamefont
  {LeBlanc}}\ and\ \bibinfo {author} {\bibfnamefont {J.~H.}\ \bibnamefont
  {Thywissen}},\ }\href {\doibase 10.1103/PhysRevA.75.053612} {\bibfield
  {journal} {\bibinfo  {journal} {Phys. Rev. A}\ }\textbf {\bibinfo {volume}
  {75}},\ \bibinfo {pages} {053612} (\bibinfo {year} {2007})}\BibitemShut
  {NoStop}%
\bibitem [{\citenamefont {Jesacher}\ \emph {et~al.}(2004)\citenamefont
  {Jesacher}, \citenamefont {F\"{u}rhapter}, \citenamefont {Bernet},\ and\
  \citenamefont {Ritsch-Marte}}]{Alexander2004}%
  \BibitemOpen
  \bibfield  {author} {\bibinfo {author} {\bibfnamefont {A.}~\bibnamefont
  {Jesacher}}, \bibinfo {author} {\bibfnamefont {S.}~\bibnamefont
  {F\"{u}rhapter}}, \bibinfo {author} {\bibfnamefont {S.}~\bibnamefont
  {Bernet}}, \ and\ \bibinfo {author} {\bibfnamefont {M.}~\bibnamefont
  {Ritsch-Marte}},\ }\href {\doibase 10.1364/OPEX.12.004129} {\bibfield
  {journal} {\bibinfo  {journal} {Opt. Express}\ }\textbf {\bibinfo {volume}
  {12}},\ \bibinfo {pages} {4129} (\bibinfo {year} {2004})}\BibitemShut
  {NoStop}%
\bibitem [{\citenamefont {Ao}\ and\ \citenamefont {Chui}(1998)}]{Ao1998}%
  \BibitemOpen
  \bibfield  {author} {\bibinfo {author} {\bibfnamefont {P.}~\bibnamefont
  {Ao}}\ and\ \bibinfo {author} {\bibfnamefont {S.~T.}\ \bibnamefont {Chui}},\
  }\href {\doibase 10.1103/PhysRevA.58.4836} {\bibfield  {journal} {\bibinfo
  {journal} {Phys. Rev. A}\ }\textbf {\bibinfo {volume} {58}},\ \bibinfo
  {pages} {4836} (\bibinfo {year} {1998})}\BibitemShut {NoStop}%
\bibitem [{\citenamefont {Kasamatsu}\ \emph {et~al.}(2005)\citenamefont
  {Kasamatsu}, \citenamefont {Tsubota},\ and\ \citenamefont
  {Ueda}}]{KASAMATSU2005}%
  \BibitemOpen
  \bibfield  {author} {\bibinfo {author} {\bibfnamefont {K.}~\bibnamefont
  {Kasamatsu}}, \bibinfo {author} {\bibfnamefont {M.}~\bibnamefont {Tsubota}},
  \ and\ \bibinfo {author} {\bibfnamefont {M.}~\bibnamefont {Ueda}},\ }\href
  {\doibase 10.1142/S0217979205029602} {\bibfield  {journal} {\bibinfo
  {journal} {Int. J. Mod. Phys. B}\ }\textbf {\bibinfo {volume} {19}},\
  \bibinfo {pages} {1835} (\bibinfo {year} {2005})}\BibitemShut {NoStop}%
\bibitem [{\citenamefont {Ville}\ \emph {et~al.}(2018)\citenamefont {Ville},
  \citenamefont {Saint-Jalm}, \citenamefont {Le~Cerf}, \citenamefont
  {Aidelsburger}, \citenamefont {Nascimb\`ene}, \citenamefont {Dalibard},\ and\
  \citenamefont {Beugnon}}]{Ville2018}%
  \BibitemOpen
  \bibfield  {author} {\bibinfo {author} {\bibfnamefont {J.~L.}\ \bibnamefont
  {Ville}}, \bibinfo {author} {\bibfnamefont {R.}~\bibnamefont {Saint-Jalm}},
  \bibinfo {author} {\bibfnamefont {E.}~\bibnamefont {Le~Cerf}}, \bibinfo
  {author} {\bibfnamefont {M.}~\bibnamefont {Aidelsburger}}, \bibinfo {author}
  {\bibfnamefont {S.}~\bibnamefont {Nascimb\`ene}}, \bibinfo {author}
  {\bibfnamefont {J.}~\bibnamefont {Dalibard}}, \ and\ \bibinfo {author}
  {\bibfnamefont {J.}~\bibnamefont {Beugnon}},\ }\href {\doibase
  10.1103/PhysRevLett.121.145301} {\bibfield  {journal} {\bibinfo  {journal}
  {Phys. Rev. Lett.}\ }\textbf {\bibinfo {volume} {121}},\ \bibinfo {pages}
  {145301} (\bibinfo {year} {2018})}\BibitemShut {NoStop}%
\bibitem [{\citenamefont {Chomaz}\ \emph {et~al.}(2015)\citenamefont {Chomaz},
  \citenamefont {Corman}, \citenamefont {Bienaim\'e}, \citenamefont
  {Desbuquois}, \citenamefont {Weitenberg}, \citenamefont {Nascimb\'ene},
  \citenamefont {Beugnon},\ and\ \citenamefont {Dalibard}}]{Chomaz2015}%
  \BibitemOpen
  \bibfield  {author} {\bibinfo {author} {\bibfnamefont {L.}~\bibnamefont
  {Chomaz}}, \bibinfo {author} {\bibfnamefont {L.}~\bibnamefont {Corman}},
  \bibinfo {author} {\bibfnamefont {T.}~\bibnamefont {Bienaim\'e}}, \bibinfo
  {author} {\bibfnamefont {R.}~\bibnamefont {Desbuquois}}, \bibinfo {author}
  {\bibfnamefont {C.}~\bibnamefont {Weitenberg}}, \bibinfo {author}
  {\bibfnamefont {S.}~\bibnamefont {Nascimb\'ene}}, \bibinfo {author}
  {\bibfnamefont {J.}~\bibnamefont {Beugnon}}, \ and\ \bibinfo {author}
  {\bibfnamefont {J.}~\bibnamefont {Dalibard}},\ }\href {\doibase
  10.1038/ncomms7162} {\bibfield  {journal} {\bibinfo  {journal} {Nat.
  Commun.}\ }\textbf {\bibinfo {volume} {6}},\ \bibinfo {pages} {6162}
  (\bibinfo {year} {2015})}\BibitemShut {NoStop}%
\bibitem [{\citenamefont {Ma}\ \emph {et~al.}(2010)\citenamefont {Ma},
  \citenamefont {Carretero-Gonz\'alez}, \citenamefont {Kevrekidis},
  \citenamefont {Frantzeskakis},\ and\ \citenamefont {Malomed}}]{Ma2010}%
  \BibitemOpen
  \bibfield  {author} {\bibinfo {author} {\bibfnamefont {M.}~\bibnamefont
  {Ma}}, \bibinfo {author} {\bibfnamefont {R.}~\bibnamefont
  {Carretero-Gonz\'alez}}, \bibinfo {author} {\bibfnamefont {P.~G.}\
  \bibnamefont {Kevrekidis}}, \bibinfo {author} {\bibfnamefont {D.~J.}\
  \bibnamefont {Frantzeskakis}}, \ and\ \bibinfo {author} {\bibfnamefont
  {B.~A.}\ \bibnamefont {Malomed}},\ }\href {\doibase
  10.1103/PhysRevA.82.023621} {\bibfield  {journal} {\bibinfo  {journal} {Phys.
  Rev. A}\ }\textbf {\bibinfo {volume} {82}},\ \bibinfo {pages} {023621}
  (\bibinfo {year} {2010})}\BibitemShut {NoStop}%
\bibitem [{\citenamefont {Nore}\ \emph
  {et~al.}(1997{\natexlab{a}})\citenamefont {Nore}, \citenamefont {Abid},\ and\
  \citenamefont {Brachet}}]{nore1997kolmogorov}%
  \BibitemOpen
  \bibfield  {author} {\bibinfo {author} {\bibfnamefont {C.}~\bibnamefont
  {Nore}}, \bibinfo {author} {\bibfnamefont {M.}~\bibnamefont {Abid}}, \ and\
  \bibinfo {author} {\bibfnamefont {M.}~\bibnamefont {Brachet}},\ }\href
  {\doibase 10.1103/PhysRevLett.78.3896} {\bibfield  {journal} {\bibinfo
  {journal} {Phys. Rev. Lett.}\ }\textbf {\bibinfo {volume} {78}},\ \bibinfo
  {pages} {3896} (\bibinfo {year} {1997}{\natexlab{a}})}\BibitemShut {NoStop}%
\bibitem [{\citenamefont {Saffman}(1971)}]{Saffman_1971}%
  \BibitemOpen
  \bibfield  {author} {\bibinfo {author} {\bibfnamefont {P.~G.}\ \bibnamefont
  {Saffman}},\ }\href {\doibase https://doi.org/10.1002/sapm1971504377}
  {\bibfield  {journal} {\bibinfo  {journal} {Stud. Appl. Math.}\ }\textbf
  {\bibinfo {volume} {50}},\ \bibinfo {pages} {377} (\bibinfo {year}
  {1971})}\BibitemShut {NoStop}%
\bibitem [{\citenamefont {McCarron}\ \emph {et~al.}(2011)\citenamefont
  {McCarron}, \citenamefont {Cho}, \citenamefont {Jenkin}, \citenamefont
  {K\"oppinger},\ and\ \citenamefont {Cornish}}]{McCarron2011}%
  \BibitemOpen
  \bibfield  {author} {\bibinfo {author} {\bibfnamefont {D.~J.}\ \bibnamefont
  {McCarron}}, \bibinfo {author} {\bibfnamefont {H.~W.}\ \bibnamefont {Cho}},
  \bibinfo {author} {\bibfnamefont {D.~L.}\ \bibnamefont {Jenkin}}, \bibinfo
  {author} {\bibfnamefont {M.~P.}\ \bibnamefont {K\"oppinger}}, \ and\ \bibinfo
  {author} {\bibfnamefont {S.~L.}\ \bibnamefont {Cornish}},\ }\href {\doibase
  10.1103/PhysRevA.84.011603} {\bibfield  {journal} {\bibinfo  {journal} {Phys.
  Rev. A}\ }\textbf {\bibinfo {volume} {84}},\ \bibinfo {pages} {011603}
  (\bibinfo {year} {2011})}\BibitemShut {NoStop}%
\bibitem [{\citenamefont {Zeng}\ \emph {et~al.}(1995)\citenamefont {Zeng},
  \citenamefont {Zhang},\ and\ \citenamefont {Lin}}]{Zeng1995}%
  \BibitemOpen
  \bibfield  {author} {\bibinfo {author} {\bibfnamefont {H.}~\bibnamefont
  {Zeng}}, \bibinfo {author} {\bibfnamefont {W.}~\bibnamefont {Zhang}}, \ and\
  \bibinfo {author} {\bibfnamefont {F.}~\bibnamefont {Lin}},\ }\href {\doibase
  10.1103/PhysRevA.52.2155} {\bibfield  {journal} {\bibinfo  {journal} {Phys.
  Rev. A}\ }\textbf {\bibinfo {volume} {52}},\ \bibinfo {pages} {2155}
  (\bibinfo {year} {1995})}\BibitemShut {NoStop}%
\bibitem [{\citenamefont {Pethick}\ and\ \citenamefont
  {Smith}(2008)}]{Pethick2008}%
  \BibitemOpen
  \bibfield  {author} {\bibinfo {author} {\bibfnamefont {C.}~\bibnamefont
  {Pethick}}\ and\ \bibinfo {author} {\bibfnamefont {H.}~\bibnamefont
  {Smith}},\ }\href@noop {} {\emph {\bibinfo {title} {Bose-Einstein
  Condensation of Dilute Gases}}}\ (\bibinfo  {publisher} {Cambridge University
  Press},\ \bibinfo {address} {Cambridge},\ \bibinfo {year} {2008})\BibitemShut
  {NoStop}%
\bibitem [{\citenamefont {Bandyopadhyay}\ \emph {et~al.}(2017)\citenamefont
  {Bandyopadhyay}, \citenamefont {Roy},\ and\ \citenamefont
  {Angom}}]{Soumik2017}%
  \BibitemOpen
  \bibfield  {author} {\bibinfo {author} {\bibfnamefont {S.}~\bibnamefont
  {Bandyopadhyay}}, \bibinfo {author} {\bibfnamefont {A.}~\bibnamefont {Roy}},
  \ and\ \bibinfo {author} {\bibfnamefont {D.}~\bibnamefont {Angom}},\ }\href
  {\doibase 10.1103/PhysRevA.96.043603} {\bibfield  {journal} {\bibinfo
  {journal} {Phys. Rev. A}\ }\textbf {\bibinfo {volume} {96}},\ \bibinfo
  {pages} {043603} (\bibinfo {year} {2017})}\BibitemShut {NoStop}%
\bibitem [{\citenamefont {Muruganandam}\ and\ \citenamefont
  {Adhikari}(2009)}]{muruganandam2009fortran}%
  \BibitemOpen
  \bibfield  {author} {\bibinfo {author} {\bibfnamefont {P.}~\bibnamefont
  {Muruganandam}}\ and\ \bibinfo {author} {\bibfnamefont {S.~K.}\ \bibnamefont
  {Adhikari}},\ }\href {\doibase 10.1016/j.cpc.2009.04.015} {\bibfield
  {journal} {\bibinfo  {journal} {Comput. Phys. Commun.}\ }\textbf {\bibinfo
  {volume} {180}},\ \bibinfo {pages} {1888} (\bibinfo {year}
  {2009})}\BibitemShut {NoStop}%
\bibitem [{\citenamefont {Madelung}(1927)}]{madelung1927quantentheorie}%
  \BibitemOpen
  \bibfield  {author} {\bibinfo {author} {\bibfnamefont {E.}~\bibnamefont
  {Madelung}},\ }\href {\doibase https://doi.org/10.1007/BF01400372} {\bibfield
   {journal} {\bibinfo  {journal} {Z. Phys.}\ }\textbf {\bibinfo {volume}
  {40}},\ \bibinfo {pages} {322} (\bibinfo {year} {1927})}\BibitemShut
  {NoStop}%
\bibitem [{\citenamefont {Nore}\ \emph
  {et~al.}(1997{\natexlab{b}})\citenamefont {Nore}, \citenamefont {Abid},\ and\
  \citenamefont {Brachet}}]{Nore1997}%
  \BibitemOpen
  \bibfield  {author} {\bibinfo {author} {\bibfnamefont {C.}~\bibnamefont
  {Nore}}, \bibinfo {author} {\bibfnamefont {M.}~\bibnamefont {Abid}}, \ and\
  \bibinfo {author} {\bibfnamefont {M.~E.}\ \bibnamefont {Brachet}},\ }\href
  {\doibase 10.1103/PhysRevLett.78.3896} {\bibfield  {journal} {\bibinfo
  {journal} {Phys. Rev. Lett.}\ }\textbf {\bibinfo {volume} {78}},\ \bibinfo
  {pages} {3896} (\bibinfo {year} {1997}{\natexlab{b}})}\BibitemShut {NoStop}%
\bibitem [{\citenamefont {Mithun}\ \emph {et~al.}(2021)\citenamefont {Mithun},
  \citenamefont {Kasamatsu}, \citenamefont {Dey},\ and\ \citenamefont
  {Kevrekidis}}]{Mithun2021}%
  \BibitemOpen
  \bibfield  {author} {\bibinfo {author} {\bibfnamefont {T.}~\bibnamefont
  {Mithun}}, \bibinfo {author} {\bibfnamefont {K.}~\bibnamefont {Kasamatsu}},
  \bibinfo {author} {\bibfnamefont {B.}~\bibnamefont {Dey}}, \ and\ \bibinfo
  {author} {\bibfnamefont {P.~G.}\ \bibnamefont {Kevrekidis}},\ }\href
  {\doibase 10.1103/PhysRevA.103.023301} {\bibfield  {journal} {\bibinfo
  {journal} {Phys. Rev. A}\ }\textbf {\bibinfo {volume} {103}},\ \bibinfo
  {pages} {023301} (\bibinfo {year} {2021})}\BibitemShut {NoStop}%
\bibitem [{\citenamefont {Reeves}\ \emph {et~al.}(2012)\citenamefont {Reeves},
  \citenamefont {Anderson},\ and\ \citenamefont {Bradley}}]{Reeves2012}%
  \BibitemOpen
  \bibfield  {author} {\bibinfo {author} {\bibfnamefont {M.~T.}\ \bibnamefont
  {Reeves}}, \bibinfo {author} {\bibfnamefont {B.~P.}\ \bibnamefont
  {Anderson}}, \ and\ \bibinfo {author} {\bibfnamefont {A.~S.}\ \bibnamefont
  {Bradley}},\ }\href {\doibase 10.1103/PhysRevA.86.053621} {\bibfield
  {journal} {\bibinfo  {journal} {Phys. Rev. A}\ }\textbf {\bibinfo {volume}
  {86}},\ \bibinfo {pages} {053621} (\bibinfo {year} {2012})}\BibitemShut
  {NoStop}%
\bibitem [{\citenamefont {{Kraichnan}}(1967)}]{Kraichnan1967}%
  \BibitemOpen
  \bibfield  {author} {\bibinfo {author} {\bibfnamefont {R.~H.}\ \bibnamefont
  {{Kraichnan}}},\ }\href {\doibase 10.1063/1.1762301} {\bibfield  {journal}
  {\bibinfo  {journal} {Phys. Fluids}\ }\textbf {\bibinfo {volume} {10}},\
  \bibinfo {pages} {1417} (\bibinfo {year} {1967})}\BibitemShut {NoStop}%
\bibitem [{\citenamefont {Kraichnan}(1975)}]{kraichnan_1975}%
  \BibitemOpen
  \bibfield  {author} {\bibinfo {author} {\bibfnamefont {R.~H.}\ \bibnamefont
  {Kraichnan}},\ }\href {\doibase 10.1017/S0022112075000225} {\bibfield
  {journal} {\bibinfo  {journal} {J. Fluid Mech.}\ }\textbf {\bibinfo {volume}
  {67}},\ \bibinfo {pages} {155–175} (\bibinfo {year} {1975})}\BibitemShut
  {NoStop}%
\bibitem [{\citenamefont {Sreenivasan}(1999)}]{Sreenivasan1999}%
  \BibitemOpen
  \bibfield  {author} {\bibinfo {author} {\bibfnamefont {K.~R.}\ \bibnamefont
  {Sreenivasan}},\ }\href {\doibase 10.1103/RevModPhys.71.S383} {\bibfield
  {journal} {\bibinfo  {journal} {Rev. Mod. Phys.}\ }\textbf {\bibinfo {volume}
  {71}},\ \bibinfo {pages} {S383} (\bibinfo {year} {1999})}\BibitemShut
  {NoStop}%
\bibitem [{\citenamefont {Batchelor}(1969)}]{Batchelor1969}%
  \BibitemOpen
  \bibfield  {author} {\bibinfo {author} {\bibfnamefont {G.~K.}\ \bibnamefont
  {Batchelor}},\ }\href {\doibase 10.1063/1.1692443} {\bibfield  {journal}
  {\bibinfo  {journal} {Phys. Fluids}\ }\textbf {\bibinfo {volume} {12}},\
  \bibinfo {pages} {II} (\bibinfo {year} {1969})}\BibitemShut {NoStop}%
\bibitem [{\citenamefont {Parker}\ and\ \citenamefont
  {Adams}(2005)}]{Parker2005}%
  \BibitemOpen
  \bibfield  {author} {\bibinfo {author} {\bibfnamefont {N.~G.}\ \bibnamefont
  {Parker}}\ and\ \bibinfo {author} {\bibfnamefont {C.~S.}\ \bibnamefont
  {Adams}},\ }\href {\doibase 10.1103/PhysRevLett.95.145301} {\bibfield
  {journal} {\bibinfo  {journal} {Phys. Rev. Lett.}\ }\textbf {\bibinfo
  {volume} {95}},\ \bibinfo {pages} {145301} (\bibinfo {year}
  {2005})}\BibitemShut {NoStop}%
\bibitem [{\citenamefont {Numasato}\ \emph {et~al.}(2010)\citenamefont
  {Numasato}, \citenamefont {Tsubota},\ and\ \citenamefont
  {L'vov}}]{Numasato2010}%
  \BibitemOpen
  \bibfield  {author} {\bibinfo {author} {\bibfnamefont {R.}~\bibnamefont
  {Numasato}}, \bibinfo {author} {\bibfnamefont {M.}~\bibnamefont {Tsubota}}, \
  and\ \bibinfo {author} {\bibfnamefont {V.~S.}\ \bibnamefont {L'vov}},\ }\href
  {\doibase 10.1103/PhysRevA.81.063630} {\bibfield  {journal} {\bibinfo
  {journal} {Phys. Rev. A}\ }\textbf {\bibinfo {volume} {81}},\ \bibinfo
  {pages} {063630} (\bibinfo {year} {2010})}\BibitemShut {NoStop}%
\bibitem [{\citenamefont {Reeves}\ \emph {et~al.}(2017)\citenamefont {Reeves},
  \citenamefont {Billam}, \citenamefont {Yu},\ and\ \citenamefont
  {Bradley}}]{Reeves2017}%
  \BibitemOpen
  \bibfield  {author} {\bibinfo {author} {\bibfnamefont {M.~T.}\ \bibnamefont
  {Reeves}}, \bibinfo {author} {\bibfnamefont {T.~P.}\ \bibnamefont {Billam}},
  \bibinfo {author} {\bibfnamefont {X.}~\bibnamefont {Yu}}, \ and\ \bibinfo
  {author} {\bibfnamefont {A.~S.}\ \bibnamefont {Bradley}},\ }\href {\doibase
  10.1103/PhysRevLett.119.184502} {\bibfield  {journal} {\bibinfo  {journal}
  {Phys. Rev. Lett.}\ }\textbf {\bibinfo {volume} {119}},\ \bibinfo {pages}
  {184502} (\bibinfo {year} {2017})}\BibitemShut {NoStop}%
\bibitem [{\citenamefont {Brachet}\ \emph {et~al.}(1988)\citenamefont
  {Brachet}, \citenamefont {Meneguzzi}, \citenamefont {Politano},\ and\
  \citenamefont {Sulem}}]{brachet1988}%
  \BibitemOpen
  \bibfield  {author} {\bibinfo {author} {\bibfnamefont {M.~E.}\ \bibnamefont
  {Brachet}}, \bibinfo {author} {\bibfnamefont {M.}~\bibnamefont {Meneguzzi}},
  \bibinfo {author} {\bibfnamefont {H.}~\bibnamefont {Politano}}, \ and\
  \bibinfo {author} {\bibfnamefont {P.~L.}\ \bibnamefont {Sulem}},\ }\href
  {\doibase 10.1017/S0022112088003015} {\bibfield  {journal} {\bibinfo
  {journal} {J. Fluid Mech.}\ }\textbf {\bibinfo {volume} {194}},\ \bibinfo
  {pages} {333–349} (\bibinfo {year} {1988})}\BibitemShut {NoStop}%
\bibitem [{\citenamefont {Kovalev}\ \emph {et~al.}(2016)\citenamefont
  {Kovalev}, \citenamefont {Kotlyar},\ and\ \citenamefont
  {Porfirev}}]{Kovalev2016}%
  \BibitemOpen
  \bibfield  {author} {\bibinfo {author} {\bibfnamefont {A.~A.}\ \bibnamefont
  {Kovalev}}, \bibinfo {author} {\bibfnamefont {V.~V.}\ \bibnamefont
  {Kotlyar}}, \ and\ \bibinfo {author} {\bibfnamefont {A.~P.}\ \bibnamefont
  {Porfirev}},\ }\href {\doibase 10.1103/PhysRevA.93.063858} {\bibfield
  {journal} {\bibinfo  {journal} {Phys. Rev. A}\ }\textbf {\bibinfo {volume}
  {93}},\ \bibinfo {pages} {063858} (\bibinfo {year} {2016})}\BibitemShut
  {NoStop}%
\bibitem [{\citenamefont {Proukakis}\ and\ \citenamefont
  {Jackson}(2008)}]{Proukakis_2008}%
  \BibitemOpen
  \bibfield  {author} {\bibinfo {author} {\bibfnamefont {N.~P.}\ \bibnamefont
  {Proukakis}}\ and\ \bibinfo {author} {\bibfnamefont {B.}~\bibnamefont
  {Jackson}},\ }\href {\doibase 10.1088/0953-4075/41/20/203002} {\bibfield
  {journal} {\bibinfo  {journal} {J. Phys. B: At. Mol. and Opt. Phys.}\
  }\textbf {\bibinfo {volume} {41}},\ \bibinfo {pages} {203002} (\bibinfo
  {year} {2008})}\BibitemShut {NoStop}%
\bibitem [{\citenamefont {Aikawa}\ \emph {et~al.}(2012)\citenamefont {Aikawa},
  \citenamefont {Frisch}, \citenamefont {Mark}, \citenamefont {Baier},
  \citenamefont {Rietzler}, \citenamefont {Grimm},\ and\ \citenamefont
  {Ferlaino}}]{Aikawa2012}%
  \BibitemOpen
  \bibfield  {author} {\bibinfo {author} {\bibfnamefont {K.}~\bibnamefont
  {Aikawa}}, \bibinfo {author} {\bibfnamefont {A.}~\bibnamefont {Frisch}},
  \bibinfo {author} {\bibfnamefont {M.}~\bibnamefont {Mark}}, \bibinfo {author}
  {\bibfnamefont {S.}~\bibnamefont {Baier}}, \bibinfo {author} {\bibfnamefont
  {A.}~\bibnamefont {Rietzler}}, \bibinfo {author} {\bibfnamefont
  {R.}~\bibnamefont {Grimm}}, \ and\ \bibinfo {author} {\bibfnamefont
  {F.}~\bibnamefont {Ferlaino}},\ }\href {\doibase
  10.1103/PhysRevLett.108.210401} {\bibfield  {journal} {\bibinfo  {journal}
  {Phys. Rev. Lett.}\ }\textbf {\bibinfo {volume} {108}},\ \bibinfo {pages}
  {210401} (\bibinfo {year} {2012})}\BibitemShut {NoStop}%
\bibitem [{\citenamefont {Lahaye}\ \emph {et~al.}(2009)\citenamefont {Lahaye},
  \citenamefont {Menotti}, \citenamefont {Santos}, \citenamefont {Lewenstein},\
  and\ \citenamefont {Pfau}}]{Lahaye2009}%
  \BibitemOpen
  \bibfield  {author} {\bibinfo {author} {\bibfnamefont {T.}~\bibnamefont
  {Lahaye}}, \bibinfo {author} {\bibfnamefont {C.}~\bibnamefont {Menotti}},
  \bibinfo {author} {\bibfnamefont {L.}~\bibnamefont {Santos}}, \bibinfo
  {author} {\bibfnamefont {M.}~\bibnamefont {Lewenstein}}, \ and\ \bibinfo
  {author} {\bibfnamefont {T.}~\bibnamefont {Pfau}},\ }\href {\doibase
  10.1088/0034-4885/72/12/126401} {\bibfield  {journal} {\bibinfo  {journal}
  {Rep. Prog. Phys.}\ }\textbf {\bibinfo {volume} {72}},\ \bibinfo {pages}
  {126401} (\bibinfo {year} {2009})}\BibitemShut {NoStop}%
\bibitem [{\citenamefont {Cao}\ \emph {et~al.}(2017)\citenamefont {Cao},
  \citenamefont {Bolsinger}, \citenamefont {Mistakidis}, \citenamefont
  {Koutentakis}, \citenamefont {Kr{\"o}nke}, \citenamefont {Schurer},\ and\
  \citenamefont {Schmelcher}}]{Cao2017}%
  \BibitemOpen
  \bibfield  {author} {\bibinfo {author} {\bibfnamefont {L.}~\bibnamefont
  {Cao}}, \bibinfo {author} {\bibfnamefont {V.}~\bibnamefont {Bolsinger}},
  \bibinfo {author} {\bibfnamefont {S.}~\bibnamefont {Mistakidis}}, \bibinfo
  {author} {\bibfnamefont {G.}~\bibnamefont {Koutentakis}}, \bibinfo {author}
  {\bibfnamefont {S.}~\bibnamefont {Kr{\"o}nke}}, \bibinfo {author}
  {\bibfnamefont {J.}~\bibnamefont {Schurer}}, \ and\ \bibinfo {author}
  {\bibfnamefont {P.}~\bibnamefont {Schmelcher}},\ }\href {\doibase
  10.1063/1.4993512} {\bibfield  {journal} {\bibinfo  {journal} {J. Chem.
  Phys.}\ }\textbf {\bibinfo {volume} {147}},\ \bibinfo {pages} {044106}
  (\bibinfo {year} {2017})}\BibitemShut {NoStop}%
\end{thebibliography}%
\end{document}